\documentstyle[12pt]{article}
\begin{document}

\begin{center}
{\Large
POINCARE INVARIANCE OF HAMILTONIAN SEMICLASSICAL FIELD THEORY
}
\\
{{\large  O.Yu.Shvedov} \\
{\it
Sub-Dept. of Quantum Statistics and Field Theory},\\
{\it Dept. of Physics, Moscow State University},\\
{\it 119899, Moscow, Vorobievy Gory, Russia}
}
\end{center}

\def\qp{
\mathrel{\mathop{\bf x}\limits^2},
\mathrel{\mathop{-i\frac{\partial}{\partial {\bf x}}}\limits^1} 
}

\setcounter{page}{0}

\begin{flushright}
hep-th/0103076
\end{flushright}

\section*{Abstract}


Semiclassical Hamiltonian   field  theory  is  investigated  from  the
axiomatic point  of  view.  A  notion  of  a  semiclassical  state  is
introduced.  An "elementary" semiclassical state is specified by a set
of classical field configuration and quantum state  in  this  external
field. "Composed" semiclassical states viewed as formal superpositions
of "elementary" states are nontrivial only  if  the  Maslov  isotropic
condition is satisfied;  the inner product of "composed" semiclassical
states is degenerate. The mathematical proof of Poincare invariance of
semiclassical field theory is obtained for "elementary" and "composed"
semiclassical states. The notion of semiclassical field is introduced;
its Poincare invariance is also mathematically proved.


\footnotetext{e-mail:  shvedov@qs.phys.msu.su}

\newcounter{eqn}[section]
\renewcommand{\theeqn}{\thesection.\arabic{eqn}}
\def\lab{\refstepcounter{eqn}\eqno(\thesection.\arabic{eqn})}
\def\l#1{\lab\label{#1}}
\def\r#1{(\ref{#1})}
\def\c#1{\cite{#1}}
\def\i#1{\bibitem{#1}}

\newpage

\section{Introduction}

Different approaches to semiclassical  field  theory  have been
developed.
Most of them were based on the functional integral technique: physical
quantities were  expressed  via  functional   integrals   which   were
evaluated with the help of saddle-point or stationary-phase technique.
Since energy spectrum and $S$-matrix elements can be  found  from  the
functional integral \c{DHN,R}, this approach appeared to be useful for
the soliton quantization theory \c{DHN,R,J2,J,FK}.

Another important partial case of the semiclassical  field  theory  is
the theory  of  quantization in a strong external background classical
field \c{GMM} or in  curved  space-time  \c{BD}:  one  decomposes  the
field as  a  sum  of  a  classical  c-number  component  and a quantum
component. Then the theory is quantized.

The one-loop  approximation  \c{B1,B2,B11,B12},   the   time-dependent
Hartree-Fock   approximation   \c{B1,B2,HF1,HF2}   and  the  Gaussian
approximation developed in  \c{G1,G2,G3,G4}  may  be  also  viewed  as
examples of applications of semiclassical conceptions.

On the  other  hand,  the axiomatic field theory \c{A1,A2,A3} tells us
that main objects of QFT are  states  and  observables.  The  Poincare
group is  represented  in the Hilbert state space,  so that evolution,
boosts and  other  Poincare  transformations  are  viewed  as  unitary
operators.

The purpose of this paper is to introduce the semiclassical analogs of
such QFT notions as states,  fields and Poincare transformations.  The
analogs of  Wightman  Poincare invariance and field
axioms for the semiclassical field theory are to
be formulated and checked.

Unfortunately, "exact"  QFT  is  mathematically  constructed   for   a
restricted class     of     models    only    (see,    for    example,
\c{H,GJ,Ar1,Ar2}). Therefore,  formal  approximate  methods  such   as
perturbation theory  seem  to  be  ways  to  quantize the field theory
rather than to construct approximations for the exact solutions of QFT
equations. The   conception   of   field   quantization   within   the
perturbation framework is popular \c{BS,SF}.  One can expect that  the
semiclassical approximation plays an analogous role.

To construct  the  semiclassical  formalism  based  on the notion of a
state, one should use the equation-of-motion formulation of QFT rather
than the   usual   $S$-matrix   formulation.  It  is  well-known  that
additional difficulties such as Stueckelberg divergences  \c{Stu}  and
problems associated with the
Haag theorem  \c{Haag,A2,A3} arise in the equation-of-motion approach.
There are some ways to overcome them.  The vacuum divergences  can  be
eliminated in  the  perturbation  theory  with the help of the Faddeev
transformation \c{F}. Stueckelberg divergences can  be  treated
analogously \c{MS-F}   (exactly   solvable  models  with  Stueckelberg
divergences have   been   suggested   recently   \c{Sh1,Sh2}).   These
investigations are  important  for the semiclassical Hamiltonian field
theory \c{MS-FT}.

The semiclassical approaches are formally applicable  to  the  quantum
field theory  models if the Lagrangian depends on the fields $\varphi$
and the small parameter $\lambda$ as follows (see, for example, \c{J}):
$$
{\cal L} = \frac{1}{2} \partial_{\mu} \varphi \partial_{\mu} \varphi -
\frac{m^2}{2} \varphi^2  -  \frac{1}{\lambda}  V (\sqrt{\lambda}
\varphi),
\l{0}
$$
where $V$ is an interaction potential.
To illustrate  the  {\it  formal}  semiclassical  ansatz for the state
vector, use  the  functional  Schrodinger  representation  (see,   for
example, \c{HF1,HF2,G3,G4}). States at fixed
moment  of  time are represented as functionals $\Psi[\varphi(\cdot)]$
depending on fields $\varphi({\bf x})$,  ${\bf x} \in {\bf R}^d$,  the
field   operator   $\hat{\varphi}({\bf   x})$   is   the  operator  of
multiplication by $\varphi({\bf x})$, while the canonically conjugated
momentum  $\hat{\pi}({\bf  x})$  is  represented  as a differentiation
operator $-i\delta/\delta    \varphi({\bf    x})$.    The   functional
Schrodinger equation reads
$$
i\frac{d\psi^t}{dt} = {\cal H} \psi^t,
\l{1}
$$
where
$$
{\cal H} = \int d{\bf x} \left[
-\frac{1}{2} \frac{\delta^2}{\delta     \varphi({\bf     x})    \delta
\varphi({\bf x})}  +  \frac{1}{2}  (\nabla   \varphi)^2({\bf   x})   +
\frac{m^2}{2} \varphi^2({\bf x}) +
\frac{1}{\lambda} V(\sqrt{\lambda} \varphi({\bf x}))
\right]
$$
The simplest semiclassical state corresponds to the Maslov  theory  of
complex germ  in  a  point  \c{M1,M2,MS3}.  It  depends  on  the small
parameter $\lambda$ as
$$
\psi^t[\varphi(\cdot)] = e^{\frac{i}{\lambda}S^t} e^{\frac{i}{\lambda}
\int d{\bf  x}  \Pi^t({\bf  x})  [\varphi({\bf   x})\sqrt{\lambda}   -
\Phi^t({\bf x})]}                 f^t\left(\varphi(\cdot)                 -
\frac{\Phi^t(\cdot)}{\sqrt{\lambda}}\right)
\equiv (K_{S^t,\Pi^t,\Phi^t} f^t)[\varphi(\cdot)],
\l{2}
$$
where $S^t$,  $\Pi^t({\bf x})$, $\Phi^t({\bf x})$, $t\in {\bf R}$, ${\bf
x}\in {\bf R}^d$ are smooth real functions which rapidly damp with all
their derivatives as ${\bf x}\to\infty$,  $f^t[\phi(\cdot)]$ is a
$t$-dependent functional.

As $\lambda\to  0$,  the  substitution  \r{2} satisfies eq.\r{1}
in the leading order in  $\lambda$  if  the  following  relations  are
obeyed. First, for the "action" $S^t$ one finds,
$$
\frac{dS^t}{dt} = \int d{\bf x} [\Pi^t({\bf x}) \dot{\Phi}^t({\bf  x})
- \frac{1}{2}(\Pi^t({\bf  x}))^2  -  \frac{1}{2}  (\nabla  \Phi^t({\bf
x}))^2 - \frac{m^2}{2}  (\Phi^t({\bf  x}))^2  -  V(\Phi^t({\bf
x}))],
\l{3a}
$$
Second, $\Pi^t$, $\Phi^t$ obeys the classical Hamiltonian system
$$
\dot{\Phi}^t = \Pi^t,
- \dot{\Pi}^t = (-\Delta + m^2) \Phi^t + V'(\Phi^t),
\l{3}
$$
Finally, the functional
$f^t$ satisfies the functional Schrodinger equation with the quadratic
Hamiltonian
$$
i\dot{f}^t[\phi(\cdot)] = \int d{\bf x} \left[
- \frac{1}{2}    \frac{\delta^2}{\delta    \phi({\bf   x})   \delta
\phi({\bf x})}   +   \frac{1}{2}   (\nabla\phi({\bf   x}))^2   +
\frac{m^2}{2}\phi^2({\bf x})   +  \frac{1}{2}  V''(\Phi^t({\bf  x}))
\phi^2({\bf x})
\right] f^t[\phi(\cdot)].
\l{4}
$$
There are   more   complicated   semiclassical   states   that  also
approximately satisfy the functional Schrodinger equation \r{1}. These
ansatzes correspond  to the Maslov theory of Lagrangian manifolds with
complex germs \c{M1,M2,MS3}. They are discussed in section 5.

However, the QFT divergences lead to the following difficulties.

It is not evident how one should  specify  the  class  of  possible
functionals $f$  and  introduce  the inner product on such a space via
functional integral. This class was constructed in \c{MS-FT}.
In particular, it was found when the Gaussian functional
$$
f[\phi(\cdot)] =  const  \exp(\frac{i}{2}  \int   d{\bf   x}d{\bf   y}
\phi({\bf x}) \phi({\bf y}) {\cal R}({\bf x},{\bf y}))
\l{5}
$$
belongs to  this  class.  The condition on the quadratic form $\cal R$
which was obtained in \c{MS-FT} depends on $\Phi$, $\Pi$ and
differs from  the  analogous condition in the free theory.  This is in
agreement with  the  statement  of  \c{GM,Shir}   that   nonequivalent
representations of  the  canonical  commutation relations at different
moments of time should be considered if QFT  in  the  strong  external
field is investigated in the leading order in $\lambda$. However, this
does not lead to non-unitarity of the exact theory: the simple example
has been presented in \c{Sh2}.

Another problem  is  to formulate the semiclassical theory in terms of
the axiomatic field theory. Section 2 deals with formulation of axioms
of relativistic invariance and field for
the  semiclassical theory.  Section 3 is devoted to construction of
Poincare transformations.  In section 4 the  notion  of  semiclassical
field is  investigated.  More  complicated  semiclassical  states  are
constructed in section 5. Section 6 contains concluding remarks.

\section{Axioms of semiclassical field theory}

In the Wightman axiomatic approach the main object of QFT is a  notion
of a state space \c{A1,A2,A3}.  Formula \r{2} shows  us  that  in  the
semiclassical  field  theory a state at fixed moment of time should be
viewed as a set $(S,  \Pi(\cdot),  \Phi(\cdot),  f[\phi(\cdot)])$ of a
real number $S$, real functions $\Pi({\bf x})$, $\Phi({\bf x})$, ${\bf
x}\in {\bf R}^d$ and a functional $f[\phi(\cdot)]$  from  some  class.
This class depends on $\Pi$ and $\Phi$. Superposition of semiclassical
states $(S_1,\Pi_1,\Phi_1,f_1)$ and $(S_2,\Pi_2,\Phi_2,f_2)$ is of the
semiclassical  type  \r{2} if and only if $S_1=S_2$,  $\Phi_1=\Phi_2$,
$\Pi_1=\Pi_2$.

Thus, one  introduces  \c{Shv1,Shv2}  the  structure of a vector bundle
(called as a  "semiclassical  bundle"  in  \c{Shv2})  on  the  set  of
semiclassical states  of  the type \r{2}.  The base of the bundle being
a space of sets $(S,\Pi,\Phi)$   ("extended phase space"
\c{Shv1}) will be denoted as $\cal X$.
The fibers are classes of functionals which depend on
$\Phi$ and $\Pi$.  Making use of the result concerning  the  class  of
functionals \c{MS-FT}, one makes the bundle trivial as follows.
Consider the $\Phi$, $\Pi$- dependent mapping $V$
which defines a correspondence between functionals
$f$ and elements of the Fock space $\cal F$:
$$
V: \Psi \mapsto f,\qquad \Psi \in {\cal F}, \qquad f=f[\phi(\cdot)].
$$
as follows. Let  $\tilde{\cal  R}({\bf x},{\bf y})$ be an
$\Phi$, $\Pi$  -  dependent symmetric function such that its imaginary
part is a kernel of a positively definite operator and  the  condition
of ref. \c{MS-FT} (see eq.\r{p1} of subsection 3.6) is satisfied. .
By $\hat{\cal R}$ we denote the operator with kernel
$\tilde{\cal  R}$,  while
$\hat{\Gamma}$ has a kernel $i^{-1}(\tilde{\cal R} - \tilde{\cal R}^*)$.
The vacuum vector of  the  Fock  space  corresponds  to  the  Gaussian
functional \r{5}.  The  operator  $V$  is  uniquely  defined  from the
relations
$$
\begin{array}{c}
V^{-1} \phi({\bf x}) V = i(\hat{\Gamma}^{-1/2}(A^+-A^-))({\bf x}),\\
V^{-1} \frac{1}{i} \frac{\delta}{\delta\phi({\bf x})} V = i (\hat{\cal
R} \hat{\Gamma}^{-1/2} A^+ - \hat{\cal R}^*
\hat{\Gamma}^{-1/2} A^-) ({\bf x}).
\end{array}
\l{6*}
$$
Here $A^{\pm}({\bf x})$ are creation and annihilation operators in the
Fock space.

{\bf Definition 2.1.}  {\it A semiclassical state is a point
on the trivial bundle ${\cal X} \times {\cal F} \to {\cal
X}$.}

An important postulate of QFT is Poincare invariance.  This means that
a representation  of  the  Poincare group in the state space should be
specified. For each Poincare transformation of the form
$$
x'{}^{\mu} =  \Lambda^{\mu}_{\nu} x^{\nu} + a^{\mu},  \qquad \mu,\nu =
\overline{0,d}
\l{6**}
$$
which is   denoted  as  $(a,\Lambda)$,  the  unitary  operator  ${\cal
U}_{a,\Lambda}$ should be specified. The group property
$$
{\cal U}_{(a_1,\Lambda_1)} {\cal U}_{(a_2,\Lambda_2)}
= {\cal U}_{(a_1,\Lambda_1)(a_2,\Lambda_2)}
$$
with
$$
(a_1,\Lambda_1)(a_2,\Lambda_2) =          (a_1+          \Lambda_1a_2,
\Lambda_1\Lambda_2).
$$
should be satisfied.

Formulate an analog of the Poincare invariance
axiom for   the   semiclassical  theory.  Suppose  that  the  Poincare
transformation ${\cal U}_{a,\Lambda}$ takes  any  semiclassical  state
$(X,f)$ to   a  semiclassical  state  $(\tilde{X},\tilde{f})$  in  the
leading order    in    $\lambda^{1/2}$.    Denote    $\tilde{X}     =
u_{a,\Lambda}X$, $\tilde{f} = U(u_{a,\Lambda}X \gets X) f$.

{\bf Axiom 1 (Poincare invariance)}
\\
{\it
( ) the mappings $u_{a,\Lambda}: {\cal X} \to {\cal X}$ are specified,
the group properties for them
$u_{a_1,\Lambda_1}
u_{a_2,\Lambda_2} = u_{(a_1,\Lambda_1)(a_2,\Lambda_2)}$ are satisfied;\\
(¡) for all  $X\in   {\cal   X}$  the unitary operators
$U_{a,\Lambda} (u_{a,\Lambda}X  \gets  X):  {\cal  F}  \to  {\cal F}$,
obeying the group property
$$
\begin{array}{c}
U_{a_1,\Lambda_1} (u_{(a_1,\Lambda_1)(a_2,\Lambda_2)}X  \gets
u_{(a_2,\Lambda_2)}X)
U_{a_2,\Lambda_2} (u_{(a_2,\Lambda_2)}X  \gets  X)
=\\
U_{(a_1,\Lambda_1)(a_2,\Lambda_2)}
(u_{(a_1,\Lambda_1)(a_2,\Lambda_2)}X  \gets X)
\end{array}
$$
are specified.
}

An important  feature  of QFT is the notion of a field:  it is assumed
that an  operator  distribution  $\hat{\varphi}({\bf  x},t)$
is  specified.
Investigate  it  in  the  semiclassical theory.  Applying the operator
$\varphi({\bf x})$ to the semiclassical  state  \r{2},  we  obtain  an
analogous state:
$$
e^{\frac{i}{\lambda}S^t} e^{\frac{i}{\lambda}
\int d{\bf  x}  \Pi^t({\bf  x})  [\varphi({\bf   x})\sqrt{\lambda}   -
\Phi^t({\bf x})]}
\tilde{f}^t(\varphi(\cdot)                 -
\frac{\Phi^t(\cdot)}{\sqrt{\lambda}}),
$$
where
$$
\tilde{f}^t[\phi(\cdot)] = (\lambda^{-1/2} \Phi^t({\bf x}) + \phi({\bf
x}) ) f^t[\phi(\cdot)]
$$
As $\lambda\to 0$, one has
$$
\hat{\varphi}({\bf x},t) =
\lambda^{-1/2} \Phi^t({\bf x}) + \hat{\phi}({\bf x},t:X),
$$
where $\hat{\phi}({\bf  x},t:X)$  is a $\Pi,\Phi$-dependent operator
in ${\cal F}$,  $\Phi^t({\bf x})\equiv \Phi(x:X)$ is a solution to the
Cauchy problem for eq.\r{3}.
The  field  axiom  can be reformulated as
follows.

{\bf Axiom 2}.  {\it For each $X\in  {\cal  X}$  the
operator distribution  $\phi({\bf  x},t;X):  {\cal F} \to {\cal F}$ is
specified.}

An important  feature  of the relativistic quantum field theory is the
property of Poincare invariance of fields.  The operator  distribution
$\hat{\varphi}({\bf x},t)$ should obey the following property
$$
{\cal U}_{a,\Lambda}  \hat{\varphi}(x) = \hat{\varphi} (\Lambda x + a)
{\cal U}_{a,\Lambda}.
$$
Apply this identity to  a  semiclassical  state  $(X,f)$.  In  leading
orders in $\lambda^{1/2}$, one obtains:
$$
\begin{array}{c}
\lambda^{-1/2} \Phi(x:X)
(u_{a,\Lambda}X,  U_{a,\Lambda}(u_{a,\Lambda} X \gets X) f)
+
(u_{a,\Lambda}X,  U_{a,\Lambda}(u_{a,\Lambda} X \gets X) \hat{\phi}(x:X)
f)
= \\
\lambda^{-1/2} \Phi(\Lambda x+a: u_{a,\Lambda}X)
(u_{a,\Lambda}X,  U_{a,\Lambda}(u_{a,\Lambda} X \gets X) f)
\\
+
(u_{a,\Lambda}X,  \hat{\phi}(\Lambda x + a: u_{a,\Lambda}X)
U_{a,\Lambda}(u_{a,\Lambda} X \gets X) f).
\end{array}
$$
Therefore, we formulate the following axiom.

{\bf Axiom  3.  (Poincare  invariance  of fields).} {\it The following
properties are satisfied:
$$
\Phi(x:X) = \Phi(\Lambda x+a:u_{a,\Lambda}X);
\l{aa1}
$$
$$
\hat{\phi}(\Lambda x + a: u_{a,\Lambda}X)
U_{a,\Lambda}(u_{a,\Lambda} X \gets X)
= U_{a,\Lambda}(u_{a,\Lambda} X \gets X)
\hat{\phi}(x: X).
\l{aa2}
$$
}

\section{Construction of the Poincare transformations}

This section  is  devoted to the problem of relativistic invariance of
the semiclassical field theory.  The axiom  1  will  be  checked.  The
mappings  $u_{a,\Lambda}$ and unitary operators $U_{a,\Lambda}$ are to
be specified, the group property is to be justified.

\subsection{Heuristic definition}

Consider some special cases of Poincare transformations $(a,\Lambda)$.
The transformation $(a,1)$ is called translation.  If $a^0=0$, this is
a spatial  translation,  while the ${\bf a}=0$-case corresponds to the
time translation or evolution.  The transformation  $(0,\Lambda)\equiv
\Lambda$ is   called  as  a  Lorentz  transformation.  If  $\Lambda  =
\Lambda_{\tau}$,
$$
\Lambda_{\tau} =  \left(\begin{array}{cc}
\begin{array}{cc}
\cosh\tau & -\sinh\tau \\ -\sinh\tau & \cosh\tau
\end{array}
& 0 \\ 0 & 1
\end{array}
\right)
\l{e1}
$$
the Lorentz transformation is called as $x^1$-boost. If
$$
\Lambda =
\left(
\begin{array}{cc}
1 & 0 \\ 0 & L
\end{array}
\right)
$$
the Lorentz transformation  is  called  as  a  spatial  rotation.  Let
$L_{\bf n}$ be such a spatial rotation that
$$
L_{\bf n}{\bf n} = {\bf e}_1,
\l{e3}
$$
where ${\bf e}_1$ is a spatial vector of the form  $(1,0,0,...)$.  The
transformation
$$
\Lambda_{\phi {\bf n}} = L_{\bf n}^{-1} \Lambda_{\phi} L_{\bf n}
\l{e3a}
$$
will be called as ${\bf n}$-boost.  It does not depend  on  choice  of
$L_{{\bf n}}$.  Namely, let $L_{\bf n}^{(1)}$ and $L_{\bf n}^{(2)}$ be
spatial rotations obeying eq.\r{e3}. Then
$L_{\bf n}^{(1)} (L_{\bf n}^{(2)})^{-1} {\bf e}_1 = {\bf e}_1$.
Therefore, the transformations
$\Lambda_{\phi}$ and
$L_{\bf n}^{(1)} (L_{\bf n}^{(2)})^{-1}$ commute. Thus,
$
(L_{\bf n}^{(1)})^{-1} \Lambda_{\phi} L_{\bf n}^{(1)}
= (L_{\bf n}^{(2)})^{-1} \Lambda_{\phi} L_{\bf n}^{(2)}$.

{\bf Lemma  3.1.} \c{A3}
{\it Let $(a,\Lambda)$ be a Poincare transformation.
It is uniquely presented as
$$
(a,\Lambda) = (a,1) \Lambda_{\vec{\phi}} L,
$$
where $\Lambda_{\vec{\phi}}$ is a boost, $L$ is a spatial rotation.
}

{\it Proof.} It follows from \r{6**} that $(a,\Lambda)= (a,1)\Lambda$.
Let us   show   that  $\Lambda  =  \Lambda_{\vec{\phi}}  L$  and  this
decomposition is unique. Consider the transformation $x'=\Lambda x$:
$$
\begin{array}{c}
x^{'0} = ax^0 + \beta_i x^i, \\
x^{'i} = \gamma^i x^0 + \Lambda^i_k x^k.
\end{array}
$$
Let $L^{(1)}$   be   such   a  rotation  that  $L^{(1)}\vec{\gamma}  =
||\vec{\gamma}|| {\bf e}_1$.  Consider the  transformation  $y^{'i}  =
(L^{(1)})^i_j x^{'j}$, $y^{'0} =x^{'0}$. One has
$$
\begin{array}{c}
y^{'0} = ax^0 + \beta_ix^i;\\
y^{'1} = ||\vec{\gamma}|| x^0 + A^1_k x^k;\\
y^{'\beta} = A^{\beta}_{\rho} x^{\rho}.
\end{array}
$$
where $\beta,  \rho=\overline{2,d}$, $i,k=\overline{1,d}$. Since $a^2-
||\vec{\gamma}||^2 =1$, set $a=\cosh\phi$, $||\vec{\gamma}|| = \sinh \phi$.
Denote $z' = \Lambda_{\phi}^{-1}y'$. One has
$$
\begin{array}{c}
z^{'0} = x^0 + \tilde{\beta}_ix^i, \\
z^{'i} = B^i_j x^j.
\end{array}
$$
Therefore, $\tilde{\beta}_i=0$.  This is a  rotation  $L^{(2)}$  then.
Thus, $\Lambda  = (L^{(1)})^{-1} \Lambda_{\phi} L^{(1)} (L^{(1)})^{-1}
L^{(2)} = \Lambda_{\phi {\bf n}} L$ with ${\bf n}= \vec{\gamma}/
||\vec{\gamma}||$. This decomposition is unique. Lemma 3.1 is proved.

To construct mappings $u_{a,\Lambda}$ and  operators  $U_{a,\Lambda}$,
one may  consider  first  the  partial cases (time evolution,  spatial
translations, $x^1$ or ${\bf n}$-boost,  spatial rotations)  and  then
use the group property.

First of all, let us consider the representation of the Poincare group
$\tilde{U}_{a,\Lambda}$ in the functional Schrodinger  representation.
Formally, they are related with $U_{a,\Lambda}$ by the relation
$$
\tilde{U}_{a,\Lambda} (u_{a,\Lambda}X \gets X) =
V_{u_{a,\Lambda}X} U_{a,\Lambda} (u_{a,\Lambda}X \gets X) V_X^{-1}.
\l{b0}
$$
To construct    operators    $\tilde{U}_{a,\Lambda}$    and   mappings
$u_{a,\Lambda}$, let  us  use  formal  expressions  for  the  Poincare
transformations in  the "exact" field theory.  Namely,  the (formally)
unitary operator ${\cal U}_{a,\Lambda}$ corresponding to the  Poincare
transformation
$$
(a,\Lambda) =    (a^0,0)    ({\bf    a},0)    \exp(\alpha^k    l^{0k})
\exp(\frac{1}{2} \theta_{sm} l^{sm})
$$
where $\theta_{sm}=-theta_{ms}$,
$$
(l^{\lambda\mu})^{\alpha}_{\beta}   =   -  g^{\lambda
\alpha} \delta^{\mu}_{\beta} + g^{\mu\alpha} \delta^{\lambda}_{\beta},
$$
has the form
$$
{\cal U}_{a,\Lambda} =
\exp[i{\cal P}^{0}a^0]
\exp[-i{\cal P}^{j}a^j]
\exp[{i}\alpha^k{\cal M}^{0k}]
\exp[\frac{i}{2}{\cal M}^{lm} \theta_{lm}].
\l{b1}
$$
The momentum and angular momentum operators entering to formula \r{b1}
have the well-known form (see,  for example, \c{BS})
$$
{\cal P}^{\mu} = \int d{\bf x} T^{\mu 0}({\bf x}),
\qquad
{\cal M}^{\mu\lambda} = \int d{\bf x} [x^{\mu} T^{\lambda 0}({\bf x})
- x^{\lambda} T^{\mu 0}({\bf x})],
\l{et1}
$$
where formally
$$
T^{00} = \frac{1}{2} \hat{\pi}^2 + \frac{1}{2} \partial_i \hat{\varphi}
\partial_i \hat{\varphi}    +    \frac{m^2}{2}    \hat{\varphi}^2    +
\frac{1}{\lambda} V(\sqrt{\lambda} \hat{\varphi}),
\qquad
T^{k0} = - \partial_k\hat{\varphi} \hat{\pi}.
$$
We are  going  to apply the operator \r{b1} to the semiclassical state
\r{2}. Note  that  the   operators   ${\cal   P}^{\mu}$   and   ${\cal
M}^{\mu\nu}$ \r{et1}  depend  on  field  $\hat{\varphi}$  and  momentum
$\hat{\pi}$ semiclassically,
$$
{\cal P}^{\mu}    =    \frac{1}{\lambda}    P^{\mu}    (\sqrt{\lambda}
\hat{\pi}(\cdot), \sqrt{\lambda}\hat{\varphi}(\cdot)),
\qquad
{\cal M}^{\mu\nu} =    \frac{1}{\lambda}    M^{\mu\nu}    (\sqrt{\lambda}
\hat{\pi}(\cdot), \sqrt{\lambda} \hat{\varphi}(\cdot)),
$$
It is convenient to consider the more general problem  (cf.\c{M2}).Let
us find as $\lambda\to 0$ the state
$$
\exp(-i{\cal A}) K_{S^0,\Pi^0,\Phi^0} f^0,
\l{e2}
$$
where $K_{S,\Pi,\Phi}$ has the form \r{2},
$$
{\cal A}   =    \frac{1}{\lambda}    A(\sqrt{\lambda}\hat{\pi}(\cdot),
\sqrt{\lambda} \hat{\varphi}(\cdot)).
$$
Note that  the  state functional \r{e2} may be viewed as a solution to
the Cauchy problem of the form
$$
\begin{array}{c}
i \frac{\partial\Psi^{\tau}}{\partial\tau}     =     \frac{1}{\lambda}
A(\frac{\sqrt{\lambda}}{i} \frac{\delta}{\delta        \varphi(\cdot)},
\sqrt{\lambda} \varphi(\cdot)) \Psi^{\tau}, \\
\Psi^0[\varphi(\cdot)] = (K_{S^0,\Pi^0,\Phi^0} f^0)[\varphi(\cdot)]
\end{array}
\l{et3}
$$
at $\tau=1$.  Let us look for the asymptotic solution to eq.\r{et3}  in
the following form:
$$
\Psi^{\tau}[\varphi(\cdot)]                                          =
(K_{S^{\tau},\Pi^{\tau},\Phi^{\tau}}f^{\tau})[\varphi(\cdot)].
\l{e4}
$$
Substitution of  functional \r{e4} to eq.\r{et3} gives us the following
relation:
$$
\begin{array}{c}
[-\frac{1}{\lambda} (\dot{S}^{\tau}  -  \int  d{\bf x} \Pi^{\tau}({\bf
x}) \dot{\Phi}^{\tau}({\bf x})) - \frac{1}{\sqrt{\lambda}} \int  d{\bf
x} (\dot{\Pi}^{\tau}({\bf  x})  \phi({\bf x}) + \dot{\Phi}^{\tau}({\bf
x}) i    \frac{\delta}{\delta     \phi({\bf     x})}     )     +     i
\frac{\partial}{\partial\tau} ] f^{\tau}[\phi(\cdot)] =\\
\frac{1}{\lambda} A(\Pi^{\tau}(\cdot)        -         i\sqrt{\lambda}
\frac{\delta}{\delta \phi(\cdot)}, \Phi^{\tau}(\cdot) + \sqrt{\lambda}
\phi(\cdot)) f^{\tau}[\phi(\cdot)].
\end{array}
\l{e5}
$$
Considering the    terms    of    the    orders     $O(\lambda^{-1})$,
$O(\lambda^{-1/2})$ and $O(1)$ in eq.\r{e5}, we obtain
$$
\dot{S}^{\tau} =     \int     d{\bf     x}     \Pi^{\tau}({\bf     x})
\dot{\Phi}^{\tau}({\bf x}) - A(\Pi^{\tau}(\cdot),\Phi^{\tau}(\cdot)),
\l{e6}
$$
$$
\dot{\Phi}^{\tau}({\bf x})  =   \frac{\delta   A   (\Pi^{\tau}(\cdot),
\Phi^{\tau}(\cdot))}{\delta \Pi({\bf x})},
\qquad
\dot{\Pi}^{\tau}({\bf x})  = - \frac{\delta   A   (\Pi^{\tau}(\cdot),
\Phi^{\tau}(\cdot))}{\delta \Phi({\bf x})},
\l{e7}
$$
$$
\begin{array}{c}
i \frac{\partial f^{\tau}[\phi(\cdot)]}{\partial\tau} = \left(
\int d{\bf x} d{\bf y} \left[
\frac{1}{2} \frac{1}{i} \frac{\delta}{\delta \phi({\bf x})}
\frac{\delta^2A}{\delta\Pi({\bf x}) \delta\Pi({\bf y})}
\frac{1}{i} \frac{\delta}{\delta \phi({\bf y})}
+ \right. \right. \\
\left. \left.
\phi({\bf x})
\frac{\delta^2A}{\delta\Phi({\bf x}) \delta\Pi({\bf y})}
\frac{1}{i} \frac{\delta}{\delta \phi({\bf y})}
+
\frac{1}{2} \phi({\bf x})
\frac{\delta^2A}{\delta\Phi({\bf x}) \delta\Phi({\bf y})}
\phi({\bf y})
\right] + A_1 \right) f^{\tau}[\phi(\cdot)].
\end{array}
\l{e8}
$$
Here $A_1$ is a c-number quantity which depends on the ordering of the
operators $\hat{\varphi}$  and  $\hat{\pi}$  and  is  relevant  to the
renormalization problem.

We see that for the cases ${\cal A} = - {\cal P}^0a^0$,  ${\cal  A}  =
{\cal P}^ja^j$,  ${\cal  A}  =  -\alpha^k {\cal M}^{0k}$,  ${\cal A} =
\frac{1}{2} \theta_{sm}{\cal  M}^{sm}$  the  mapping   $u_{a,\Lambda}$
takes the  initial  condition  for  the  system \r{e6},  \r{e7} to the
solution of the Cauchy  problem  for  this  system  at  $\tau=1$.  The
operators $\tilde{U}_{a,\Lambda}$ transforms the initial condition for
eq.\r{e8} to the solution at $\tau=1$.

\subsection{Poincare invariance of the classical theory}

The purpose of this subsection is to find explicit forms  of  mappings
$u_{a,\Lambda}$. Consider some special cases.

\subsubsection{Spatial rotations}

For this case, $a=0$, $\Lambda=L= \exp(\frac{1}{2}l^{sm}\theta_{sm})$,
so that $L^k_l = (\exp\theta)^k_l$, where $\theta$ is an antisymmetric
matrix with    elements   $\theta_{kl}$.   One   has   ${\cal   A}   =
-\frac{1}{2}\theta_{sm} {\cal M}^{sm}$, so that
$$
A[\Pi,\Phi] = \frac{1}{2} \int d{\bf x} \Pi({\bf x}) \theta_{sm}  (x^s
\partial_m - x^m \partial_s) \Phi({\bf x})
\l{eq1}
$$
with $\partial_m \equiv \frac{\partial}{\partial x^m}$. System \r{e6},
\r{e7} takes the form
$$
\begin{array}{c}
\dot{\Phi}^{\tau}({\bf x}) = \frac{1}{2} \theta_{sm} (x^s\partial_m  -
x^m \partial_s) \Phi^{\tau}({\bf x}),
\\
\dot{\Pi}^{\tau}({\bf x}) = \frac{1}{2} \theta_{sm} (x^s\partial_m  -
x^m \partial_s) \Pi^{\tau}({\bf x}),\\
\dot{S}^{\tau} =0.
\end{array}
\l{e9}
$$
Eqs.\r{e9} can be represented as
$$
\frac{\partial}{\partial \tau}         \Phi^{\tau}(\exp(\frac{\tau}{2}
\theta_{sm} l^{sm}) {\bf x}) = 0,
\qquad
\frac{\partial}{\partial \tau}         \Pi^{\tau}(\exp(\frac{\tau}{2}
\theta_{sm} l^{sm}) {\bf x}) = 0.
$$
Therefore, $\Phi^1(L{\bf x}) = \Phi^0({\bf x})$,  $\Pi^1(L{\bf  x})  =
\Pi^0({\bf x})$, so that
$$
\Phi^1({\bf x}) = \Phi^0(L^{-1}{\bf x}), \qquad
\Pi^1({\bf x}) = \Pi^0(L^{-1}{\bf x}), \qquad S^1=S^0.
\l{e10}
$$

\subsubsection{Spatial translations}

For this case,  $a^0=0$,  $\Lambda=1$,  so that ${\cal A} = {\cal P}^j
a^j$ and
$$
A[\Pi,\Phi] = - \int d{\bf x} a^k \partial_k  \Phi({\bf  x})  \Pi({\bf
x}).
\l{eq2}
$$
System \r{e6}, \r{e7} takes the form
$$
\dot{\Phi}^{\tau}({\bf x}) = - a^k \partial_k \Phi^{\tau}({\bf x}),
\qquad
\dot{\Pi}^{\tau}({\bf x}) = - a^k \partial_k \Pi^{\tau}({\bf x}),
\qquad \dot{S}^{\tau} = 0.
\l{w27}
$$
Therefore,
$$
\Phi^{\tau}({\bf x}) = \Phi({\bf x} - {\bf a}\tau), \qquad
\Pi^{\tau}({\bf x}) = \Pi({\bf x} - {\bf a}\tau), \qquad
{S}^{\tau} = S^0.
\l{e11}
$$

\subsubsection{Evolution transformation}

Let $a^0=-t$,  ${\bf  a}=0$,  $\Lambda=1$.  Then  $A[\Pi,\Phi]$  is  a
classical Hamiltonian,  so that $u_{-t,0,1}$ is a mapping  taking  the
initial condition  for  system  \r{3a},  \r{3}  to the solution of the
corresponding Cauchy problem.

\subsubsection{The ${\bf n}$-boost}

Let $\Lambda=\Lambda_{\tau {\bf n}}$ have  the  form  \r{e3a},  $a=0$.
Then ${\cal A} = n^k {\cal M}^{k0}$, so that
$$
A[\Pi,\Phi] =  \int  d{\bf  x}  n^kx^k  [\frac{1}{2}\Pi^2({\bf  x})  +
\frac{1}{2} (\nabla \Phi)^2({\bf x}) + \frac{m^2}{2} \Phi^2({\bf x}) +
V_i(\Phi({\bf x}))]
\l{eq3}
$$
System \r{e6}, \r{e7} takes the form:
$$
\begin{array}{c}
\dot{\Phi}^{\tau}({\bf x}) = n^kx^k \Pi^{\tau}({\bf x}),
\\
- \dot{\Pi}^{\tau}({\bf  x}) = - \nabla x^kn^k \nabla \Phi^{\tau}({\bf
x}) +  x^kn^k  (m^2\Phi^{\tau}({\bf  x})   +   V_i{}'(\Phi^{\tau}({\bf
x}))),\\
\dot{S}^{\tau} =    \int     d{\bf     x}     [\Pi^{\tau}({\bf     x})
\dot{\Phi}^{\tau}({\bf x})   -  x^kn^k  [\frac{1}{2}  (\Pi^{\tau}({\bf
x}))^2 + \frac{1}{2} (\nabla \Phi^{\tau}({\bf x}))^2  +  \frac{m^2}{2}
(\Phi^{\tau}({\bf x}))^2 + V(\Phi^{\tau}({\bf x}))].
\end{array}
\l{e12}
$$

\subsubsection{General formulas}

Let $(a,\Lambda)$be an arbitrary Poincare transformation.  It  happens
that the  mapping  $u_{a,\Lambda}:  (S,\Pi,\Phi)  \mapsto  (\tilde{S},
\tilde{\Pi}, \tilde{\Phi})$ has the  following  form.  Let  $\Phi({\bf
x},t) \equiv \Phi(x)$ be a solution of the Cauchy problem
$$
\begin{array}{c}
\partial_{\mu} \partial^{\mu} \Phi(x) + m^2\Phi(x) + V_i{}'(\Phi(x)) =
0, \\
\Phi({\bf x},0) = \Phi({\bf x}), \qquad
\frac{\partial}{\partial t} \Phi({\bf x},t)|_{t=0} = \Pi({\bf x}).
\end{array}
\l{e13}
$$
Denote
$$
\breve{\Phi}(x) = \Phi(\Lambda^{-1}(x-a)).
$$
It appears that
$$
\begin{array}{c}
\tilde{\Phi}({\bf x}) = \breve{\Phi}({\bf x},0),
\qquad
\tilde{\Pi}({\bf x}) = \frac{\partial}{\partial t}
\breve{\Phi}({\bf x},t)|_{t=0},\\
\tilde{S} = S  +  \int  dx  [\theta(x^0)  \theta(-(\Lambda  x+a)^0)  -
\theta(-x^0) \theta((\Lambda x+a)^0)]\\
\times
[\frac{1}{2} \partial_{\mu}   \Phi(x)   \partial^{\mu}    \Phi(x)    -
\frac{{m^2}}{2} \Phi^2(x) - V_i(\Phi(x)).
\end{array}
\l{e14}
$$
First of all,  show that eqs.\r{e14} are correct for the partial cases
mentioned above.  For  spatial  translations  and  rotations  formulas
\r{e14} give $\tilde{S}=S$,
$\tilde{\Phi}({\bf x}) = \Phi(L^{-1}({\bf x}-{\bf a}))$,
$\tilde{\Pi}({\bf x}) = \Pi(L^{-1}({\bf x}-{\bf a}))$.  This coincides
with eqs.\r{e10},  \r{e11}.  For evolution transformation, eqs.\r{e14}
are in agreement with eqs.\r{3a}, \r{3}. Let us check formulas \r{e14}
for the $x^1$-boost, $\Lambda=\Lambda_{\tau}$. One has
$$
\begin{array}{c}
\tilde{\Phi}_{\tau} ({\bf  x})  = \Phi(x^1 \cosh\tau + x^0
\sinh \tau,  x^2,
..., x^d, x^0 \cosh\tau + x^1 \sinh\tau)|_{x^0=0},
\\
\tilde{\Pi}_{\tau} ({\bf  x})  = \frac{\partial}{\partial x^0}
\Phi(x^1 \cosh\tau + x^0 \sinh \tau,  x^2,
..., x^d, x^0 \cosh\tau + x^1 \sinh\tau)|_{x^0=0},
\end{array}
$$
The functions $\Phi_{\tau}$, $\Pi_{\tau}$
obey system \r{e12}.  For the integral for $\tilde{S}$,  consider  the
substitution $x^0  =  y^1  \sinh\tilde{\tau}$,  $x^1=y^1 \cosh\tilde{\tau}$,
$x^2=y^2$,..., $x^d=y^d$. One finds
$$
\tilde{S}^{\tau} = S + \int_0^{\tau} d\tilde{\tau} y^1 d{\bf y} [
\frac{1}{2} (\tilde{\Pi}_{\tau}({\bf y}))^2 -
\frac{1}{2} (\nabla \tilde{\Phi}_{\tau}({\bf y}))^2 -
\frac{m^2}{2} \tilde{\Phi}_{\tau}^2({\bf             y})             -
V(\tilde{\Phi}_{\tau}({\bf y}))
]
$$
This agrees with \r{e12}.

Any Poincare transformation  can  be  obtained  as  a  composition  of
considered partial  cases.  To  check formulas \r{e14} for the general
case, it is sufficient to prove the following lemma.

{\bf Lemma 3.2.} {\it For transformation \r{e14},  the group  property
is satisfied. }

{\it Proof.} Let $u_{a_1,\Lambda_1}$,  $u_{a_2,\Lambda_2}$ be Poincare
transformations. Show that
$u_{(a_1,\Lambda_1)(a_2,\Lambda_2)} =
u_{a_1,\Lambda_1}u_{a_2,\Lambda_2}$. Let
$u_{(a_2,\Lambda_2)}
(S,\Pi,\Phi) = (\tilde{S},\tilde{\Pi},\tilde{\Phi})$,
$u_{(a_1,\Lambda_1)}
(\tilde{S},\tilde{\Pi},\tilde{\Phi})
 = (\overline{S},\overline{\Pi},\overline{\Phi})$,
$u_{(a_1,\Lambda_1)(a_2,\Lambda_2)}
({S},{\Pi},{\Phi})
 = (\underline{S},\underline{\Pi},\underline{\Phi})$.
One has  ${\tilde{\Phi}}({\bf x}) = \Phi(\Lambda_2^{-1}(x-a_2))$
because of    Poincare    invariance    of     \r{e13}.     Therefore,
$\breve{\tilde{\Phi}}(x) =  \Phi(\Lambda_2^{-1} (\Lambda_1^{-1} (x-a_1)
- a_2)) =  \Phi((\Lambda_1\Lambda_2)^{-1}  (x-a_1-\Lambda_1a_2))$.  We
see that
$\underline{\Pi}({\bf x}) = \overline{\Pi}({\bf x})$,
$\underline{\Phi}({\bf x})   =  \overline{\Phi}({\bf  x})$.  Lemma  is
proved.

We obtain the following corollaries.

{\bf Corollary 1.} {\it For arbitrary Poincare transformation,
$u_{a,\Lambda} (S,\Pi,\Phi)  =  (\tilde{S},\tilde{\Pi},\tilde{\Phi})$,
where $(\tilde{S},\tilde{\Pi},\tilde{\Phi})$ has the form \r{e14}.}

{\bf Corollary 2.} {\it Property \r{aa1} is satisfied.}

Let us make more precise the definition of the space $\cal X$.

{\bf Definition 3.1.} {\it $\cal X$ is a space of sets  $(S,\Pi,\Phi)$
of a  number  $S$  and functions $\Pi,\Phi \in S({\bf R}^d)$ such that
there exists a unique solution of the Cauchy problem \r{e13} such that
the functions $\Phi(\Lambda x+a)|_{x^0=0}$ and
$\partial_{\mu}\Phi(\Lambda x+a)|_{x^0=0}$ are of  the  class  $S({\bf
R}^d)$ for all $a.\Lambda$. }

We see that the transformation $u_{a,\Lambda}:  {\cal X} \to {\cal X}$
is defined.

\subsubsection{Infinitesimal properties}

According to formula \r{x1} of  Appendix  A,  one  can  introduce  the
operators $\delta[A]$  on  the space of differentiable functionals $F$
of $S$,  $\Pi$,  $\Phi$ for each element $A$ of the Poincare  algebra.
This operator  plays  an  important  role  in  analysis  of  algebraic
properties of the representation $U_{a,\Lambda}$.

Elements of Lie algebra of the Poincare group can be  identified  with
sets $(b^{\mu}, \theta^{\mu\nu})$, $\theta^{\nu\mu}=-\theta^{\mu\nu}$.
The curve on the Poincare group with the tangent  vector  $(b,\theta)$
can be chosen to be
$$
(a(\tau),\Lambda(\tau)) =   (\tau  b,  \exp(\tau  \theta^{0k}  l_{0k})
\exp(\frac{\tau}{2} \theta^{km}l_{km})).
$$
The operator $\delta[(b,\theta)]$ is a linear combination of $b,\theta$:
$$
\delta[(b,\theta)] =  \frac{1}{2}  \theta_{lm}  \delta_M^{lm}  -   b^k
\delta_P^k - \theta_{0m} \delta_B^m + b^0\delta_H.
$$
Let us  find the coefficients from eqs.\r{e9}, \r{e11}, \r{3}, \r{e12}.
Let $F$ be a differentiable functional
of $S$, $\Pi$, $\Phi$.

1. Let $\theta=0$, $b^0=0$, ${\bf b} \ne 0$. Then
$$
-b^k\delta_P^k F = - b^k\int d{\bf x} \left(
\frac{\delta F}{\delta \Phi({\bf x})} \partial_k \Phi({\bf x})
+ \frac{\delta F}{\delta \Pi({\bf x})} \partial_k \Pi({\bf x})
\right)
$$

2. For $\theta^{0k}=0$, $b=0$, one has
$$
\frac{1}{2} \theta_{lm}  \delta_M^{lm} F = \frac{1}{2}\theta_{lm} \int
d{\bf x}
(
(x^l\partial_m -  x^m  \partial_l)\Phi({\bf  x}) \frac{\delta F}{\delta
\Phi({\bf x})}
+ (x^l\partial_m -  x^m  \partial_l)\Pi({\bf  x}) \frac{\delta F}{\delta
\Pi({\bf x})}).
$$

3. For $b^0\ne 0$, ${\bf b}=0$, $\theta=0$
$$
\begin{array}{c}
b^0 \delta_H F = - b^0 \int d{\bf x} \left[
\Pi({\bf x})   \frac{\delta  F}{\delta  \Phi  ({\bf  x})}  -  (-\Delta
\Phi({\bf x}) + m^2 \Phi({\bf x}) +  V'(\Phi({\bf  x})))  \frac{\delta
F}{\delta \Pi({\bf x})}
\right] -
\\
b^0 \frac{\partial  F}{\partial  S}   \int   d{\bf   x}
[\frac{1}{2}
\Pi^2({\bf x})  - \frac{1}{2} (\nabla \Phi({\bf x}))^2 - \frac{m^2}{2}
\Phi^2({\bf x}) - V(\Phi({\bf x}))].
\end{array}
$$

4. For  boost  transformation  $\theta^{0k}  \ne 0$,  $\theta^{lm}=0$,
$b=0$ and
$$
\begin{array}{c}
-\theta_{0m} \delta_B^m  F  =  \theta_{0m} \int d{\bf x} [x^m \Pi({\bf
x}) \frac{\delta F}{\delta \Phi({\bf x})} -  (-\partial_ix^m\partial_i
\Phi({\bf x})  +  x^m  m^2\Phi({\bf  x})  +  x^m  V'(\Phi({\bf  x})) )
\frac{\delta F}{\delta \Pi({\bf x})}] + \\
\theta_{0m} \frac{\partial F}{\partial S} \int d{\bf x} x^m
[\frac{1}{2}
\Pi^2({\bf x})  - \frac{1}{2} (\nabla \Phi({\bf x}))^2 - \frac{m^2}{2}
\Phi^2({\bf x}) - V(\Phi({\bf x}))].
\end{array}
$$

The introduced operators obey usual properties of the Poincare algebra
\r{x4}:
$$
[i\delta_{P}^k, i\delta_H] = 0;\qquad
[i\delta_{M}^{lm}, i\delta_{P}^s]   =   i(g^{ms}i\delta_{P}^l    -    g^{ls}
i\delta_{P}^m); \qquad [i\delta_{P}^k, i\delta_{P}^l] =0.
$$
$$
[i\delta_{M}^{kl}, i\delta_{H}]=0; \qquad
[i\delta_{B}^{k}, i\delta_{P}^s] = -ig^{ks} i\delta_{H};
$$
$$
[i\delta_{M}^{lm}, i\delta_{M}^{rs}]    =    -i(g^{lr}i\delta_{M}^{ms}     -
g^{mr}i\delta_{M}^{ls} +      g^{ms}     i\delta_{M}^{lr}     -     g^{ls}
i\delta_{M}^{mr});
$$
$$
[i\delta_{M}^{lm},i\delta_{B}^{k}] =       -i(g^{lk}i\delta_{B}^{m}      -
g^{mk}i\delta_{B}^{l});
$$
$$
[i\delta_{B}^{k}, i\delta_{H}] = -\delta_{P}^k, \qquad
[i\delta_{B}^{k}, i\delta_{B}^{l}] = \delta_{M}^{kl}.
$$

\subsection{Semiclassical Poincare transformations in  the  functional
representation}

We have formally found the operators $\tilde{U}_{a,\Lambda}$. However,
it is not easy to check the group property.  Therefore,  construct the
representation of the Poincare algebra according to Appendix  A.  Then
we will  check  the  algebraic property.  The group property will be a
corollary of the results of Appendix A.

Let us construct  the  operators  $\tilde{H}((b,\theta):  S,\Pi,\Phi)$
\r{x4a} for elements of the Poincare algebra:
$$
\tilde{H}((b,\theta): S,\Pi,\Phi)    =   -   \frac{1}{2}   \theta_{lm}
\tilde{M}^{lm} + b^k \tilde{P}^k + \theta_{0m} \tilde{B}^m - b^0 H.
$$
Consider some cases.

\subsubsection{Spatial rotations}

For this case,  $a=0$,  $\Lambda = L = \exp(\frac{\tau}{2} \theta_{sm}
l^{sm})$, ${\cal  A}  = - \frac{1}{2} \theta_{sm} {\cal M}^{sm}$,  $A$
has the form \r{eq1}. Therefore, eq.\r{e8} takes the form
$$
i \dot{f}^{\tau}   =   \frac{1}{2}   \theta_{sm}   \int    d{\bf    x}
(x^s\partial_m -    x^m    \partial_s)   \phi({\bf   x})   \frac{1}{i}
\frac{\delta}{\delta \phi({\bf x})} f^{\tau}.
$$
It follows from eq.\r{x4a} that
$$
\tilde{M}^{sm} = - \int d{\bf x}  [(x^s  \partial_m  -  x^m\partial_s)
\phi({\bf x})] \frac{1}{i} \frac{\delta}{\delta \phi({\bf x})}.
\l{f1}
$$

\subsubsection{Spatial translations}

Let $a^0=0$,  ${\bf  a}= {\bf b}\tau$,  $\Lambda=1$.  Then ${\cal A} =
{\cal P}^k a^k$, $A$ has the form \r{eq2}. Eq.\r{e8} takes the form
$$
i\dot{f}^{\tau} = - b^{k} \int d{\bf x}  \partial_k  \phi({\bf  x})
\frac{1}{i} \frac{\delta}{\delta \phi({\bf x})} f^{\tau}.
$$
Therefore,
$$
\tilde{P}^k =  -  \int  d{\bf  x}  \partial_k\phi({\bf x}) \frac{1}{i}
\frac{\delta}{\delta \phi({\bf x})}.
\l{f2}
$$

\subsubsection{Evolution transformation}

Let $a^0=-\tau$,  ${\bf a}=0$,  $\Lambda=1$.  Then $A$ is a  classical
Hamiltonian, eq.\r{e8} takes the form \r{4}. $\tilde{H}$ takes the form
$$
\tilde{H} = \int d{\bf x} \left[
- \frac{1}{2} \frac{\delta^2}{\delta \phi({\bf  x})  \delta  \phi({\bf
x})} +   \frac{1}{2}   (\nabla   \phi)^2({\bf   x})   +  \frac{m^2}{2}
\phi^2({\bf x}) + \frac{1}{2} V''(\Phi({\bf x})) \phi^2({\bf x}).
\right]
\l{f3}
$$

\subsubsection{The ${\bf n}$-boost}

Let $a=0$,  $\Lambda  = \Lambda_{\tau {\bf n}}$ have the form \r{e3a},
so that $\theta_{k0} = - n^k = - \theta_{0k}$.  Then ${\cal A}  =  n^k
{\cal M}^{k0}$,  $A$ has the form \r{eq3}, so that eq.\r{e8} takes the
form:
$$
i \dot{f}^{\tau} = \int d{\bf x} n^k x^k \left[
- \frac{1}{2}  \frac{\delta^2}{\delta  \phi({\bf  x}) \delta \phi({\bf
x})} +  \frac{1}{2}   (\nabla   \phi)^2({\bf   x})   +   \frac{m^2}{2}
\phi^2({\bf x}) + \frac{1}{2} V''(\Phi({\bf x})) \phi^2({\bf x})
\right] f^{\tau}.
$$
Therefore,
$$
\tilde{B}^m = \int d{\bf x} x^m
\left[
- \frac{1}{2}  \frac{\delta^2}{\delta  \phi({\bf  x}) \delta \phi({\bf
x})} +  \frac{1}{2}   (\nabla   \phi)^2({\bf   x})   +   \frac{m^2}{2}
\phi^2({\bf x}) + \frac{1}{2} V''(\Phi({\bf x})) \phi^2({\bf x})
\right].
\l{f4}
$$

Note that the divergences in these operators are to be  eliminated  by
adding $c$-number quantities to them.

\subsubsection{Properties of infinitesimal transformations}

For operators
$$
\breve{\tilde{M}}^{ms} = \tilde{M}^{ms} + i \delta_M^{ms},
\qquad
\breve{\tilde{P}}^{m} = \tilde{P}^{m} + i \delta_P^{m},
\qquad
\breve{\tilde{P}}^{0} = \tilde{H}^{m} + i \delta_H,
\qquad
\breve{\tilde{M}}^{k0} = \tilde{B}^{k} + i \delta_B^{k}
$$
the commutation relations of the Poincare algebra
$$
[\breve{\tilde{P}}^{\lambda}, \breve{\tilde{P}}^{\mu}] = 0;\qquad
[\breve{\tilde{M}}^{\lambda \mu}, \breve{\tilde{P}}^{\sigma}]
=  i(g^{\mu \sigma}\breve{\tilde{P}}^{\lambda}    -    g^{\lambda \sigma}
\breve{\tilde{P}}^{\mu})
$$
$$
[\breve{M}^{\lambda\mu}, \breve{M}^{\rho\sigma}]
=    -i(g^{\lambda \rho}\breve{M}^{\mu\sigma}     -
g^{\mu\rho}\breve{M}^{\lambda \sigma}
+      g^{\mu\sigma}     \breve{M}^{\lambda \rho}     -
g^{\lambda\sigma} \breve{M}^{\mu\rho}).
\l{f5}
$$
should be satisfied (eq.\r{x7*}).  The formal check of these relations
is straightforward.  However,   the   functional   representation   is
ill-defined, so  that  one  should  use  the  Fock representations and
perform a renormalization.

Find a relationship between operators  $H(A:X)$  and  $\tilde{H}(A:X)$
being generators   of   representation  $U_{a,\Lambda}$  in  Fock  and
Schrodinger pictures.  Let $g(\tau)$ be a curve on the  Poincare  group
with tangent vector $A$. Eq.\r{b0} implies
$$
\tilde{U}_{g(\tau)} [X] = V_{u_{g(\tau)}X} U_{g(\tau)} [X] V_X^{-1}.
$$
Differentiating this relation by $\tau$ at $\tau=0$, we find
$$
- i \tilde{H}(A:X) = \delta[A] V_X V_X^{-1} - iV_X H(A:X) V_X^{-1}.
$$
Therefore,
$$
H(A:X) - i \delta[A] = V_X^{-1} (\tilde{H}(A:X) - i\delta[A])V_X.
$$
We see that operators
$$
\breve{H}(A:X) \equiv  H(A:X) - i\delta[A] = - \frac{1}{2} \theta_{lm}
\breve{M}^{lm} +  b^k  \breve{P}^k   +   \theta_{0m}   \breve{B}^m   -
b^0\breve{H}
$$
with
$$
\begin{array}{c}
\breve{{M}}^{ms} = V_X^{-1} (\tilde{M}^{ms} + i \delta_M^{ms}) V_X,
\qquad
\breve{{P}}^{m} = V_X^{-1} (\tilde{P}^{m} + i \delta_P^{m}) V_X, \\
\breve{{P}}^{0} = V_X^{-1} (\tilde{H}^{m} + i \delta_H) V_X,
\qquad
\breve{{M}}^{k0} = V_X^{-1} (\tilde{B}^{k} + i \delta_B^{k}) V_X
\end{array}
$$
formally obey commutation relations \r{f5}.  However,  the divergences
and renormalization problem should be taken into account.

\subsection{Poincare transformations in the Fock representation}

The purpose    of   this   subsection   is   to   construct   Poincare
transformations $U_{a,\Lambda}(u_{a,\Lambda}X\gets  X)$  in  the  Fock
space. First  of  all,  we  calculate the explicit form of generators.
Then we will  renormalize  the  obtained  expressions  and  check  the
conditions of   Appendix   A.   Then   we   will  construct  operators
$U_{a,\Lambda}$ and check the group property.

First of all, investigate some properties of the operator $V_X$.

\subsubsection{Some properties of the operator $V$}

Remind that  the operator $V$ taking the Fock space vector $\Psi\in {\cal
F}$ to the functional $f[\phi(\cdot)]$ is defined from the relation
$$
V: |0>  \mapsto  c\exp[\frac{i}{2}\int d{\bf x} d{\bf y}
\tilde{\cal R}({\bf x},{\bf y}) \phi({\bf x}) \phi({\bf y})]
\l{b11}
$$
and from formulas \r{6*} which can be rewritten as
$$
\begin{array}{c}
VA^{+}({\bf x}) V^{-1} = {\cal A}^{+}({\bf x}) \equiv
(\hat{\Gamma}^{-1/2}\hat{\cal R}^*\phi
-      \hat{\Gamma}^{-1/2}     \frac{1}{i}
\frac{\delta}{\delta\phi}) ({\bf x}), \\
VA^{-}({\bf x}) V^{-1} = {\cal A}^{-}({\bf x}) \equiv
(\hat{\Gamma}^{-1/2}\hat{\cal R}\phi
-      \hat{\Gamma}^{-1/2}     \frac{1}{i}
\frac{\delta}{\delta\phi}) ({\bf x}).
\end{array}
\l{b12}
$$
$|c|$  can be formally found from the normalization condition
$$
|c|^2 \int D\phi |\exp[\frac{i}{2}\int d{\bf x} d{\bf y} \phi({\bf x})
\tilde{\cal R}({\bf x},{\bf y}) \phi({\bf y})]|^2 = 1
\l{b13}
$$
The argument can be chosen to be arbitrary, for example,
$$
Arg c = 0.
\l{b14}
$$

{\bf Proposition 3.3.} {\it The operator  $V$ is defined form
the relations \r{b11} - \r{b14} uniquely.}

Namely, any element of the Fock space can be presented
\c{Ber} via its components, vacuum state an creation operators as
$$
\Psi = \sum_{n=0}^{\infty} \frac{1}{\sqrt{n!}}  \int  d{\bf  x}_1  ...
d{\bf x}_n \Psi_n({\bf x}_1,...,{\bf x}_n) A^+({\bf x}_1) ... A^+({\bf
x}_n) |0>
$$
Specify \footnotemark
$$
V\Psi = \sum_{n=0}^{\infty} \frac{1}{\sqrt{n!}}  \int  d{\bf  x}_1  ...
d{\bf x}_n  \Psi_n({\bf  x}_1,...,{\bf x}_n) {\cal A}^+({\bf x}_1) ...
{\cal A}^+({\bf
x}_n) V|0>.
$$
\footnotetext{The problem of divergence of the series is related with
the problem   of   correctness   of   the    functional    Schrodinger
representation. It is not investigated here}

Since the  operators ${\cal A}^{\pm}({\bf x})$ satisfy usual canonical
commutation relations and
${\cal A}^-({\bf x})|0>=0$,
we obtain $VA^{\pm}({\bf x}) = {\cal A}^{\pm}({\bf x})V$.

The operator $V$  depend on  $\cal  R$.  It is useful to find
an explicit form of the operator
$V^{-1}\delta V$.

{\bf Proposition 4.4.} {\it The following property is satisfied:
$$
\begin{array}{c}
V^{-1}\delta V = - \frac{i}{2} A^+ \hat{\Gamma}^{-1/2} \delta \hat{\cal R}
\hat{\Gamma}^{-1/2} A^+ - \frac{i}{2} A^- \hat{\Gamma}^{-1/2}
\delta  \hat{\cal  R}^*
\hat{\Gamma}^{-1/2} A^-   +
\\
  A^+   [\hat{\Gamma}^{1/2}   \delta  \hat{\Gamma}^{-1/2}  +
i\hat{\Gamma}^{-1/2} \delta \hat{\cal R}^* \hat{\Gamma}^{-1/2}] A^-
+  \frac{i}{4}  Tr
[\delta(\hat{\cal R} +  \hat{\cal R}^*) \hat{\Gamma}^{-1}]
\end{array}
\l{b15}
$$
}

The notations of the type
$A^+\hat{\cal  B}A^-$  are used for the operators like
$\int  d{\bf  x}  d{\bf y} A^+({\bf x}) \tilde{\cal B}({\bf
x},{\bf y}) A^-({\bf y}) $,  where $\tilde{\cal B}({\bf x},{\bf  y})$  is  a
kernel of the operator $\hat{\cal B}$.

To check formula \r{b15}, consider the variation of the formula
\r{6*} if $\cal R$ is varied:
$$
\begin{array}{c}
[A^{\pm}({\bf x})   ;   V^{-1}\delta   V]   =   (\hat{\Gamma}^{1/2}
\delta
\hat{\Gamma}^{-1/2} A^{\pm})({\bf  x}) \\ -  i(\hat{\Gamma}^{-1/2}
\delta\hat{\cal   R}
\hat{\Gamma}^{-1/2} A^+)  ({\bf  x})  +  i  (\hat{\Gamma}^{-1/2}
\delta\hat{\cal R}^*
\hat{\Gamma}^{-1/2} A^-)({\bf x}).
\end{array}
$$
Therefore, formula   \r{b15} is correct up to an additive constant.
To find it, note that
$$
\delta V  |0>  =  [\frac{i}{2}  \int  d{\bf  x} d{\bf y} \phi({\bf x})
\delta \tilde{\cal R}({\bf x},{\bf y}) \phi({\bf y}) + \delta ln c] V|0>.
$$
This relation and formula \r{6*} imply
$$
<0|V^{-1} \delta V|0> = \frac{i}{2} Tr(\delta \hat{\cal  R}
\hat{\Gamma}^{-1})  +
\delta ln c.
$$
It follows from the normalization conditions
\r{b13}  and  \r{b14} that $c=(det\hat{\Gamma})^{1/4}$.
Therefore, $\delta ln c = \frac{1}{4} Tr \delta \hat{\Gamma}
\hat{\Gamma}^{-1}$. Thus,   $<0|V^{-1}\delta  V|0>  =  \frac{i}{4}  Tr
\delta(\hat{\cal R}+\hat{\cal R}^*)\Gamma^{-1}$. Formula \r{b15} is checked.

\subsubsection{Explicit forms of Poincare generators}

Proposition 3.4  allows  us  to  find  an  explicit  forms of Poincare
generators. For  the  simplicity,  we  consider  the  case  when   the
quadratic form  $\cal  R$  is invariant under spatial translations and
rotations:
$$
\tilde{\cal R} ({\bf x},{\bf y}: u_{({\bf a},L)} X) =
\tilde{\cal R} (L^{-1}({\bf
x}- {\bf a}), L^{-1}({\bf y}-{\bf a}):X).
\l{f6}
$$
This property implies that
$$
[\partial_k; \hat{\cal R}] = \delta_P^k \hat{\cal R};
\qquad
[\partial_k; \hat{\Gamma}^{1/2}] = \delta_P^k \hat{\Gamma}^{1/2};
$$
$$
[(x^k\partial_l - x^l\partial_k); \hat{\cal R}] = \delta_M^{kl}
\hat{\cal R};
\qquad
[(x^k\partial_l -  x^l\partial_k);   \hat{\Gamma}^{1/2}]   =
\delta_M^{kl} \hat{\Gamma}^{1/2}.
\l{f7}
$$

{\bf 1. Spatial translations and rotations.}

Eqs.\r{f7}, proposition 3.4 and relation \r{6*} imply that formally
$$
\breve{P}^k = -iA^+ \partial_k A^- + i\delta_P^k; \qquad
\breve{M}^{kl} =  -  i  A^+  (x^k\partial_l  -  x^l\partial_k)  A^-  +
i\delta_M^{kl}.
\l{f8}
$$
These operators do not contain any divergences.

{\bf 2. Time evolution.}

Proposition 3.4 and eqs.\r{6*} imply that the operator  $\breve{H}(X)$
is also quadratic with respect to creation and annihilation operators,
$$
\breve{H}(X) =  i\delta_H  +  \frac{1}{2} A^-{\cal H}^{--}(X)A^- + A^+
(\hat{\omega} + {\cal H}(X))A^- + \frac{1}{2} A^+ {\cal H}^{++}(X) A^+ +
\overline{H}.
$$
Here ${\cal  H}^{\pm\pm}(X)$  and  ${\cal  H}(X)$  are  the  following
operators:
$$
\begin{array}{c}
{\cal H}^{++}(X) =
\hat{\Gamma}^{-1/2} [\delta_H \hat{\cal R} - \hat{\cal R} \hat{\cal R}
- (-\Delta  +  m^2  +  V''(\Phi({\bf   x}))]   \hat{\Gamma}^{-1/2};
\qquad
{\cal H}^{--}(X) = ({\cal H}^{++})^+;
\\
{\cal H}(X) =
\hat{\Gamma}^{-1/2} (\hat{\cal  R}\hat{\cal R}^*
+ (-\Delta + m^2 + V''(\Phi({\bf
x})) - \frac{1}{2} \delta_H(\hat{\cal  R}  +  \hat{\cal  R}^*)
+  \frac{i}{2}
[\delta_H \hat{\Gamma}^{1/2}; \hat{\Gamma}^{1/2}])\hat{\Gamma}^{-1/2}
- \hat{\omega}
\end{array}
$$
while
$$
\hat{\omega} = \sqrt{-\Delta +m^2}
$$
is a $(\Pi,\Phi)$-independent self-adjoint operator.

The divergent number $\overline{H}$ is formally equal to
$$
\overline{H} =  \frac{1}{2}  Tr  [\hat{\Gamma}^{-1/2}
(\hat{\cal  R}\hat{\cal  R}^* +
(-\Delta + m^2  +  V''(\Phi({\bf  x}))  -  \frac{1}{2}  \delta_H(\hat{\cal
R}+\hat{\cal R}^*) )\hat{\Gamma}^{-1/2} ].
\l{b25}
$$

{\bf 3. The {\bf n}-boost}

For boost transformation, we obtain that
$$
\breve{B}^k(X) =  i\delta_B^k  +  \frac{1}{2} A^-{\cal B}^{k --}(X)A^- + A^+
(L_k + {\cal B}^k(X))A^- + \frac{1}{2} A^+ {\cal B}^{k ++}(X) A^+ +
\overline{B^k}.
$$
Here
$$
\begin{array}{c}
{\cal B}^{k++}(X) =
\hat{\Gamma}^{-1/2} [\delta_k^B \hat{\cal R} - \hat{\cal R}
x^k \hat{\cal R} - (-\partial_ix^k \partial_i + x^km^2 + x^k V''(\Phi({\bf
x})))]\hat{\Gamma}^{-1/2};
\\
{\cal B}^{k--} = ({\cal B}^{k++})^+;
\\
{\cal B}^k =
\hat{\Gamma}^{-1/2} [\hat{\cal R} x^k \hat{\cal R}^* +
(-\partial_ix^k \partial_i + x^km^2 + x^k V''(\Phi({\bf x})))
- \frac{1}{2}   \delta^k_B(\hat{\cal   R}  +  \hat{\cal  R}^*)  + \\
\frac{i}{2}
[\delta_k^B \hat{\Gamma}^{1/2}, \hat{\Gamma}^{1/2}]
] \hat{\Gamma}^{-1/2} - L_k,
\end{array}
\l{b28}
$$
while
$$
L_k = \frac{1}{2} \hat{\omega}^{-1/2} [\hat{\omega} x^k \hat{\omega} +
(-\partial_i x^k \partial_i + x^k m^2) ] \hat{\omega}^{-1/2}
$$
is a boost generator in  the  free  theory  which  is  a  self-adjoint
operator.

The divergent term is
$$
\overline{B^k} = \frac{1}{2} Tr \hat{\Gamma}^{-1}
[ \hat{\cal R} x^k \hat{\cal  R}^*
+ (-\partial_ix^k \partial_i + x^km^2 + x^k V''(\Phi({\bf x})))
- \frac{1}{2}   \delta^k_B(\hat{\cal   R}  +  \hat{\cal  R}^*)]
\l{b29}
$$

\subsubsection{Check of algebraic conditions and renormalization}

Let us write down the requirements which are sufficient for satisfying
the properties H1-H6 of the Appendix A.  Since the Poincare generators
are quadratic with respect to creation and annihilation operators,  we
will use the results of Appendix B.

{\bf 1.} Let
$$
\hat{K} = \hat{\omega}^{-1/4} ({\bf x}^2+1)^{-1} \hat{\omega}^{-1/4}.
$$
This is  a  bounded  self-adjoint  operator  without zero eigenvalues.
Therefore, $\hat{K}^{-1}   \equiv   T^{1/2}$   is   a    (non-bounded)
self-adjoint operator and
$$
T = \hat{\omega}^{1/4}  ({\bf x}^2 + 1) \hat{\omega}^{1/2}
({\bf x}^2 + 1) \hat{\omega}^{1/4}.
$$
By ${\cal D} \subset
{\cal F}$ we denote the domain $\{\psi\in {\cal F}  |  ||\psi||^T_1  <
\infty\}$.

{\bf Lemma 3.5.} {\it For self-adjoint operators
$$
A_k = L_k,  \qquad
A_{d+k} = - i\partial_k, \qquad
A_{2d+kd+l} = - i(x^k\partial_l - x^l\partial_k),
\qquad A_{2d+d^2+1} = \hat{\omega}
$$
the following properties are satisfied:
1. $||T^{-1/2}A_jT^{-1/2}|| <\infty$, $||A_jT^{-1}|| < \infty$.\\
2. $||T^{1/2} e^{iA_jt} T^{-1/2}|| \le C$,  $||Te^{-iA_jt} T^{-1}|| \le
C$, $t \in [0,t_1]$.
}

{\bf Proof.}  The  first  part  of lemma is justified as follows.  One
should check that the following norms are finite:
$$
\begin{array}{c}
||\hat{\omega}^{-1/4} ({\bf x}^2+1)^{-1} \hat{\omega}^{-1/4}
\hat{\omega} \hat{\omega}^{-1/4}          ({\bf           x}^2+1)^{-1}
\hat{\omega}^{-1/4} ||; \\
||\hat{\omega}^{-1/4} ({\bf x}^2+1)^{-1} \hat{\omega}^{-1/4}
\hat{\omega} x^s \hat{\omega}^{-1/4}          ({\bf           x}^2+1)^{-1}
\hat{\omega}^{-1/4} ||; \\
||\hat{\omega}^{-1/4} ({\bf x}^2+1)^{-1} \hat{\omega}^{-1/4}
(\hat{k}^j x^s - \hat{k}^s x^j)
\hat{\omega}^{-1/4}          ({\bf           x}^2+1)^{-1}
\hat{\omega}^{-1/4} ||; \\
||\hat{\omega}^{-1/4} ({\bf x}^2+1)^{-1} \hat{\omega}^{-1/4}
\hat{k}^j \hat{\omega}^{-1/4}          ({\bf           x}^2+1)^{-1}
\hat{\omega}^{-1/4} ||; \\
||\hat{\omega}\hat{\omega}^{-1/4} ({\bf x}^2+1)^{-1} \hat{\omega}^{-1/4}
 \hat{\omega}^{-1/4}          ({\bf           x}^2+1)^{-1}
\hat{\omega}^{-1/4} ||; \\
||\hat{\omega} x^s \hat{\omega}^{-1/4} ({\bf x}^2+1)^{-1} \hat{\omega}^{-1/4}
 \hat{\omega}^{-1/4}          ({\bf           x}^2+1)^{-1}
\hat{\omega}^{-1/4} ||; \\
||
(\hat{k}^j x^s - \hat{k}^s x^j)
\hat{\omega}^{-1/4} ({\bf x}^2+1)^{-1} \hat{\omega}^{-1/4}
\hat{\omega}^{-1/4}          ({\bf           x}^2+1)^{-1}
\hat{\omega}^{-1/4} ||; \\
||\hat{k}^j
\hat{\omega}^{-1/4} ({\bf x}^2+1)^{-1} \hat{\omega}^{-1/4}
\hat{\omega}^{-1/4}          ({\bf           x}^2+1)^{-1}
\hat{\omega}^{-1/4} ||,
\end{array}
$$
where $\hat{k}^j = - i \partial/\partial x^j$.
This statement is a corollary of the following lemma.

{\bf Lemma 3.6.} {\it  The operators
$$
[\hat{\omega}^{\alpha}; ({\bf x}^2+1)^{-1} ];
\qquad
[\hat{\omega}^{\alpha}; x^s({\bf x}^2+1)^{-1} ];
\qquad
[\hat{\omega}^{\alpha}; x^lx^s({\bf x}^2+1)^{-1} ]
$$
are bounded if $\alpha \le 1$.
}

This lemma is a corollary of Lemma C.29 of Appendix C.

To prove the second statement  of  lemma  3.5,  represent  it  in  the
following form:
$$
||e^{iA_jt}T^{1/2}e^{-iA_jt} T^{-1/2}||    \equiv    ||   T_j^{1/2}(t)
T^{-1/2}|| \le C; \qquad ||T_j(t)T^{-1}|| \le C.
\l{f11}
$$
It is necessary to investigate the Poincare transformation  properties
of the operators $\hat{x}^j$ and $\hat{k}^j$.

{\bf Lemma 3.7.} {\it The following relations are satisfied:
$$
\begin{array}{c}
e^{i\hat{\omega}t} \hat{x}^l   e^{-i\hat{\omega}t}   =   \hat{x}^l   +
\hat{k}^l \hat{\omega}^{-1} t,
\qquad
e^{i\hat{\omega}t} \hat{k}^l   e^{-i\hat{\omega}t}   =   \hat{k}^l;
\\
e^{i\hat{k}^sa^s} \hat{x}^l  e^{-  i\hat{k}^sa^s}  =  \hat{x}^l + a^l;
\qquad
e^{i\hat{k}^sa^s} \hat{k}^l  e^{-  i\hat{k}^sa^s}  =  \hat{k}^l;
\\
e^{\frac{i\tau}{2} \theta_{ms}    (\hat{x}^m\hat{k}^s    -   \hat{x}^s
\hat{k}^m)}
\hat{x}^l
e^{- \frac{i\tau}{2} \theta_{ms}    (\hat{x}^m\hat{k}^s    -   \hat{x}^s
\hat{k}^m)} =
(e^{-\tau\theta} \hat{x})^l;
\\
e^{\frac{i\tau}{2} \theta_{ms}    (\hat{x}^m\hat{k}^s    -   \hat{x}^s
\hat{k}^m)}
\hat{k}^l
e^{- \frac{i\tau}{2} \theta_{ms}    (\hat{x}^m\hat{k}^s    -   \hat{x}^s
\hat{k}^m)} =
(e^{-\tau\theta} \hat{k})^l;
\\
e^{iL^1\tau} \hat{k}^l e^{-iL^1\tau} = \hat{k}^l , \quad  l \ge 2;
\qquad
e^{iL^1\tau} \hat{k}^1   e^{-iL^1\tau}   =   \hat{k}^1   \cosh   \tau   -
\hat{\omega} \sinh\tau.
\end{array}
$$
The operators $\hat{X}^l(\tau) = e^{iL^1\tau} \hat{x}^l e^{-iL^1\tau}$
have the following Weyl symbols:
$$
X^1 = \frac{\omega_{\bf k}}{\omega_{\bf  k}\cosh\tau  -  k^1\sinh\tau}  x^1;
\qquad
X^{\alpha} = x^{\alpha} + \frac{k^{\alpha}\sinh\tau x^1}
{\omega_{\bf  k}\cosh\tau  -  k^1\sinh\tau}
$$
}

To check  the  properties,  it  is  sufficient  to  show that they are
satisfied at $\tau=0$ and show that the derivatives of  left-hand  and
right-hand sides of these relations coincide.

Making use    of   commutation   relations   $[x^s,f(\hat{k})]   =   i
\frac{\partial f }{\partial k^s}(\hat{k})$ and result of lemma  3.6,  we
find that  operators  \r{f11} are bounded uniformly with respect to $t
\in [0,t_1]$. Lemma 3.5 is proved.

Let the following conditions on $\cal R$ be imposed.

Let $h(\alpha)$ be an arbitrary smooth curve on the Poincare group.

P1, {\it The property \r{f6} is satisfied.}

P2. {\it The  $\alpha$-dependent  operator  functions  $T  {\cal   B}^{k++}
(u_{h(\alpha)}X)$ and $T{\cal H}^{++}(u_{h(\alpha)}X)$ are continuous in
the Hilbert-Schmidt topology $||\cdot||_2$.}

P3. {\it The     $\alpha$-dependent     operator      functions      ${\cal
B}^{k++}(u_{h(\alpha)}X)$ and   ${\cal   H}^{++}(u_{h(\alpha)}X)$  are
continuously differentiable   with   respect   to   $\alpha$   in   the
Hilbert-Schmidt topology.}

P4. {\it The      $\alpha$-dependent      operator     functions     ${\cal
B}^k(u_{h(\alpha)}X)$, ${\cal       H}(u_{h(\alpha)}X)$,
$T{\cal B}^k(u_{h(\alpha)}X)T^{-1}$,
$T^{1/2}{\cal B}^k(u_{h(\alpha)}X)T^{-1/2}$,
$T{\cal H}(u_{h(\alpha)}X)T^{-1}$,
$T^{1/2}{\cal H}(u_{h(\alpha)}X)T^{-1/2}$ are strongly continuous.}

P5. {\it The  $\alpha$-dependent   operator   functions
$T^{-1/2}   {\cal H}(u_{h(\alpha)}X)T^{-1/2}$,
$T^{-1/2}   {\cal B}^k(u_{h(\alpha)}X)T^{-1/2}$,
${\cal H}(u_{h(\alpha)}X)T^{-1}$,
${\cal B}^k(u_{h(\alpha)}X)T^{-1}$
are continuously   differentiable  with  respect  to  $\alpha$  in  the
operator norm $||\cdot||$ topology.}

P6. {\it The functions $\overline{H}(u_{h(\alpha)}X)$ and
 $\overline{B^k}(u_{h(\alpha)}X)$ are continuous.}

{\bf Lemma   3.6.}  {\it Let  the  properties  P1-P6  be  satisfied.  Then
properties H1, H2, H4-H6 are also satisfied.}

{\bf Proof.} Property H1 is a corollary of  estimations  performed  in
lemmas B.1,  B.2,  B.3  of  Appendix B.  Property H4 is a corollary of
theorem B.15. Properties H2 and H5 are obtained from lemma B.4. Making
use of the results of lemmas B.1,  B.2, B.3 and property $||U_B^{\tau}
\Psi - \Psi||^T_1 \to_{\tau\to 0} 0$ obtained in theorem B.15, we find
that property H6 is satisfied. Lemma is proved.

{\bf 2.} Let us check the commutation relations \r{x6},  i.e. property
H3. Note that the divergences  arise  in  terms  $\overline{B^k}$  and
$\overline{H}$ only,  so that we suppose them to be arbitrary and then
find the conditions that provide Poincare invariance.

Let
$$
\breve{H}_k = \frac{1}{2} A^+ {\cal H}_k^{++} A^+ + A^+ {\cal  H}^{+-}
A^- + \frac{1}{2} A^-{\cal H}^{--} A^- + \overline{H_k} + i\delta_k
$$
be arbitrary quadratic Hamiltonians.  Then the property $[\breve{H}_1,
\breve{H}_2] = \breve{H}_3$ under condition $[i\delta_1,  i\delta_2] =
i\delta_3$  means that
$$
{\cal H}_3^{++}  =  -  i  [
{\cal  H}_1^{+-}  {\cal  H}_2^{++}  +
{\cal H}_2^{++} ({\cal H}_1^{+-})^* -
{\cal H}_1^{++} ({\cal H}_2^{+-})^* -
{\cal  H}_2^{+-}  {\cal  H}_1^{++}]
+ \delta_1 {\cal H}_2^{++} - \delta_2 {\cal H}_1^{++}.
\l{g1}
$$
$$
{\cal H}_3^{+-} = -i\{
{\cal H}_2^{++} ({\cal H}_1^{++})^*
- {\cal H}_1^{++} ({\cal H}_2^{++})^*
+ [{\cal H}_1^{+-}; ({\cal H}_2^{+-})^*]\}
+ \delta_1 {\cal H}_2^{+-} - \delta_2 {\cal H}_1^{+-},
\l{g2}
$$
$$
\overline{H_3} = - \frac{i}{2} Tr [
{\cal H}_2^{++} ({\cal H}_1^{++})^* -
{\cal H}_1^{++}  ({\cal  H}_2^{++})^*]  +  \delta_1  \overline{H_2} -
\delta_2 \overline{H_1}.
\l{g3}
$$
Relations \r{g1},  \r{g2},  \r{g3} are treated in  sense  of  bilinear
forms on $D(T)$.

Consider now the commutation relations.

{\bf 1.} The relations
$$
[\breve{P}^k, \breve{P}^l] = 0, \qquad [\breve{M}^{lm}, \breve{P}^s] =
i(g^{ms}\breve{P}^l - g^{ls} \breve{P}^m]
$$
are satisfied automatically since
$$
[\partial_k, \partial_l] = 0, \qquad
- [x^l\partial_m - x^m\partial_l ,  \partial_s ] = g^{ms} \partial_l -
g^{ls}\partial_m.
$$

{\bf 2.} The relation
$$
[\breve{M}^{lm}, \breve{M}^{rs}]  = -i (g^{lr} \breve{M}^{ms} - g^{mr}
\breve{M}^{ls} + g^{ms} \breve{M}^{lr} - g^{ls} \breve{M}^{mr} )
$$
is also satisfied.

{\bf 3.} For the relation
$$
[\breve{P}^k, \breve{P}^0] = 0
$$
eqs \r{g1}- \r{g3} takes the form
$$
\delta_P^k {\cal H}^{++} - [\partial_k; {\cal H}^{++}] =0,
\qquad
\delta_P^k {\cal H}^{+-} - [\partial_k; {\cal H}^{+-}] =0,
\l{g4}
$$
$$
\delta_P^k \overline{H} =0.
\l{g4a}
$$

4. For the relation
$$
[\breve{M}^{kl}, \breve{P}^0] =0,
$$
eqs.\r{g1} - \r{g3} are written as
$$
\delta_M^{kl} {\cal H}^{++} -
[ x^k \partial_l - x^l \partial_k ; {\cal H}^{++} ] = 0;
\qquad
\delta_M^{kl} {\cal H}^{+-} -
[ x^k \partial_l - x^l \partial_k ; {\cal H}^{+-} ] = 0;
\l{g5}
$$
$$
\delta_M^{kl} \overline{H} = 0.
\l{g5a}
$$

5. Consider the relation
$$
[\breve{M}^{k0}, \breve{P}^s] = -ig^{ks}\breve{P}^0.
$$
We write eqs.\r{g1} - \r{g3} as follows:
$$
[\partial_s ,  {\cal B}^{k++} ] - \delta_P^s {\cal B}^{k++} = - g^{ks}
{\cal H}^{++},
\qquad
[\partial_s ,  {\cal B}^{k+-} ] - \delta_P^s {\cal B}^{k+-} = - g^{ks}
{\cal H}^{+-},
\l{g6}
$$
$$
\delta_P^s \overline{B^k} = g^{ks} \overline{H}.
\l{g6a}
$$

6. The commutation relation
$$
[\breve{M}^{lm}, \breve{M}^{k0} ] = -i (g^{lk} \breve{M}^{m0} - g^{mk}
\breve{M}^{l0})
$$
is equivalent to
$$
\begin{array}{c}
[x^l\partial_m -  x^m  \partial_l  ;  {\cal B}^{k++} ] - \delta_M^{lm}
{\cal B}^{k++} = g^{lk} {\cal B}^{m++} - g^{mk} {\cal B}^{l++},
\end{array}
$$
$$
\begin{array}{c}
[x^l\partial_m -  x^m  \partial_l  ;  {\cal B}^{k+-} ] - \delta_M^{lm}
{\cal B}^{k+-} = g^{lk} {\cal B}^{m+-} - g^{mk} {\cal B}^{l+-},
\end{array}
\l{g7}
$$
$$
- \delta_M^{kl}   \overline{B^k}  =  g^{lk}  \overline{B^m}  -  g^{mk}
\overline{B^l}.
\l{g7a}
$$

7. The most nontrivial commutation relations are
$$
[\breve{M}^{k0}; \breve{P}^0] = i \breve{P}^k,
\qquad
[\breve{M}^{k0}; \breve{M}^{l0}] = -i\breve{M}^{kl}
$$
They can be rewritten as follows:
$$
\begin{array}{c}
0 = -i \{
{\cal B}^{k+-} {\cal H}^{++} +
{\cal H}^{++} ({\cal B}^{k+-})^*
-
{\cal B}^{k++} ({\cal H}^{+-})^* -
{\cal H}^{+-} ({\cal B}^{k++})
\} + \delta_B^k {\cal H}^{++} - \delta_H {\cal B}^{k++};
\\
-i \partial_k = -i \{
{\cal H}^{++} {\cal B}^{k++}
- {\cal B}^{k++} ({\cal H}^{++})^* +
[{\cal B}^{k+-}; {\cal H}^{+-}]
\} + \delta_B^k {\cal H}^{+-} - \delta_H {\cal B}^{k+-};
\end{array}
\l{g8}
$$
$$
0 = -  \frac{i}{2}  Tr  [{\cal  H}^{++}  ({\cal  B}^{k++})^*  -  {\cal
B}^{k++} ({\cal  H}^{++})^*]  +  \delta_B^k  \overline{H}  -  \delta_H
\overline{B^k}
\l{g8a}
$$
and
$$
\begin{array}{c}
0 = -i \{
{\cal B}^{k+-} {\cal B}^{l++} +
{\cal B}^{l++} ({\cal B}^{k+-})^*
-
{\cal B}^{k++} ({\cal B}^{l+-})^* -
{\cal B}^{l+-} ({\cal B}^{k++})
\} + \delta_B^k {\cal B}^{l++} - \delta_B^l {\cal B}^{k++};
\\
i(x^k \partial_l - x^l\partial_k) = -i \{
{\cal B}^{l++} {\cal B}^{k++}
- {\cal B}^{k++} ({\cal B}^{l++})^* +
[{\cal B}^{k+-}; {\cal B}^{l+-}]
\} + \delta_B^k {\cal B}^{l+-} - \delta_B^l {\cal B}^{k+-};
\end{array}
\l{g9}
$$
$$
0 = -  \frac{i}{2}  Tr  [{\cal  B}^{l++}  ({\cal  B}^{k++})^*  -  {\cal
B}^{k++} ({\cal B}^{l++})^*] + \delta_B^k \overline{B^l} - \delta_B^l
\overline{B^k}.
\l{g9a}
$$

{\bf 3.}  Properties \r{g4},  \r{g5},  \r{g6},  \r{g7} are obvious
corollaries of relations \r{f7}.  Properties  \r{g8}  and  \r{g9}  are
checked by nontrivial but also direct computations.

To justify the properties \r{g4a}, \r{g5a}, \r{g6a}, \r{g7a}, \r{g8a},
\r{g9a}, let  us   extract   divergences   from   $\overline{H}$   and
$\overline{B^k}$. Notice that formally
$$
\overline{H} = \overline{H}_{reg} + \frac{1}{4} Tr \hat{\Gamma},
\qquad
\overline{B^k} = \overline{B^k}_{reg} + \frac{1}{4} Tr x^k \hat{\Gamma}
$$
with
$$
\overline{H}_{reg} = - \frac{1}{4} Tr [{\cal H}^{++} + {\cal H}^{--}],
\qquad
\overline{B^k}_{reg} = - \frac{1}{4} Tr [{\cal B}^{k++}
+ {\cal B}^{k--}].
\l{g10}
$$
Expressions \r{g10}  are  well-defined  if  we  impose  the  following
additional condition.

P7. {\it The operators ${\cal B}^{k++}$ and  ${\cal  H}^{++}$  are  of  the
trace class and $Tr {\cal B}^{k++}(u_{h(\alpha)}X)$
and $Tr {\cal  H}^{++}(u_{h(\alpha)}X)$  are  continuous  functions  of
$\alpha$.}

To perform  a renormalization of the semiclassical theory,  one should
substitute the  divergent  terms  $Tr  \hat{\Gamma}$   and   $Tr   x^k
\hat{\Gamma}$  by
finite terms  which  will  be  denoted  as  $Tr_R  \hat{\Gamma}$  and  $Tr_R
x^k\hat{\Gamma}$,
$$
\overline{H} = \overline{H}_{reg} + \frac{1}{4} Tr_R \hat{\Gamma},
\qquad
\overline{B^k} =   \overline{B^k}_{reg}   +   \frac{1}{4}   Tr_R   x^k
\hat{\Gamma} .
$$
Relations
\r{g4a}, \r{g5a}, \r{g6a}, \r{g7a}, \r{g8a},
\r{g9a} are straightforwardly checked under the following condition.

P8. {\it The quantities $Tr_R \hat{\Gamma}$ and $Tr_R x^k\hat{\Gamma}$
obey  the following properties:
$$
\begin{array}{c}
\delta^P_k Tr_R \hat{\Gamma} = 0; \qquad
\delta^M_{kl} Tr_R\hat{\Gamma} = 0;
\\
\delta^P_l Tr_R x^k \hat{\Gamma} = - \delta^{kl} Tr_R \hat{\Gamma};
\qquad
\delta^M_{kl} Tr_R x^k\hat{\Gamma} = \delta^{kl} Tr_R x^m \hat{\Gamma}
- \delta^{mk} Tr_R x^l \hat{\Gamma};\\
Tr [x^l  (\delta^B_k  \hat{\Gamma}  -  \hat{\cal  A}  x^k  \hat{\Gamma}  -
\hat{\Gamma} x^k \hat{\cal A})
- x^k  (\delta^B_l  \hat{\Gamma}  -  \hat{\cal  A}  x^l \hat{\Gamma} -
\hat{\Gamma} x^l \hat{\cal A})]
+ \delta_l^B Tr_R x^k\hat{\Gamma} - \delta_k^B Tr_R x^l \hat{\Gamma} = 0; \\
Tr [x^l  (\delta^H  \hat{\Gamma}  -  \hat{\cal   A}   \hat{\Gamma}   -
\hat{\Gamma} \hat{\cal A})
- (\delta^B_l  \hat{\Gamma}  -  \hat{\cal  A}   x^l   \hat{\Gamma}   -
\hat{\Gamma} x^l \hat{\cal A})]
+ \delta_l^B Tr_R \hat{\Gamma} - \delta^H Tr_R x^l \hat{\Gamma} = 0,
\end{array}
$$
where $\hat{\cal A}  = \frac{1}{2} (\hat{\cal R} + \hat{\cal R}^*)$.
}

Note also  that  property  P6  can  be  substituted  by  the following
property.

P9. {\it The functions $Tr_R \Gamma(u_{h(\alpha)}X)$ and
$Tr_R x^k\Gamma(u_{h(\alpha)}X)$ are continuous.}

Thus, we   have   formulated  the  conditions  of  invariance  of  the
semiclassical field theory under Poincare algebra.

\subsection{Construction of Poincare transformations}

Let us construct now the operators $U_{a,\Lambda}$.

{\bf 1.} First of all,  consider the case of spatial rotations, $a=0$,
$\Lambda=L$. Denote  $U_{0,L}  \equiv  {\cal V}_L$ for this case.  Let
$L(\tau) = \exp(\frac{\tau}{2} \theta_{sm} l^{sm})$. Then the operator
${\cal V}_{L(\tau)}$ transforms the initial condition for the equation
$$
i \dot{\Psi}^{\tau}  =  \frac{i}{2} \theta_{sm} A^+ (x^s\partial_m -
x^m \partial_s)A^- \Psi^{\tau}, \qquad \Psi^{\tau} \in {\cal D}
\l{w30}
$$
to the solution. The operator ${\cal V}_{L(\tau)}$ is uniquely defined
from the relations
$$
{\cal V}_{L} |0> = |0>, \qquad
{\cal V}_L A^{\pm}({\bf x}) {\cal V}_L^{-1} = A^{\pm} (L{\bf x}).
\l{h1}
$$
The group property for operators ${\cal V}_L$ is obviously satisfied.

{\bf 2.}   Let  $(a,\Lambda)$  is  an  $x^1$-boost:  $a=0$,  $\Lambda=
\Lambda_{\tau}$ has   the   form    \r{e1}.    Then    the    operator
$U_{0,\Lambda_{\tau}}\equiv W_{\tau}$  takes the initial condition for
the Cauchy problem for the equation
$$
i\dot{\Psi}^{\tau} =    B^1(S^{\tau},     \Phi^{\tau},     \Pi^{\tau})
\Psi^{\tau}
\l{w32}
$$
to the  solution  of this Cauchy problem.  Here $S^{\tau},\Phi^{\tau},
\Pi^{\tau}$ are obtained from system \r{e12}.

{\bf 3.} Let $L$ be such a rotation that $L{\bf  e}_1  =  {\bf  e}_1$.
Then there  exists  a  matrix  smooth  function  $L(t)$  such  that
$L(0)=1$, $L(1)=L$,  $L(t){\bf e}_1 = {\bf e}_1$.  One has $L(t)
\Lambda_{t\tau} L(\tau)^{-1}        \Lambda_{t\tau}^{-1}=1$
for all $\tau \in {\bf R}$, so that condition
\r{x11} of lemma A.8 is satisfied. Therefore,
$$
W_{\tau} {\cal V}_L = {\cal V}_L W_{\tau},
\l{h2}
$$
provided that $L{\bf e}_1 = {\bf e}_1$.

{\bf 4}.  Let $|{\bf n}|=1$,  $L_{\bf n}$ be such a  spatial  rotation
that $L_{\bf  n}{\bf  n}= {\bf e}_1$.  Introduce the operator $W_{\phi
{\bf n}}$ as follows,
$$
W_{\phi{\bf n}} = {\cal V}_L^{-1} W_{\phi} {\cal V}_L.
\l{h3}
$$
It corresponds to the Lorentz transformation
$$
\Lambda_{\phi {\bf n}} = L_{\bf n}^{-1} \Lambda_{\phi} L_{\bf n}.
$$
Show that definition \r{h3} is correct. Let
$L_{\bf n}^{(1)} {\bf n} = {\bf e}_1$,
$L_{\bf n}^{(2)} {\bf n} = {\bf e}_1$. One has $L_{\bf n}^{(1)} (L_{\bf
n}^{(2)})^{-1} {\bf e}_1 = {\bf e}_1$,  so that formula \r{h2} implies
that
$$
({\cal V}_L^{(1)})^{-1} W_{\phi} {\cal V}_L^{(1)} =
({\cal V}_L^{(2)})^{-1} W_{\phi} {\cal V}_L^{(2)}.
$$
Note that the following property is satisfied.

Let $L$ be a spatial rotation. Then
$$
L \Lambda_{\vec{\phi}} L^{-1} = \Lambda_{L\vec{\phi}},
\qquad
{\cal V}_L W_{\vec{\phi}} {\cal V}_L^{-1} = W_{L\vec{\phi}},
\l{h4}
$$
Namely, let  $L_{\bf  n}$  be  such a rotation that $L_{\bf n}{\bf n}=
{\bf e}_1$, where ${\bf n}= \vec{\phi}/||\vec{\phi}||$. Then
$$
\begin{array}{c}
L \Lambda_{\vec{\phi}} L^{-1} = LL_{\bf n}^{-1} \Lambda_{\phi}  L_{\bf
n}L^{-1} = \Lambda_{L\vec{\phi}},\\
{\cal V}_L W_{\vec{\phi}} {\cal V}_L^{-1} =
{\cal V}_L {\cal V}^{-1}_{L{\bf n}}
W_{\vec{\phi}} {\cal V}_{L{\bf n}} {\cal V}_L^{-1} =
W_{L\vec{\phi}}.
\end{array}
$$

{\bf 5.} Let $\Lambda$ be an arbitrary Lorentz  transformation.  Lemma
3.1 implies that it can be uniquely decomposed as follows,  $\Lambda =
\Lambda_{\vec{\phi}} L$. Set
$$
U_{\Lambda} = W_{\vec{\phi}} {\cal V}_L.
\l{h4*}
$$
Let us check the group property. Let
$$
\Lambda_1= \Lambda_{\vec{\phi}_1} L_1,
\qquad
\Lambda_2= \Lambda_{\vec{\phi}_2'} L_2
$$
be Lorentz transformations. Then
$$
\Lambda_1\Lambda_2 = \Lambda_{\vec{\phi}_1} \Lambda_{L_1\vec{\phi}_2'}
L_1L_2 = \Lambda_{\vec{\phi}_3(\vec{\phi}_1, L_1\vec{\phi}_2')}
L(\vec{\phi}_1, L_1\vec{\phi}_2') L_1 L_2,
$$
where $\vec{\phi}_3$  and  $L$  are  defined  from the construction of
proof of lemma 3.1. One should check that
$$
U_{\Lambda_1} U_{\Lambda_2} = U_{\Lambda_1\Lambda_2},
$$
i.e.
$$
W_{\vec{\phi}_1} W_{\vec{\phi}_2} =
W_{\vec{\phi}_3(\vec{\phi}_1, \vec{\phi}_2)}
{\cal V}_{L(\vec{\phi}_1, \vec{\phi}_2)},
\l{h5}
$$
where $\vec{\phi}_2 = L_1 \vec{\phi}_2'$,
$$
\Lambda_{\vec{\phi}_1} \Lambda_{\vec{\phi}_2}
= \Lambda_{\vec{\phi}_3(\vec{\phi}_1, \vec{\phi}_2)}
{L(\vec{\phi}_1, \vec{\phi}_2)}
\l{h6}
$$
Consider the functions $\alpha \vec{\phi}_1$ and $\alpha \vec{\phi}_2$
instead of fixed vectors $\vec{\phi}_1$ and $\vec{\phi}_2$. Then
$$
\Lambda_{\alpha \vec{\phi}_1} \Lambda_{\alpha \vec{\phi}_2}
\Lambda_{\vec{\phi}_3(\alpha \vec{\phi}_1, \alpha \vec{\phi}_2)}^{-1}
{L(\alpha\vec{\phi}_1, \alpha\vec{\phi}_2)}^{-1} = 1
\l{h7}
$$
so that we can apply the result of lemma A8:
$$
W_{\alpha\vec{\phi}_1} W_{\alpha\vec{\phi}_2}
W_{\vec{\phi}_3(\alpha\vec{\phi}_1, \alpha\vec{\phi}_2)}^{-1}
{\cal V}_{L(\alpha\vec{\phi}_1, \alpha\vec{\phi}_2)}^{-1} = 1.
$$
Thus, eq.\r{h5} is satisfied.  We have checked the group property  for
the Lorentz transformations.

{\bf 6.}  Consider  the  spatial translations,  $a^0=0$,  $\Lambda=1$,
${\bf a}\ne 0$.  Then the operator $U_{0,{\bf a},1}$ takes the initial
condition for the equation
$$
i \dot{\Psi}^{\tau}  = a^k A^+ (-i\partial_k) A^- \Psi^{\tau},  \qquad
\Psi^{\tau} \in {\cal D}
\l{w28}
$$
to the solution of this  equation  at  $\tau=1$.  Thus,  the  operator
$U_{0,{\bf a},1}$ is uniquely defined from the relations
$$
U_{0, {\bf a},1} |0> = |0>, \qquad
U^{-1}_{0, {\bf a},1} A^{\pm}({\bf x}) U_{0, {\bf a},1} =
A^{\pm}({\bf x} + {\bf a}).
\l{h8}
$$

{\bf 7.}  For  the  evolution transformation,  $a^0=-t$,  ${\bf a}=0$,
$\Lambda=1$, the operator $U_{-t,0,1}$ takes the initial condition for
the equation
$$
i\dot{\Psi}^{\tau} =   H(S^t,\Phi^t,\Pi^t)\Psi^t,  \qquad  \Psi^t  \in
{\cal D}
\l{w31}
$$
to the solution of the Cauchy problem.  Here $S^t$,  $\Phi^t$, $\Pi^t$
obey system \r{3a}, \r{3}.

{\bf 8.} Consider an arbitrary space-time translation $(a,1)$. Set
$$
U_{(a,1)} \equiv U_{a^0,0,1} U_{0, {\bf a},1}.
\l{h8a}
$$
The group property is a corollary of the relation
$$
U_{0,\tau {\bf a},1} U_{\tau a^0,0,1}
= U_{\tau a^0,0,1} U_{0,\tau {\bf a},1}
$$
which is correct because of lemma A.8.

{\bf 9.}  Consider an arbitrary Poincare transformation $(a,\Lambda)$.
Set
$$
U_{(a,\Lambda)} \equiv U_{(a,1)} U_{(0,\Lambda)}.
\l{h8*}
$$
One should check the group property
$$
U_{(a_1+\Lambda_1a_2, \Lambda_1\Lambda_2)}    =    U_{(a_1,\Lambda_1)}
U_{(a_2,\Lambda_2)}
$$
or
$$
U_{(a_1,1)} U_{(\Lambda_1a_2,1)} U_{(0,\Lambda_1)} U_{(0,\Lambda_2)} =
U_{(a_1,1)} U_{(0,\Lambda_1)} U_{(a_2,1)}  U_{(0,\Lambda_2)}
$$
since the    group    properties    for   translations   and   Lorentz
transformations have  been  already  checked.  We  see  that   it   is
sufficient to justify the property
$$
U_{(\Lambda a, 1)} = U_{(0,\Lambda)} U_{(a,1)}  U_{(0,\Lambda)}^{-1}
\l{h10}
$$
Since the operator $U_{(0,\Lambda)}$ was defined as
$$
U_{(0,\Lambda)} = {\cal V}_{L_1} W_{{\phi}} {\cal V}_{L_2}
$$
if $\Lambda  =  L_1\Lambda_{\phi}  L_2$,  where   $L_1$,   $L_2$   are
rotations, while
$$
U_{(a,1)} = U_{(a^0,0,1)} U_{(0,{\bf a},1)},
$$
it is sufficient to check eq.\r{h10} for the following cases:

1. $a^0=0$, $\Lambda=L$;\\

2. ${\bf a}=0$, $\Lambda=L$,\\

3. $a^0=0$, $\Lambda = \Lambda_{\tau}$;\\

4. ${\bf a}=0$, $\Lambda = \Lambda_{\tau}$.

Lemma A.8 imply all these properties.

Therefore, we have proved the following statement.

{\bf Theorem 3.6.} {\it Let conditions H1-H5, H7-H9 be satisfied. Then
the operators $U_{(a,\Lambda)}(u_{a,\Lambda}X \gets X)$defined by eqs.
\r{h8*}, \r{h8a}, \r{h4*} are unitary and satisfy the group property.
}

\subsection{Choice of the operator $\cal R$}

Let us  choose operator $\cal R$ in order to satisfy properties P1-P5,
P7. We will use the notions of Appendix C (subsection C.5).  First, we
construct such    an   asymptotic   expansion   of   a   Weyl   symbol
$\underline{\cal R}_N$ that for $\underline{\cal R} =  \underline{\cal
R}_N$
$$
\begin{array}{c}
deg [\delta_B^l \underline{\cal  R}  -  \underline{\cal  R}  *  x^l  *
\underline{\cal R}   -   x^l  (\omega_k^2  +  V''(\Phi({\bf  x})))]  >
\max\{d/2, d-1\}; \\
deg [\delta_H \underline{\cal  R}  -  \underline{\cal  R}  *
\underline{\cal R}   -    (\omega_k^2  +  V''(\Phi({\bf  x})))]  >
\max\{d/2, d-1\}. \\
\end{array}
\l{p1}
$$
Next, we  will construct another asymptotic expansion of a Weyl symbol
$\underline{\cal R}$ which obeys the condition $Im \underline{\cal  R}
> 0$  and  approximately  equals  to  $\underline{\cal  R}_N$ at large
$|k|$, so that eqs.\r{p1} are satisfied.

This will imply that properties P1-P5, P7 are satisfied.

Let us define the expansions $\underline{\cal R}_N$ with the  help  of
the following recursive relations. Set
$$
\begin{array}{c}
\underline{\cal R}_0 = i \omega_k; \\
\underline{\cal S}_n    =    -   \delta_H   \underline{\cal   R}_n   +
\underline{\cal R}_n   *   \underline{\cal   R}_n   +   \omega_k^2   +
V''(\Phi({\bf x}));\\
\underline{\cal R}_{n+1} = \underline{\cal R}_n +  \frac{i}{2\omega_k}
\underline{\cal S}_{n}.
\end{array}
\l{p1a}
$$

{\bf Lemma 3.9.} {\it The following relation is satisfied:
$$
deg \underline{\cal S}_n = n.
$$
}

{\bf Proof.}  For $n=0$,  $\underline{\cal S}_0 = V''(\Phi({\bf x}))$,
so that statement of lemma is satisfied.  Suppose  that  statement  of
lemma is justified for $n<N$. Check it for $n=N$. One has
$$
\underline{\cal S}_N = \underline{\cal S}_{N-1} + \underline{\cal R}_N
* \left( \frac{i}{2\omega_k} \underline{\cal S}_{N-1} \right)
+ \left( \frac{i}{2\omega_k} \underline{\cal S}_{N-1} \right)
* \underline{\cal R}_N +
\left( \frac{i}{2\omega_k} \underline{\cal S}_{N-1} \right)
* \left( \frac{i}{2\omega_k} \underline{\cal S}_{N-1} \right)
- \frac{i}{2\omega_k} \delta_H \underline{\cal S}_{N-1}.
$$
Since
$$
deg\left[\left( \frac{i}{2\omega_k} \underline{\cal S}_{N-1} \right)
* \left( \frac{i}{2\omega_k} \underline{\cal S}_{N-1} \right)
- \frac{i}{2\omega_k} \delta_H \underline{\cal S}_{N-1}\right] \ge deg
\underline{\cal S}_{N-1} + 1 = N
$$
and
$$
\begin{array}{c}
\underline{\cal S}_N = \underline{\cal S}_{N-1} + \underline{\cal R}_N
* \left( \frac{i}{2\omega_k} \underline{\cal S}_{N-1} \right)
+ \left( \frac{i}{2\omega_k} \underline{\cal S}_{N-1} \right)
* \underline{\cal R}_N \simeq
\\
\underline{\cal S}_{N-1} + \underline{\cal R}_N
 \left( \frac{i}{2\omega_k} \underline{\cal S}_{N-1} \right)
+ \left( \frac{i}{2\omega_k} \underline{\cal S}_{N-1} \right)
 \underline{\cal R}_N = 0
\end{array}
$$
up to terms of the degree $N$, one finds
$$
deg \underline{\cal S}_N = N.
$$
Lemma 3.9 is proved.

Denote
$$
\underline{X}_n^l =    -    \delta_B^l    \underline{\cal    R}_n    +
\underline{\cal R}_n * x^l * \underline{\cal R}_n + x^l (\omega_k^2  +
V''(\Phi({\bf x}))).
$$

{\bf Lemma 3.10.} {\it The following property is obeyed:
$$
\delta_B^l \underline{\cal S}_n - \delta_H \underline{X}_n^l = -
\underline{X}_n^l * \underline{\cal R}_n -
\underline{\cal R}_n * \underline{X}_n^l + \underline{\cal S}_n *  x^l
* \underline{\cal  R}_n + \underline{\cal R}_n * x^l * \underline{\cal
S}_n.
\l{p2}
$$
}

{\bf Proof.} Denote
$$
\underline{F}_n^l =
\delta_B^l \underline{\cal S}_n - \delta_H \underline{X}_n^l
+ \underline{X}_n^l * \underline{\cal R}_n +
\underline{\cal R}_n * \underline{X}_n^l - \underline{\cal S}_n *  x^l
* \underline{\cal  R}_n - \underline{\cal R}_n * x^l * \underline{\cal
S}_n.
$$
One has
$$
\begin{array}{c}
\underline{F}_n^l = (\delta_B^l - x^l \delta_H) V''(\Phi({\bf  x}))  +
[\delta_H; \delta_B^l]  \underline{\cal  R}_n  -
[x^l  (\omega_k^2  + \\
V''(\Phi({\bf x})) - (\omega_k^2  + V''(\Phi({\bf x})) * x^l ] *
\underline{\cal R}_n +
\underline{\cal R}_n *
[x^l  (\omega_k^2  +
V''(\Phi({\bf x})) - x^l * (\omega_k^2  + V''(\Phi({\bf x}))]
\end{array}
$$
It follows from the definition of the Weyl symbol that
$$
x^l * f(x,k) = (x^l + \frac{i}{2} \frac{\partial}{\partial k^l}) f(x,k)
$$
One also has
$$
(\delta_B^l - x^l \delta_H) V''(\Phi({\bf  x})) = 0.
$$
Thus,
$$
\underline{F}_n^l = [\delta_H; \delta_B^l] \underline{\cal R}_n +
ik^l * \underline{\cal R}_n - \underline{\cal R}_n * ik^l =
\frac{\partial \underline{\cal   R}_n}{\partial  x^l  }  -  \delta_P^l
\underline{\cal R}_n.
$$
However, the property
$$
\frac{\partial \underline{\cal  R}_n}{\partial   x^l}   =   \delta_P^l
\underline{\cal R}_n
$$
which means that eq.\r{f6} is satisfied is checked by induction. Lemma
3.10 is proved.

{\bf Lemma 3.11.} {\it The following properties are satisfied:\\
1. $deg \underline{X}_n^l = n$. \\
2. $deg(\underline{X}_n^l - x^l \underline{\cal S}_n) \ge n+1$.
}

{\bf Proof.} It follows from the results of Appendix C that
$\underline{X}_n^l$ is an asymptotic expansion of a Weyl symbol. Let
$ deg \underline{X}_n^l = \alpha$.

Suppose that $\alpha <n$.  Then the left-hand side of eqs.\r{p2} is of
the degree $\alpha$, the degree of the right-hand side of eq.\r{p2}
is greater than or equal to
$\alpha-1$. In the leading order in $1/|k|$ the  right-hand  side  has
the form
one has $(-2i\omega_k
\underline{X}_n^l)$ and  its degree should be greater than or equal to
$\alpha$ .  Therefore, $deg \underline{X}_n^l \ge
\alpha + 1$. We obtain a contradiction.

Suppose $\alpha>n$.  Then  the  left-hand  side of eq.\r{p2} is of the
degree $n$,  the right-hand side in the leading order in  $1/|k|$  has
the form $2i\omega_k x^l \underline{\cal S}_n$. so that $deg
\underline{\cal S}_n$ should obey the inequality
$deg \underline{\cal S}_n \ge n+1$. We also obtain a contradiction.

Thus, $\alpha=n$. In the leading order in $1/|k|$ one has
$$
0 \simeq -2i\omega_k (\underline{X}_n^l - x^l \underline{\cal S}_n)
$$
up to terms of the degree $n$, so that
$deg (\underline{X}_n^l - x^l \underline{\cal S}_n)  \ge  n+1$.  Lemma
3.11 is proved.

We see  that  for  $N\ge  max\{d/2,d-1\}$  the  properties  \r{p1} are
satisfied.

{\bf Lemma 3.12.} {\it Let $\underline{\cal R}^{(1)}$ and
$\underline{\cal R}^{(2)}$  be  asymptotic expansions of Weyl symbols,
$deg \underline{\cal R}^{(1)} = deg \underline{\cal R}^{(2)} = -1$ and
$deg(\underline{\cal R}^{(1)} - \underline{\cal R}^{(2)}) = N+1$. Then
$$
deg (\underline{X}^{(1)l} - \underline{X}^{(2)l}) = N
$$
and
$$
deg (\underline{\cal S}^{(1)} - \underline{\cal S}^{(2)}) = N.
$$
}

{\bf Proof.}   Denote   $\underline{\cal  R}^{(1)}  -  \underline{\cal
R}^{(2)} = \underline{D}$. Then
$$
\underline{X}^{(1)l} -    \underline{X}^{(2)l}    =    -    \delta_B^l
\underline{D} +  \underline{\cal  R}^{(1)}  *  x^l  *  \underline{D} +
\underline{D} * x^l * \underline{\cal R}^{(1)} + \underline{D} *
\underline{D} * x^l * \underline{D}.
$$
We see  that  $deg(\underline{X}^{(1)l}  - \underline{X}^{(2)l}) = N$.
The second statement is checked analogously. Lemma 3.12 is proved.

Let us construct such an  asymptotic  expansion  $\underline{\cal  R}$
that $deg(\underline{\cal  R}  -  \underline{\cal R}_N) = N+1$ and $Im
\underline{\cal R} > 0$.  We will look  for  $\underline{\cal  R}$  as
follows (cf. \c{MS3}),
$$
\underline{\cal R}  =  \underline{\cal  A}  +  i \omega_k^{1/4} * \exp
\underline{\cal B}  *  \omega_k^{1/4}  *  \exp  \underline{\cal  B}  *
\omega_k^{1/4},
$$
where $\underline{\cal  A}$  and  $\underline{\cal  B}$ are {\it real}
asymptotic expansions. Then
$$
\begin{array}{c}
\underline{\Gamma^{1/2}} = \omega_k^{1/4} * \exp \underline{\cal B}  *
\omega_k^{1/4}; \\
\underline{\Gamma^{-1/2}} =  \omega_k^{-1/4}  * \exp (-\underline{\cal
B})  * \omega_k^{-1/4}
\end{array}
$$
are also asymptotic expansions of Weyl symbols. Choose
$\underline{\cal A}$ and $\underline{\cal B}$ to be polynomials,
$$
\underline{\cal A}  =   \sum_{s=1}^{S_1}   \frac{A_s({\bf   x},   {\bf
k}/\omega_{\bf k})}{\omega_{\bf k}^{2s}},
\qquad
\underline{\cal B}  =   \sum_{s=1}^{S_2}   \frac{B_s({\bf   x},   {\bf
k}/\omega_{\bf k})}{\omega_{\bf k}^{2s}},
$$
where $S_1 = [N/2]$, $S_2 = [\frac{N+1}{2}]$.

{\bf Lemma 3.13.} {\it  There  exists  unique  functions  $A_1$,  ...,
$A_{S_1}$, $B_1$, ..., $B_{S_2}$ such that
$deg (\underline{\cal R} - \underline{\cal R}_N) = N+1$.
}

{\bf Proof.} It follows from recursive relations \r{p1a} that
$$
\begin{array}{c}
Re \underline{\cal  R}_N  = \sum_{s=1}^{\infty} \frac{A_{N,s}({\bf x},
{\bf k}/\omega_{\bf k})}{\omega_{\bf k}^{2s}}, \\
Im \underline{\cal R}_N = \omega_{\bf k}  +
 \sum_{s=1}^{\infty} \frac{C_{N,s}({\bf x},
{\bf k}/\omega_{\bf k})}{\omega_{\bf k}^{2s}}.
\end{array}
$$
Therefore, $A_s  = A_{N,s}$,  so that $\underline{\cal A}$ is uniquely
defined. Denote
$$
\underline{{\cal B}_s} = \frac{B_{s}({\bf x},
{\bf k}/\omega_{\bf k})}{\omega_{\bf k}^{2s}}.
$$
Show that  $\underline{{\cal B}_s}$ is uniquely defined.  In the leading
order in $1/|{\bf k}|$, one has
$$
Im \underline{\cal R} \simeq \omega_{\bf k}  +  2\underline{{\cal  B}_1}
\omega_{\bf k},
$$
so that   $B_1   =   C_{N,1}/2$.   Suppose   that   one   can   choose
$\underline{{\cal B}_1}$,  ..., $\underline{{\cal B}_{s-1}}$ in such a way
that the degree of the asymptotic expansion of a Weyl symbol
$$
\underline{F}_{N,s} = Im \underline{\cal R}_N - \omega_{\bf k}^{1/4} *
\exp(\underline{{\cal B}_1} + ... + \underline{{\cal B}_{s-1}}) *
\omega_{\bf k}^{1/2} *
\exp(\underline{{\cal B}_1} + ... + \underline{{\cal B}_{s-1}})
* \omega_{\bf k}^{1/4}
$$
satisfies the inequality
$$
deg \underline{F}_{N,s} \ge 2s-1.
$$
Choose $\underline{{\cal B}_s}$ in such a way that
$deg \underline{F}_{N,s} \ge 2s-1$. One has
$$
\underline{F}_{N,s+1} = Im \underline{\cal R}_N - \omega_{\bf k}^{1/4}
* \sum_{l_1=0}^{\infty} \frac{( \underline{{\cal B}_1} + ...
+ \underline{{\cal B}_{s-1}}+ \underline{{\cal B}_s})^{l_1}}{l_1!}
* \omega_{\bf k}^{1/2} *
\sum_{l_2=0}^{\infty} \frac{( \underline{{\cal B}_1} + ...
+ \underline{{\cal B}_{s-1}}+ \underline{{\cal B}_s})^{l_2}}{l_2!}
* \omega_{\bf k}^{1/4}.
$$
Up to terms of the degree $2s+1$, one has
$$
\begin{array}{c}
\underline{F}_{N,s+1} \simeq Im \underline{\cal R}_N - \omega_{\bf k}^{1/4}
* (\sum_{l_1=0}^{\infty} \frac{( \underline{{\cal B}_1} + ...
+ \underline{{\cal B}_{s-1}}  )^{l_1}}{l_1!}
+ \underline{{\cal B}_s})
* \omega_{\bf k}^{1/2} *
(\sum_{l_2=0}^{\infty} \frac{( \underline{{\cal B}_1} + ...
+ \underline{{\cal B}_{s-1}} )^{l_2}}{l_2!}
+ \underline{{\cal B}_s})
* \omega_{\bf k}^{1/4}
\\
\simeq
\underline{F}_{N,s} - 2 \underline{{\cal B}_s} \omega_{\bf k}.
\end{array}
$$
Since
$$
\underline{F}_{N,s} =       \frac{1}{\omega_{\bf      k}^{2s-1}      }
\sum_{l=0}^{\infty} \frac{F_{N,s,l}  ({\bf  x},  {\bf   k}/\omega_{\bf
k})}{\omega_{\bf k}^l},
$$
one finds that
$$
\underline{{\cal B}_s} = \frac{1}{2\omega_k^{2s}}
{F_{N,s,0}  ({\bf  x},  {\bf   k}/\omega_{\bf
k})}
$$
is uniquely defined. Lemma 3.13 is proved.

Thus, we have constructed the operator $\cal R$ such that properties
\r{p1} are satisfied. We obtain the following theorem.

{\bf Theorem 3.14.} {\it Properties P1-P5, P7 are satisfied.}

This theorem  is  a  direct  corollary  of  the results of Appendix C.
Property P1 is satisfied because of construction of the operator $\cal
R$. Properties P2-P5, P7 are corollaries of Theorems C.31, C.32, C.33,
properties \r{p1} and lemmas C.8, C.9, C.19.

\subsection{Regularization and renormalization of a trace}

The purpose of this subsection is to specify functionals  $Tr_R\Gamma$
and $Tr_R  x^k\Gamma$  of arguments $\Phi$,  $\Pi$ in order to satisfy
properties P8,P9. We want the renormalized trace to satisfy properties
like these:\\
(i) $Tr_R \hat{A} = Tr \hat{A}$ if $\hat{A}$ is of the trace class;\\
(ii) $Tr_R (\hat{A}+\lambda \hat{B})  =  Tr_R  \hat{A}  +  \lambda  Tr_R
\hat{B}$; \\
(iii) $Tr_R [\hat{A}; \hat{B}]  = 0$; \\
(iv) $Tr_R \hat{A}_n \to 0$ if $A_n \to 0$ \\
for such  class of operators that is as wide as possible.  Under these
conditions, properties P8 and P9 are satisfied.  However,  one  cannot
specify such a renormalized trace. Namely, one should have
$$
Tr_R [\hat{x}_j;  {\cal W}(\frac{k_j}{\omega_{\bf k}^l} f({\bf x}))] =
0,
\l{p3a}
$$
where $f \in S({\bf R}^d)$.
${\cal W}(A)$ is a Weyl quantization of the  function  $A$  (see
appendix C). Property \r{p3a} means that
$$
Tr_R {\cal W}(i \frac{\partial}{\partial  k_i}  \frac{k_j}{\omega_{\bf
k}^l} f({\bf x})) = 0.
$$
Therefore,
$$
\delta_{ij} Tr_R {\cal W} (\frac{f({\bf x})}{\omega_{\bf k}^l})
- l Tr_R {\cal W}(\frac{k_ik_j}{\omega_{\bf k}^{l+2}} f({\bf x})) = 0.
\l{p4}
$$
Choose $l=d$.  Consider  $i=j$  in eq.\r{p4} and perform the summation
over $i$.  Making use of the relation $\omega_{{\bf k}}^2 -  k_ik_i  =
m^2$, we find
$$
Tr_R {\cal W}(m^2 \omega_{\bf k}^{-d-2} f({\bf x})) = 0.
$$
However, the operator with Weyl symbol  $m^2  f({\bf  x})  \omega_{\bf
k}^{-d-2}$ is of the trace class.  Its trace is nonzero, provided that
$\int d{\bf x} f(x) \ne 0$.

However, we can introduce a notion of  a  trace  for  {\it  asymptotic
expansions of  Weyl  symbols.} The trace will be specified not only by
operator but also by its asymptotic expansion which is not unique (see
remark after definition C.6).

Let $\underline{A} = (A, \check{A})$ be asymptotic expansion of a Weyl
symbol. Suppose that the coefficients $A_l$ of the  formal  asymptotic
expansion
$$
\check{A} \equiv  \sum_{l=0}^{\infty}  \omega_{\bf  k}^{-\alpha-l} A_l
(x, {\bf k}/\omega_{\bf k})
$$
are polynomial in ${\bf k}/\omega_{\bf k}$. One formally has
$$
Tr_R \underline{A} = \sum_{l=0}^{l_0}  \int \frac{d{\bf k}
d{\bf x} }{(2\pi)^d} \frac{1}{\omega_{\bf k}^{\alpha+l}}
A_l ({\bf x}, {\bf k}/\omega_{\bf k})
+ \int \frac{d{\bf k} d{\bf x}}{(2\pi)^d} (A({\bf x},{\bf k}) -
\sum_{l=0}^{l_0} \frac{1}{\omega_{\bf k}^{\alpha+l}}
A_l ({\bf x}, {\bf k}/\omega_{\bf k})).
\l{p5}
$$
For $\alpha + l_0 + 1 > d$,  the last integral in the right-hand  side
of eq.\r{p5}  converges.  To  specify trace,  it is sufficient then to
specify values of integrals
$$
I^{s,n}_{i_1...i_n} =   \int   \frac{d{\bf    k}}{\omega_{\bf    k}^s}
\frac{k_{i_1}}{\omega_{\bf k}} ... \frac{k_{i_n}}{\omega_{\bf k}}
\l{p6}
$$
for $s\le d$ which  are  divergent.  We  will  define  the  quantities
\r{p6}, making use of the following argumentation.

1. We are going to specify to specify trace in such a way that
$$
Tr_R \frac{\partial}{\partial k_i} \underline{A} = 0.
\l{p7}
$$
Let
$$
\underline{A} = \frac{1}{\omega_{\bf k}^{s-1}}
\frac{k_{j_1}}{\omega_{\bf k}} ... \frac{k_{j_{n+1}}}{\omega_{\bf k}}
$$
property \r{p7} implies the following recursive relations
$$
\sum_{s=1}^{n+1} \delta_{ij_s}        I^{s,n}_{j_1...j_{s-1}j_{s+1}...
j_{n+1}} = (s+n) I^{s,n+2}_{ij_1...j_{n+1}}.
\l{p8}
$$
Therefore, $I^{s,n}=0$  for  odd $n$,  while for even $n$ $I^{s,n}$ is
defined from eqs.\r{p8},  for example,  $I^{s,2}_{ij} =  \frac{1}{s}
\delta_{ij} I^{s,0}$. Therefore, it is sufficient to define integrals
$$
I^{s,0} = \int d{\bf k} \omega_{\bf k}^{-s}.
\l{p9}
$$

Let us  use  the  approach  based  on  the  dimensional regularization
\c{Wilson,Collins}.
It is based on considering integrals \r{p9}  at  arbitrary
dimensionality of  space-time.  Expression  \r{p9}  appears  to  be  a
meromorphic function of $d$.  Substracting the poles corresponding  to
sufficiently small positive integer values of $d$,  we obtain a finite
expression.

Formally, one has
$$
I^{s,0} = \frac{1}{\Gamma(s/2)} \int_0^{\infty} d\alpha \alpha^{s/2-1}
\int d{\bf       k}       e^{-\alpha      ({\bf      k}^2+m^2)}      =
\frac{\pi^{d/2}}{\Gamma(s/2)} \frac{\Gamma(\frac{s-d}{2})}{m^{s-d}}.
$$
If $\frac{s-d}{2} = -N$ is a nonpositive integer  number,  one  should
modify the  definition  of $I^{s,0}$.  Change $d\to d-2{\varepsilon}$.
One finds:
$$
\begin{array}{c}
I^{s,0} =            \frac{\pi^{d/2}}{\Gamma(s/2)             m^{s-d}}
\frac{\Gamma(1+{\varepsilon}) (\pi        m^2)^{-{\varepsilon}       }
}{(-N+{\varepsilon}) ... (-1+{\varepsilon}){\varepsilon} }
\\
\simeq \frac{\pi^{d/2} (-1)^N}{\Gamma(s/2) m^{s-d} N! {\varepsilon}} (
1+ {\varepsilon} (-ln (\pi m^2) + \Gamma'(1) + 1 + ...  + N^{-1}))  +
O({\varepsilon}).
\end{array}
$$
In the $MS$ renormalization scheme \c{Collins}, one should omit the term
$O({\varepsilon}^{-1})$. There    is    also    an     $\overline{MS}$
renormalization scheme  in  which one omits also a fixed term of order
$O(1)$. Let us omit the term $-ln( \pi m^2) + \Gamma'(1)$.  We  obtain
the following renormalized value of the integral:
$$
I^{s,0}_{ren} =          \frac{\pi^{d/2}}{\Gamma(s/2)         m^{s-d}}
\frac{(-1)^N}{N!} (1+ ... + 1/N),
$$
provided that $N= \frac{d-s}{2}$ is a nonnegative integer number.

Therefore, we have defined the renormalized
trace of an asymptotic expansion  of  a
Weyl symbol  by formaula \r{p5}, provided
that the coefficient functions are polynomials in ${\bf k}/\omega_{\bf
k}$.

Let us investigate properties of the renormalized trace.
Some properties are direct corollaries of definition \r{p5}.

{\bf Lemma 3.15.} {\it  The following properties are satisfied:\\
(i) $Tr_R (\underline{A} + \lambda \underline{B}) = 0$;\\
(ii) $Tr_R \frac{\partial \underline{A}}{\partial k_i} = 0$;
$Tr_R \frac{\partial \underline{A}}{\partial x_i} = 0;$\\
(ii) Let $E-\lim_{n\to\infty} \underline{A}_n = \underline{A}$.
Then $\lim_{n\to\infty} Tr_R \underline{A}_n = Tr_R \underline{A}$. \\
(iv) Let $deg \underline{A} > d$. Then $Tr_R \underline{A} = Tr A$.
}

{\bf Corollary.} The property AP9 is satisfied.

Let us check that $Tr_R (\underline{A} * \underline{B} - \underline{B}
* \underline{A}) = 0$. First of all, prove the following statement.

{\bf Lemma 3.16.}
$Tr_R \underline{A} * \underline{B}
= Tr_R \underline{A}\underline{B}$.

{\bf Proof.} Making use of eq.\r{k8}, we find
$$
\begin{array}{c}
(A*B)({\bf x},{\bf k}) - (AB)({\bf x},{\bf k}) =
\\
\int \frac{d{\bf p}_1
d {\bf  p}_2  d{\bf  y}_1  d{\bf  y}_2}{(2\pi)^{2d}}  \int_0^1 d\alpha
\frac{\partial}{\partial \alpha}
[A({\bf x}+{\bf y}_1; {\bf k}+ \alpha
\frac{{\bf p}_2}{2})   B({\bf  x}+  {\bf  y}_2,  {\bf  k}_2  -  \alpha
\frac{{\bf p}_1}{2})] e^{-i{\bf p}_1{\bf y}_1 - i {\bf p}_2 {\bf y}_2}
=
\\
- \frac{i}{2}
\int \frac{d{\bf p}_1
d {\bf  p}_2  d{\bf  y}_1  d{\bf  y}_2}{(2\pi)^{2d}}  \int_0^1 d\alpha
\left[
\frac{\partial}{\partial k^i}
A({\bf x}+{\bf y}_1; {\bf k}+ \alpha \frac{{\bf p}_2}{2})
\frac{\partial}{\partial x^i}
B({\bf  x}+  {\bf  y}_2,  {\bf  k}_2  -  \alpha
\frac{{\bf p}_1}{2}) -
\right. \\ \left.
\frac{\partial}{\partial x^i}
A({\bf x}+{\bf y}_1; {\bf k}+ \alpha \frac{{\bf p}_2}{2})
\frac{\partial}{\partial k^i}
B({\bf  x}+  {\bf  y}_2,  {\bf  k}_2  -  \alpha
\frac{{\bf p}_1}{2})
\right]
e^{-i{\bf p}_1{\bf y}_1 - i {\bf p}_2 {\bf y}_2}
= - \frac{i}{2} \frac{\partial C^i({\bf x},{\bf k})}{\partial k^i}
\end{array}
$$
with
$$
\begin{array}{c}
C^i({\bf x},{\bf k}) =
\int \frac{d{\bf p}_1
d {\bf  p}_2  d{\bf  y}_1  d{\bf  y}_2}{(2\pi)^{2d}}  \int_0^1 d\alpha
\left[
A({\bf x}+{\bf y}_1; {\bf k}+ \alpha \frac{{\bf p}_2}{2})
\frac{\partial}{\partial x^i}
B({\bf  x}+  {\bf  y}_2,  {\bf  k}_2  -  \alpha
\frac{{\bf p}_1}{2}) -
\right. \\ \left.
B({\bf  x}+  {\bf  y}_2,  {\bf  k}_2  -  \alpha
\frac{\partial}{\partial x^i}
A({\bf x}+{\bf y}_1; {\bf k}+ \alpha \frac{{\bf p}_2}{2})
\frac{{\bf p}_1}{2})
\right]
e^{-i{\bf p}_1{\bf y}_1 - i {\bf p}_2 {\bf y}_2}.
\end{array}
$$
One also has
$$
\check{A}*\check{B} -    \check{A}    \check{B}    =   -   \frac{i}{2}
\frac{\partial \check{C}^j}{\partial k^j}
$$
with
$$
\begin{array}{c}
\check{C}^j({\bf x},{\bf  k}) = \sum_{s=0}^{\infty} \sum_{l_1l_2\ge 0;
l_1+l_2 = s}  \frac{(-i)^{l_1}}{2^{l_1}  l_1!}  \frac{i^{l_2}}{2^{l_2}
l_2!} \frac{1}{l_1+l_2+1}
\left[
\frac{\partial^{l_1+l_2} \check{A}}{\partial   k^{i_1}   ...  \partial
k^{i_{l_1}} \partial x^{j_1} ... \partial x^{j_{l_2}}}
\frac{\partial}{\partial x^i}
\frac{\partial^{l_1+l_2} \check{B}}{\partial   k^{j_1}   ...  \partial
k^{j_{l_2}} \partial x^{i_1} ... \partial x^{i_{l_1}} }
\right. \\ \left. -
\frac{\partial^{l_1+l_2} \check{B}}{\partial   k^{j_1}   ...  \partial
k^{j_{l_2}} \partial x^{i_1} ... \partial x^{i_{l_1}}  }
\frac{\partial}{\partial x^i}
\frac{\partial^{l_1+l_2} \check{A}}{\partial   k^{i_1}   ...  \partial
k^{i_{l_1}} \partial x^{j_1} ... \partial x^{j_{l_2}}}.
\right]
\end{array}
$$
Analogously to  Appendix C,  one finds that $(C^j,  \check{C}^j)\equiv
\underline{C}^j$ is an asymptotic  expansion  of  a  Weyl  symbol.  It
follows from      lemma     3.15     that     $Tr_R     \frac{\partial
\underline{C}^j}{\partial k^j} = 0$. We obtain statement of lemma 3.16.

{\bf Lemma 3.17.} {\it For $deg \underline{B} \ge 2$,
$Tr_R x^k\omega_{\bf k} * \underline{B} =
Tr_R x^k\omega_{\bf k} \underline{B}$
and
$Tr_R \omega_{\bf k} * \underline{B} =
Tr_R \omega_{\bf k} \underline{B}$.
}

The proof is analogous.

{\bf Corollary 1.} {\it The following relations are satisfied:\\
1. $Tr_R    (\underline{A}   *   \underline{B}   -   \underline{B}   *
\underline{A}) = 0$; \\
2. $Tr_R    (x^k \omega_{\bf k} * \underline{B}   -   \underline{B}   *
x^k \omega_{\bf k})  = 0$; \\
3. $Tr_R    (\omega_{\bf k} * \underline{B}   -   \underline{B}   *
\omega_{\bf k})  = 0$.\\
}

{\bf Corollary 2.} {\it Property P8 is satisfied.}

Thus, we  have constructed functionals
$Tr_R x^k\hat{\Gamma} \equiv Tr_R x^k \underline{\Gamma}$
and
$Tr_R \hat{\Gamma}
\equiv Tr_R \underline{\Gamma}$ such that  properties  P8
and P9 are satisfied.

Note that the "finite renormalization" \c{BS} can be also be made. One
can add quantities $\Delta Tr_R x^k\hat{\Gamma}$
and $\Delta Tr_R \hat{\Gamma}$ to
renormalized traces in such a way that
$$
\begin{array}{c}
\delta^P_k \Delta Tr_R \hat{\Gamma} = 0; \qquad
\delta^M_{kl} \Delta Tr_R\hat{\Gamma} = 0;
\\
\delta^P_l \Delta Tr_R x^k \hat{\Gamma}
= - \delta^{kl} \Delta Tr_R \hat{\Gamma};
\qquad
\delta^M_{kl} \Delta Tr_R  x^k\hat{\Gamma}  =  \delta^{kl}
\Delta Tr_R   x^m   \hat{\Gamma}   -
\delta^{mk} \Delta Tr_R x^l \hat{\Gamma};\\
\delta_l^B \Delta Tr_R x^k\hat{\Gamma}
- \delta_k^B \Delta Tr_R x^l \hat{\Gamma} = 0; \\
\delta_l^B \Delta Tr_R \hat{\Gamma}
- \delta^H \Delta Tr_R x^l \hat{\Gamma} = 0.
\end{array}
$$
This corresponds to the possibility of adding the finite
one-loop counterterm to the Lagrangian.

\section{Semiclassical field}

An important feature of QFT is a notion of field.  In this section  we
introduce the  notion  of a semiclassical field and check its Poincare
invariance.

\subsection{Definition of semiclassical field}

First of  all,  introduce  the  notion  of   a   semiclassical   field
$\tilde{\phi}({\bf x},t:X)$  in  the
functional Schrodinger representation.  At $t=0$, this is the operator
of multiplication by $\phi({\bf x})$. For arbitrary $t$, one has
$$
\tilde{\phi}({\bf x},t:X)  = \tilde{U}_{-t}(X\gets u_tX) \phi({\bf x})
\tilde{U}_t (u_tX  \gets X),
$$
where $\tilde{U}_t(u_tX\gets X)$  is  the  operator  transforming  the
initial condition  for the Cauchy problem for eq.\r{4} to the solution
to the Cauchy problem.

The field  operator  in  the  Fock  representation  is  related   with
$\tilde{\phi}$ by the transformation \r{6*},
$$
\hat{\phi}({\bf x},t:X) = V_X^{-1} \tilde{\phi}({\bf x},t:X) V_X.
$$
Making use of eq.\r{b0}, one finds
$$
\hat{\phi}({\bf x},t:X)  =  (U_H^t(X))^{-1} \hat{\phi}({\bf x},  u_tX)
U_H^t(X)
\l{r0}
$$
Here $\hat{\phi}({\bf  x}:X)  = i (\Gamma^{-1/2} (A^+-A^-))({\bf x})$,
while
$$
U_H^t(X) \equiv  U_{a,\Lambda}  (u_{a,\Lambda}X   \gets   X),   \qquad
\Lambda=1, {\bf a} =0, a^0=-t.
$$

Let us define $\hat{\phi}$ mathematically as an operator distribution.

Let $S({\bf R}^d)$ be a space of complex  smooth  functions  $u:  {\bf
R}^d \to {\bf C}$ such that
$$
||u||_{l,m} = \max_{\alpha_1 + ... + \alpha_d \le l} \sup_{{\bf x} \in
{\bf R}^d} (1+|{\bf x}|)^m
|\frac{\partial^{\alpha_1 + ...  + \alpha_d}}{\partial x^{\alpha_1} ...
\partial x^{\alpha_d} u({\bf x}) }|
\to_{k\to\infty} 0.
$$
We say   that   the   sequence   $\{u_k\}  \in  S({\bf  R}^n)$,  $k  =
\overline{1,\infty}$ tends to zero if $||u_k||_{l,m}  \to_{k\to\infty}
0$ for all $l,m$.

{\bf Definition 4.1.} (cf.\c{A3}).
{\it 1. An operator distribution $\phi$ defined on ${\cal D} \in {\cal
F}$ is  a  linear  mapping taking functions $f\in S({\bf R}^d)$ to the
linear operator $\phi[f]: {\cal D} \to {\cal F}$,
$$
\phi: f\in S({\bf R}^d) \mapsto \phi[f]: {\cal D} \to {\cal F},
$$
such that   $||\phi[f_n]   \Phi||   \to_{n\to\infty}   0$    if    $f_n
\to_{n\to\infty} 0$. \\
2. A sequence of operator distributions $\phi_n$ is called  convergent
to the operator distribtion $\phi$ if
$$
||\phi_n[f] \Phi - \phi[f] \Phi|| \to_{n\to\infty} 0
$$
for all $\Phi \in {\cal D}$, $f\in S({\bf R}^d)$.
}

We will write
$$
\phi[f] \equiv \int d{\bf x} \phi({\bf x}) f({\bf x}), \qquad {\bf x }
\in {\bf R}^d.
$$

Consider the mapping $f\mapsto \phi_t\{f\}$,  $f\in S({\bf  R}^d)$  of
the form
$$
\phi_t\{f:X\} = \int d{\bf x} \hat{\phi}({\bf x},t:X) f({\bf x}).
$$

{\bf Lemma 4.1.} {\it  $\phi_t$  is  an  operator  distribution  being
continous with respect to $t$.}

{\bf Proof.} One has
$$
\phi_t\{f:X\} =   (U_H^t(X))^{-1}   i  \int  d{\bf  x}  (A^+({\bf  x})
(\hat{\Gamma}^{-1/2} f)({\bf x})
- A^-({\bf x}) (\hat{\Gamma}^{-1/2} f)({\bf  x}))
U_H^t(X).
$$
It follows from lemma B.3 and theorem B.15 that
this operator distribution is defined on
$$
{\cal D} = \{ \Psi \in {\cal F} | \quad ||\Psi||_T^1 < \infty\}
$$
and continous with respect to $t$. Lemma 4.1 is proved.

Consider the mapping $f\mapsto \phi[f]$,  $f \in  S({\bf  R}^{d+1})$ of
the foem
$$
\phi[f:X] = \int dt \phi_t\{f(\cdot,t):X\}.
$$

{\bf Lemma 4.2.} {\it $\phi$ is an operator distribution.}

The proof is analogous to lemma 4.1.

\subsection{Poincare invariance of the semiclassical field}

\subsubsection{Algebraic properties}

To check the property  of  Poincare  invariance,  notice  that  it  is
sufficient to  check  it  for  partial  cases:  spatial  translations,
rotations, evolution,  boost, since any Poincare transformation can be
presented as a composition of these transformations.  Let $g_B(\tau) =
(a(\tau),\Lambda(\tau))$ be  a  one-parametric  subgroup  of  Poincare
group corresponding  to  the element $B$ of the Poincare algebra.  The
Poincare invariance property can be rewritten as
$$
\hat{\phi} [f:X]    =     (U_B^{\tau}(X))^{+}     \phi[v_{g_B(\tau)}F:
U_{g_B(\tau)} X] U_B^{\tau}[X],
\l{r1}
$$
where
$$
(v_{g_B(\tau)} f)(x) = f(\Lambda^{-1}(\tau) (x-a(\tau))).
$$
Obviously, $v_{g_1} v_{g_2} = v_{g_1g_2}$.

Let us check relation \r{r1}.  It is convenient to  reduce  the  group
property to an algebraic property.  The formal derivative with respect
to $\tau$ of the right-hand side of eq.\r{r1} is
$$
(U_B^{|tau}(X))^+ \{
i [H(B:u_{g_B(\tau)}X) ; \phi[v_{g_B(\tau)}f: u_{g_B(\tau)}X]]
+ \frac{\partial}{\partial\tau} \phi[v_{g_B(\tau)}f: u_{g_B(\tau)}X]
\} U_B^{\tau}(X)
\l{r2}
$$
If the quantity \r{r2} vanishes, the property \r{r1} will be satisfied
since it  is  obeyed  at  $\tau=0$.  Making  use of the group property
$g_B(\tau+\delta\tau)= g_B(\delta  \tau)  g_B(\tau)$,  we  find   that
vanishing of expression \r{r2} is equivalent to the property:
$$
\frac{\partial}{\partial \tau} |_{\tau=0}         \phi[v_{g_B(\tau)}f:
u_{g_B(\tau)} X] - i [\phi[f:X] ; H(B:X)] = 0.
\l{r3}
$$
We obtain the following lemma.

{\bf Lemma 4.3.} {\it Let the bilinear form \r{r3} vanish on $\cal D$.
Then the property \r{r1} is satisfied on $\cal D$.}

{\bf Proof.} Consider the matrix element
$$
\chi^{\tau} =     (U_B^{\tau}(X)     \Psi_1,      \phi[v_{g_B(\tau)}f:
u_{g_B(\tau)}X] U_B^{\tau}(X)   \Psi_2)   -  (\Psi_1,  \hat{\phi}[f:X]
\Psi_2),
$$
where $\Psi_1, \Psi_2 \in {\cal D}$. Show it to be differentiable with
respect to $\tau$. Let us use an auxiliary lemma.

{\bf Lemma  4.4.}  {\it  Let  $\Psi  \in  {\cal  D}$.  Then the vector
$\phi[v_{g_B(\tau)}f: u_{g_B(\tau)}X] \Psi$  is  strongly  continously
differentiable with respect to $\tau$. }

{\bf Proof.} One has:
$$
\begin{array}{c}
\frac{\phi[v_{g_B(\tau+ \delta \tau)}f: u_{g_B(\tau+\delta \tau)}X]
- \phi[v_{g_B(\tau)}f: u_{g_B(\tau)}X]}{\delta \tau} \Psi =
\phi[ \frac{v_{g_B(\delta\tau)}-1}{\delta \tau}
v_{g_B(\tau)}f: u_{g_B(\tau+\delta\tau)}X] \Psi + \\
\frac{\phi[v_{g_B(\tau)}f: u_{g_B(\delta \tau) g_B(\tau)}X]
- \phi[v_{g_B(\tau)}f: u_{g_B(\tau)}X]}{\delta \tau} \Psi.
\end{array}
$$
The first term tends to
$\phi[ \frac{\partial}{\partial t}|_{t=0} v_{g_B(t)}
v_{g_B(\tau)}f: u_{g_B(\tau)}X]\Psi$
because of estimations of lemma  B.3.
The second    term    tends    to    $\delta[B]   \phi[v_{g_B(\tau)}f:
u_{g_B(\tau)}X]\Psi$ because of construction of operator
$\hat{\Gamma}^{-1/2}$
and formula \r{x9b} of Appendix A. Lemma 4.4 is proved.

To prove lemma 4.3, notice that
$$
\begin{array}{c}
\frac{\chi^{\tau+\delta \tau} - \chi^{\tau}}{\delta \tau}  =
(\phi[v_{g_B(\tau+ \delta \tau)}f: u_{g_B(\tau + \delta \tau)}X]
U_B^{\tau+\delta \tau}(X) \Psi_1;
\frac{U_B^{\tau+\delta\tau}(X) - U_B^{\tau}}{\delta \tau} \Psi_2)
\\
+
(U_B^{\tau+\delta\tau}(X) \Psi_1;
(\phi[v_{g_B(\tau+ \delta\tau)}f: u_{g_B(\tau+\delta \tau)}X]
- \phi[v_{g_B(\tau)}f: u_{g_B(\tau)}X]) U_B^{\tau} (X) \Psi_2)
\\
+
((U_B^{\tau+\delta\tau}(X) - U_B^{\tau}(X)) \Psi_1;
\phi[v_{g_B(\tau)}f: u_{g_B(\tau)}X] U_B^{\tau}(X) \Psi_2).
\end{array}
$$
This quantity tends as $\delta \tau\to 0$ to the matrix element of the
bilinear form \r{r2} and vanishes under condition \r{r3}. Lemma 4.3 is
proved.

\subsubsection{Check of invariance}

One should  check  property  \r{r1}  for  spatial   translations   and
rotations, evolution and boost transformations.

For spatial translations and rotations, property \r{r1} reads:
$$
\hat{\phi}({\bf x},t:X)  =  U^{-1}_{0,{\bf a},L} \hat{\phi} (L{\bf x}+
{\bf a}, t: u_{0,{\bf a},L}X) U_{0,{\bf a},L}
\l{r4}
$$
It follows  from  commutativity  of  $U_{0,{\bf  a},L}$  and $U_t$ and
eqs.\r{h8}, \r{h1}, \r{f6} that property \r{r4} is satisfied.

For evolution operator, property \r{r1} is rewritten as:
$$
\hat{\phi}({\bf x},t:X)   =    (U_H^{\tau}(X))^{-1}    \hat{\phi}({\bf
x},t-\tau:u_{\tau}X) U_H^{\tau}(X)
\l{r5}
$$
Relation \r{r5} is a direct corollary of definition \r{r0}  and  group
property for evolution operators.

Consider now the ${\bf n}$-boost transformation. Check property \r{r3}.
It can be presented as
$$
[\breve{B}^k; \hat{\phi}({\bf     x},t;X)]     =     -     i
(x^k
\frac{\partial}{\partial t}   +   t   \frac{\partial}{\partial   x^k})
\hat{\phi}({\bf x},t:X)
$$
or
$$
\begin{array}{c}
[B^k(X) ;   (U_H^t(X))^{-1}   \phi({\bf   x}:u_tX)   U_H^t(X)]   +   i
\delta_B^k\{(U_H^t(X))^{-1}   \phi({\bf   x}:u_tX)   U_H^t(X)\} = \\
-i (x^k
\frac{\partial}{\partial t}   +   t   \frac{\partial}{\partial   x^k})
(U_H^t(X))^{-1}   \phi({\bf   x}:u_tX)   U_H^t(X)
\end{array}
\l{r6}
$$
Let us make use of  property  \r{x9}.  For  partial  case  it  can  be
presented as
$$
B^k(X) (U_H^t(X))^{-1}   =   i(U_H^t(X))^{-1}   (\delta_B^k  U_H^t)(X)
(U_H^t(X))^{-1} + (U_H^t(X))^{-1} [B^k(u_tX) - t P^k(u_tX)]
\l{r7}
$$
or
$$
U_H^t(X) B^k(X)  = i (\delta_B^k U_H^t)(X) + [B^k(u_tX) - t P^k(u_tX)]
U_H^t(X).
\l{r8}
$$
Making use of relations \r{r7}, \r{r8}, we take relation \r{r6} to the
form
$$
[\breve{B}^k(Y) -   \breve{P}^k(Y)   t;   \phi({\bf   x}:Y)]   =   x^k
[\breve{H}(Y); \phi({\bf    x}:Y)]    -   it   \frac{\partial\phi({\bf
x}:Y)}{\partial x^k},
$$
where $Y=u_tX$. The property
$$
i \frac{\partial\phi({\bf  x}:Y)}{\partial  x^k}  =   [\breve{P}^k(Y);
\phi({\bf x}:Y)]
$$
is a corollary of relation \r{f7}. The relation
$$
[\breve{B}^k(Y) - x^k \breve{H}(Y); \phi({\bf x}:Y)] = 0
$$
is also checked by direct calculation.

Thus, we have obtained the following result.

{\bf Theorem 4.5.} {\it The invariance property \r{aa2} is satisfied.}

\section{Composed semiclassical states}

\subsection{Semiclassical states in quantum mechanics}

The most famous semiclassical approach to  quantum  mechanics  is  the
WKB-approach. It  is  the following.  One investigates the behavior of
solutions of semiclassical equation of the form
$$
i{\varepsilon} \frac{\partial      \psi_t(x)}{\partial      t}       =
H_t(\stackrel{\omega}{x},
\stackrel{\omega}{-i{\varepsilon}\frac{\partial}{\partial          x}})
\psi_t(x), \qquad t\in {\bf R}, x \in {\bf R}^d
\l{w1}
$$
as ${\varepsilon}\to 0$.  Here the Weyl quantization  is  considered.
The initial condition is chosen to be
$$
\psi_0(x) = \varphi_0(x) e^{\frac{i}{{\varepsilon}}S_0(x)}
\l{w2}
$$
where $S_0$ is a real function.  The WKB-result  \c{M1}  is  that  the
solution of  eq.\r{w1}  at time moment $t$ has the same form \r{w2} up
to $O({\varepsilon})$,
$$
||\psi_t -    \varphi_t    e^{\frac{i}{{\varepsilon}}S_t}     ||     =
O({\varepsilon}),
$$
provided that  $S_t(x)$  is  a  solution to the Cauchy problem for the
Hamilton-Jacobi equation
$$
\frac{\partial S_t}{\partial t} +  H_t(x,  \frac{\partial  S}{\partial
x}) = 0,
$$
while $\varphi_t(x)$ obeys the transport equation
$$
\frac{\partial \varphi_t(x)}{\partial      t}     +     \frac{\partial
H_t}{\partial p_i} (x, \frac{\partial S_t}{\partial x}) \frac{\partial
\varphi_t(x)}{\partial x_i}   +  \frac{1}{2}  \frac{\partial}{\partial
x_i} \{\frac{\partial     H_t}{\partial     p_i}(x,     \frac{\partial
S_t}{\partial x}) \} \varphi_t(x) = 0
$$
and initial condition $\varphi_0$.

However, we  are  not  obliged  to  choose  the  initial condition for
eq.\r{w1} in a form \r{w2}. There are other substitutions to eq.\r{w1}
that conserve  their  forms  under time evolution as ${\varepsilon}\to
0$. For example, the wave function
$$
\psi_t(x) = e^{\frac{i}{{\varepsilon}} S_t} e^{\frac{i}{{\varepsilon}}
P_t(x-Q_t)} f_t(\frac{x-Q_t}{\sqrt{\varepsilon}})               \equiv
(K^{\varepsilon}_{S_t,P_t,Q_t} f_t)(x), \qquad f_t \in S({\bf R}^d)
\l{w3}
$$
used in the Maslov complex-WKB  theory  \c{M1,M2}  also  approximately
satisfies eq.\r{w1} if
$$
\begin{array}{c}
\dot{S}_t = P_t\dot{Q}_t - H_t(Q_t,P_t), \qquad
\dot{Q}_t = \frac{\partial H}{\partial P}, \qquad
-\dot{P}_t = \frac{\partial H}{\partial Q}, \\
i\dot{f}_t(\xi) = \left[
\frac{1}{2} \frac{1}{i} \frac{\partial}{\partial \xi_i}
\frac{\partial^2 H}{\partial P_i \partial P_j}
\frac{1}{i} \frac{\partial}{\partial \xi_j}
+
\frac{1}{2} \frac{\partial^2 H}{\partial Q_i \partial P_j}
(\xi_i \frac{1}{i} \frac{\partial}{\partial \xi_j}
+ \frac{1}{i} \frac{\partial}{\partial \xi_j} \xi_i)
+
\frac{1}{2} \frac{\partial^2 H}{\partial Q_i \partial Q_j}
\xi_i \xi_j
\right] f_t(\xi).
\end{array}
\l{w3*}
$$
Therefore,
for the initial  condition  $K^{\varepsilon}_{S_0,P_0,Q_0}  f_0$
the solution  for  the  Cauchy   problem   for   eq.\r{w1}   will   be
asymptotically equal  to  $K^{\varepsilon}_{S_t,P_t,Q_t}  f_t$  up  to
terms of the order $O(\sqrt{\varepsilon})$.

The wave function  \r{w2}  rapidly  oscillates  with  respect  to  all
variables. The  wave  function  \r{w3}  rapidly  damps  at  $x-Q_t  >>
O(\sqrt{\varepsilon})$. One should come to the conclusion
that  there  exists  a  wave
function asymptotically  satisfying  eq.\r{w1}  which  oscillates with
respect to one group of variables and  damps  with  respect  to  other
variables. The  construction  of  such  states  is given in the Maslov
theory of  Lagrangian  manifolds  with  comples  germ  \c{M1,M2}.  Let
$\alpha \in {\bf R}^k$,  $(P(\alpha),Q(\alpha)) \in {\bf R}^{2d}$ be a
$k$-dimensional surface   in   the   $2d$-dimensional   phase   space,
$S(\alpha)$ be a real function, $f(\alpha,\xi)$, $\xi\in {\bf R}^d$ is
a smooth function.  Set $\psi(x)$ to be not exponentially amall if and
only if  the  distance between point $x$ and surface $Q(\alpha)$ is of
the order $\le O(\sqrt{\varepsilon})$.  Otherwise, set $\psi(x) \simeq
0$. If   $\min_{\alpha}   |x-Q(\alpha)|  =  |x-Q(\overline{\alpha})|  =
O(\sqrt{\varepsilon})$, set
$$
\psi(x) =                                              c_{\varepsilon}
e^{\frac{i}{{\varepsilon}}S(\overline{\alpha})}
e^{\frac{i}{{\varepsilon}}                        P(\overline{\alpha})
(x-Q(\overline{\alpha}))}
f(\overline{\alpha}, \frac{x-Q(\overline{\alpha})}{\sqrt{\varepsilon}}).
\l{w4}
$$

One can note that wave functions \r{w2} and \r{w3} are  partial  cases
of the   wave   function   \r{w4}.  Namely,  for  $k=0$  the  manifold
$(P(\overline{\alpha}), Q(\overline{\alpha})$ is a point,  so that the
functions \r{w4} coincide with \r{w3}.
Let $k=n$.      If      the      surface       $(P(\overline{\alpha}),
Q(\overline{\alpha})$ is  in  the  general  position,  for $x$ in some
domain one has $x= Q(\overline{\alpha})$ for some $\overline{\alpha}$.
Therefore,
$$
\psi(x) =          c_{\varepsilon}          e^{\frac{i}{{\varepsilon}}
S(\overline{\alpha})}  f(\overline{\alpha},0).
$$
We obtain the WKB-wave function.  Thus, WKB and wave-packet asymptotic
formulas \r{w2}  and  \r{w3}  are  partial  cases of the wave function
\r{w4} appeared in the theory of  Lagrangian  manifolds  with  complex
germ.

The lack    of    formula   \r{w4}   is   that   the   dependence   on
$\overline{\alpha}$ on $x$ is implicit and too  complicated.  However,
under certain    conditions    formula    \r{w4}   is   invariant   if
$\overline{\alpha}$ is  shifted   by   a   quantity   of   the   order
$O(\sqrt{\varepsilon})$. In  this case,  the point $\overline{\alpha}$
can be chosen in arbitrary way such  that  the  distance  of  $x$  and
$Q(\overline{\alpha})$ is of the order $O(\sqrt{\varepsilon})$.

Namely,
$$
\begin{array}{c}
e^{\frac{i}{{\varepsilon}} S(\overline{\alpha}  +   \sqrt{\varepsilon}
\beta)}
e^{\frac{i}{{\varepsilon}}
P(\overline{\alpha}  +   \sqrt{\varepsilon} \beta)
(x - Q(\overline{\alpha}  +   \sqrt{\varepsilon} \beta)) }
f(\overline{\alpha}  +   \sqrt{\varepsilon} \beta,
\frac{x-
Q(\overline{\alpha}  +   \sqrt{\varepsilon} \beta)
}{\sqrt{\varepsilon}}
) \\
\simeq
e^{\frac{i}{{\varepsilon}} S(\overline{\alpha})}
e^{\frac{i}{{\varepsilon}}
P(\overline{\alpha})
(x - Q(\overline{\alpha})) }
f(\overline{\alpha},
\frac{x-
Q(\overline{\alpha})
}{\sqrt{\varepsilon}} )
\end{array}
\l{w5}
$$
if
$$
\frac{\partial S}{\partial \overline{\alpha}_i} =
P \frac{\partial Q}{\partial \overline{\alpha}_i}
\l{w6}
$$
$$
e^{i \beta (\xi \frac{\partial P}{\partial \overline{\alpha}}
- \frac{1}{i} \frac{\partial}{\partial \xi}
\frac{\partial P}{\partial \overline{\alpha}})} f = f
\l{w7}
$$
To obtain eqs.\r{w6} and \r{w7},  one should expand left-hand side  of
eq.\r{w5}. Considering   rapidly   oscillating   factors,   we  obtain
eq.\r{w6}. To obtain eq.\r{w7}, it is sufficient to consider the limit
${\varepsilon}\to 0$.

Conditions \r{w6},  \r{w7}  simplify  the  check  \c{M1} that the wave
function \r{w4} approximately satisfies  eq.\r{w1}  if  the  functions
$S,P,Q,f$ are time-dependent. They should satisfy eqs.\r{w3*}. One can
show that conditions \r{w6}, \r{w7} are invariant under time evolution.

The form \r{w4} of the semiclassical state appeared in the  theory  of
Lagrangian manifolds   with   complex   germ  is  not  convenient  for
generalization to systems of infinite number of degrees of freedom. It
is much  more  convenient to consider to consider wave function \r{w3}
as an ''elementary'' semiclassical state and wave function \r{w4} as a
''composed'' semiclassical  state  presented  as  a  superposition  of
elementary semiclassical states:
$$
\psi(x) = C_{\varepsilon} \int d\alpha
e^{\frac{i}{{\varepsilon}} S(\alpha)}
e^{\frac{i}{{\varepsilon}}P(\alpha) (x-Q(\alpha))}
g(\alpha, \frac{x-Q(\alpha)}{\sqrt{\varepsilon}}),
\l{w8}
$$
where $g(\alpha,\xi)$  is  a  rapidly  damping  function  as   $\xi\to
\infty$. Superpositions  of  such  type were considered in \c{P,K,KV};
the general  case  was  investigated  in  \c{MS3,MS4}.  The   composed
semiclassical states   for  the  abstract  semiclassical  theory  were
studied in \c{Shv2}.

To show that expression \r{w8} is in agreement  with  formula  \r{w4},
notice that  the  wave  function  \r{w8} is exponentially small if the
distance between  $x$  and  the  surface  $Q(\alpha)$  is   of   order
$>O(\sqrt{\varepsilon})$. Let     $min_{\alpha}     |x-Q(\alpha)|    =
O(\sqrt{\varepsilon})$ and         $|x-Q(\overline{\alpha})|         =
O(\sqrt{\varepsilon})$. Consider    the    substitution    $\alpha   =
\overline{\alpha} + \sqrt{\varepsilon} \beta$. We find
$$
\psi(x) = C_{\varepsilon} {\varepsilon}^{k/2} \int d\beta
e^{\frac{i}{{\varepsilon}}
S(\overline{\alpha}  +   \beta \sqrt{\varepsilon})}
e^{\frac{i}{{\varepsilon}}
P(\overline{\alpha}  +   \beta \sqrt{\varepsilon})
(x- Q(\overline{\alpha}  +   \beta \sqrt{\varepsilon}))
}
g(\overline{\alpha}  +   \beta \sqrt{\varepsilon},
\frac{x- Q(\overline{\alpha}                  +                  \beta
\sqrt{\varepsilon})}{\sqrt{\varepsilon}}
)
$$
If the condition \r{w6} is not satisfied,  this is an  integral  of  a
rapidly oscillating   function.   It  is  exponentially  small.  Under
condition \r{w6} one can consider a  limit  ${\varepsilon}\to  0$  and
obtain the expression \r{w4}, provided that
$$
c_{\varepsilon}  = C_{\varepsilon} {\varepsilon}^{k/2}
$$
and
$$
\begin{array}{c}
f(\overline{\alpha}, \xi) = \int d\beta
e^{i\beta_s (\frac{\partial P_m}{\partial\overline{\alpha}_s} \xi_m  -
\frac{\partial Q_m}{\partial \overline{\alpha}_s} \frac{1}{i}
\frac{\partial}{\partial \xi_m})} g(\overline{\alpha},\xi)
= \\
(2\pi)^k \prod_{s=1}^k
\delta(\frac{\partial P_m}{\partial \overline{\alpha}_s} \xi_m -
\frac{\partial Q_m}{\partial      \overline{\alpha}_s}     \frac{1}{i}
\frac{\partial}{\partial \xi_m}) g(\overline{\alpha},\xi).
\end{array}
\l{w9}
$$
Integral representation \r{w8} simplifies  substitution  of  the  wave
function to   eq.\r{w1}   and  estimation  of  accuracy.  Namely,  the
integrand entering  to  eq.\r{w8}  is  an   asymptotic   solution   to
eq.\r{w1}, provided   that   eqs,\r{w3*}   are  satisfied.  Using  the
linearity property,  we  obtain  that   the   wave   function   \r{w8}
approximately satisfies eq.\r{w1} \c{MS3}.  Properties \r{w6},  \r{w9}
are shown to be invariant under time evolution \c{MS3}.

It follows from eq.\r{w9} that the function $f$ is invariant under the
following change of the function $g$ ("gauge transformation"):
$$
g(\alpha,\xi) \to g(\alpha,\xi) +
(\frac{\partial P_m}{\partial   \alpha_s}   \xi_m   -   \frac{\partial
Q_m}{\partial \alpha_s} \frac{1}{i} \frac{\partial}{\partial \xi_m})
\chi_s(\alpha,\xi).
\l{w10}
$$
Thus, the  semiclassical  state  is  specified  at  fixed $S(\alpha)$,
$P(\alpha)$, $Q(\alpha)$ not by the function $g$ but by the  class  of
equivalence of functions $g$: two functions are equivalent if they are
related by the transformation \r{w10}.

This fact can be also illustrated if we  evaluate  the  inner  product
$||\psi||^2$ as ${\varepsilon}\to 0$:
$$
\begin{array}{c}
||\psi||^2 = C_{\varepsilon}^2 \int d\alpha d\gamma \int dx
e^{-\frac{i}{{\varepsilon}} S(\alpha)}
e^{-\frac{i}{{\varepsilon}}P(\alpha) (x-Q(\alpha)}
g^*(\alpha, \frac{x-Q(\alpha)}{\sqrt{\varepsilon}})
\\
e^{\frac{i}{{\varepsilon}} S(\gamma)}
e^{\frac{i}{{\varepsilon}}P(\gamma) (x-Q(\gamma))}
g(\gamma, \frac{x-Q(\gamma)}{\sqrt{\varepsilon}}),
\end{array}
$$
The integral  over  $x$ is not exponentially small if $\alpha-\gamma =
O(\sqrt{\varepsilon})$. After substitution $\gamma =  \alpha  +  \beta
\sqrt{\varepsilon}$, $x-Q(\alpha) = \xi\sqrt{{\varepsilon}}$
and considering the limit ${\varepsilon}\to 0$, we find
$$
||\psi||^2 \simeq C_{\varepsilon}^2 {\varepsilon}^{\frac{k+n}{2}} \int
d\alpha (g(\alpha,\cdot), \prod_{s=1}^k 2\pi \delta
(\frac{\partial P_m}{\partial   \alpha_s}   \xi_m   -   \frac{\partial
Q_m}{\partial \alpha_s} \frac{1}{i} \frac{\partial}{\partial \xi_s}
) g(\alpha,\cdot)).
\l{w11}
$$

The $k$-dimensional   surface    $\{(S(\alpha),P(\alpha),Q(\alpha))\}$
("isotropic manifols")  in  the extended phase space has the following
physical meaning.  Consider  the  average  value  of  a  semiclassical
observable $A(x,-i{\varepsilon}\partial/\partial        x)$.        As
${\varepsilon}\to 0$, one has
$$
\begin{array}{c}
(\psi, A(x, - i{\varepsilon} \partial/\partial x) \psi) \simeq
C_{\varepsilon}^2 {\varepsilon}^{\frac{k+n}{2}} \int
d\alpha A(Q(\alpha),P(\alpha))
(g(\alpha,\cdot), \\
\prod_{s=1}^k 2\pi \delta
(\frac{\partial P_m}{\partial   \alpha_s}   \xi_m   -   \frac{\partial
Q_m}{\partial \alpha_s} \frac{1}{i} \frac{\partial}{\partial \xi_s}
) g(\alpha,\cdot)).
\end{array}
$$
We see that only values of the corresponding classical  observable  on
the surface $\{(Q(\alpha), P(\alpha))\}$ are relevant for calculations
fo average  values  as  ${\varepsilon}\to  0$.  This  means  that  the
Blokhintsev-Wigner density   function  (Weyl  symbol  of  the  density
matrix) corresponding  to  the   composed   semiclassical   state   is
proportional to the delta function on the manifold
$\{(Q(\alpha), P(\alpha))\}$.

Therefore, elementary semiclassical states  describe  evolution  of  a
point particle,   while   composed   semiclassical  states  (including
WKB-states) describe evolution  of  the  more  complicated  objects  -
isotropic manifolds.

\subsection{Composed semiclassical states in quantum field theory}

\subsubsection{Construction of semiclassical states}

Analogously to   qunatum   mechanical  formula  \r{w8},  consider  the
superposition of the "elementary" quantum field  semiclassical  states
\r{2} of the form
$$
\psi [\varphi(\cdot)]    =    \int    \frac{d\alpha   }{\lambda^{k/4}}
e^{\frac{i}{\lambda} S(\alpha)}  e^{\frac{i}{\lambda}  \Pi(\alpha;{\bf
x}) [\varphi({\bf   x})   \sqrt{\lambda}  -  \Phi(\alpha,  {\bf  x})]}
g(\alpha, \varphi(\cdot) - \frac{\Phi(\alpha,\cdot)}{\sqrt{\lambda}})
\l{w12}
$$
where $\alpha\in   {\bf  R}^k$,  $S(\alpha)$,  $\Pi(\alpha;{\bf  x})$,
$\Phi(\alpha;{\bf x})$ are smooth functions.  Calculate (formally) the
functional integral for $(\psi,\psi)$:
$$
\begin{array}{c}
(\psi,\psi) = \int \frac{d\alpha d\gamma}{\lambda^{k/2}} \int D\varphi
e^{-\frac{i}{\lambda} S(\alpha)}  e^{-\frac{i}{\lambda}  \Pi(\alpha;{\bf
x}) [\varphi({\bf   x})   \sqrt{\lambda}  -  \Phi(\alpha,  {\bf  x})]}
g^*(\alpha, \varphi(\cdot) - \frac{\Phi(\alpha,\cdot)}{\sqrt{\lambda}})
\\
e^{\frac{i}{\lambda} S(\gamma)}  e^{\frac{i}{\lambda}  \Pi(\gamma;{\bf
x}) [\varphi({\bf   x})   \sqrt{\lambda}  -  \Phi(\gamma,  {\bf  x})]}
g(\gamma, \varphi(\cdot) - \frac{\Phi(\gamma,\cdot)}{\sqrt{\lambda}})
\end{array}
$$
After substitution  $\gamma  =   \alpha   +   \sqrt{\lambda}   \beta$,
$\varphi(\cdot) =      \frac{\Phi(\alpha,\cdot)}{\sqrt{\lambda}}     +
\phi(\cdot)$ we obtain as $\lambda\to 0$:
$$
\begin{array}{c}
(\psi,\psi) \simeq \int d\alpha d\beta
e^{\frac{i}{\sqrt{\lambda}} \beta_s
(\frac{\partial S}{\partial \alpha_s} - \int d{\bf x} \Pi(\alpha,{\bf x})
\frac{\partial \Phi(\alpha,{\bf x})}{\partial \alpha_s}
)}
e^{\frac{i}{2} \beta_s \frac{\partial}{\partial \alpha_s}
(\frac{\partial S}{\partial \alpha_l} - \int d{\bf x} \Pi(\alpha,{\bf x})
\frac{\partial \Phi(\alpha,{\bf x})}{\partial \alpha_l}) \beta_l
}\\
\int D\phi g^*(\alpha,\phi(\cdot))
e^{i\beta_l \int d{\bf x}
(\frac{\partial\Pi(\alpha,{\bf x})}{\partial\alpha_l} \phi({\bf x})
- \frac{\partial\Phi(\alpha,  {\bf x})}{\partial \alpha_l} \frac{1}{i}
\frac{\delta}{\delta \phi({\bf x})})}
g(\alpha,\phi(\cdot))
\end{array}
\l{w13}
$$
The condition
$$
\frac{\partial S}{\partial \alpha_s} =
\int d{\bf x} \Pi(\alpha,{\bf x})
\frac{\partial \Phi(\alpha,{\bf x})}{\partial \alpha_s}
\l{w14}
$$
should be   satisfied.   Otherwise,   the  integral  \r{w13}  will  be
exponentially small as $\lambda\to 0$,  so that state \r{w12} will  be
trivial. Under condition \r{w14}, one has
$$
(\psi,\psi) \to_{\lambda\to 0} \int d\alpha d\beta
\int D\phi g^*(\alpha,\phi(\cdot))
e^{i\beta_l \int d{\bf x}
(\frac{\partial\Pi(\alpha,{\bf x})}{\partial\alpha_l} \phi({\bf x})
- \frac{\partial\Phi(\alpha,  {\bf x})}{\partial \alpha_l} \frac{1}{i}
\frac{\delta}{\delta \phi({\bf x})})}
g(\alpha,\phi(\cdot))
\l{w15}
$$

To specify   the   composed  semiclassical  state  in  the  functional
representation, one should:

(i) specify the smooth  functions  $(S(\alpha),  \Pi(\alpha,{\bf  x}),
\Phi(\alpha,{\bf x}))  \equiv X(\alpha)$ obeying eq.\r{w14} (determine
the $k$-dimensional isotropic manifold in  the  extended  phase  space
$\cal X$);

(ii) specify         the         $\alpha$-dependent         functional
$g(\alpha,\phi(\cdot))$.

The inner  product  of  composed  semiclassical  states  is  given  by
expression \r{w15}.

Since the inner product \r{w15} may vanish for nonzero $g$, one should
factorize the space of composed semiclassical states. Such functionals
$g$ that obey the property
$$
\int d\alpha
( g^*(\alpha,\cdot)
\prod_l \delta[\int d{\bf x}
(\frac{\partial\Pi(\alpha,{\bf x})}{\partial\alpha_l} \phi({\bf x})
- \frac{\partial\Phi(\alpha,  {\bf x})}{\partial \alpha_l} \frac{1}{i}
\frac{\delta}{\delta \phi({\bf x})})]
g(\alpha,\cdot) = 0
\l{w16}
$$
should be set to be equal to zero, $g\sim 0$.

One can  define  the   Poincare   tarnsformation   of   the   composed
semiclassical state  as  follows.  The  transformation of
$(S(\alpha), \Pi(\alpha,\cdot), \Phi(\alpha,\cdot))$ is
$u_{a,\Lambda}(S(\alpha), \Pi(\alpha,\cdot), \Phi(\alpha,\cdot))$. The
transformation of $g(\alpha,\phi(\cdot))$ is
$$
\tilde{U}_{a,\Lambda} (u_{a,\Lambda}(S,\Pi,\Phi)  \gets  (S,\Pi,\Phi))
g(\alpha,\phi(\cdot)).
$$
One should  check  that  the  inner product entering to eq.\r{w16} is
invariant under Poincare transformations.  This will also  imply  that
equivalent states are taken to equivalent.

Since the  functional  Schrodinger representation is not well-defined,
let us consider the Fock representation.  One should then specify  the
$\alpha$-dependent Fock   vector   $Y(\alpha)=  V^{-1}g(\alpha,\cdot)$
instead of the $\alpha$-dependent  functional  $g(\alpha,\phi(\cdot)$.
Making use of formulas \r{6*},  we find that the inner product \r{w15}
takes the form
$$
\left(
\left(
\begin{array}{c}
\Lambda^k \\ Y(\cdot)
\end{array}
\right),
\left(
\begin{array}{c}
\Lambda^k \\ Y(\cdot)
\end{array}
\right)
\right)
= \int d\alpha d\beta
(Y(\alpha), e^{ \beta_s \int d{\bf x} (B_s(\alpha,{\bf x}) A^+({\bf
x}) - B_s^*(\alpha,{\bf x}) A_s^-({\bf x}))} Y(\alpha))
\l{w17}
$$
where
$$
B_s(\alpha,\cdot) =    \hat{\Gamma}^{-1/2}
(\hat{\cal    R}    \frac{\partial
\Phi(\alpha,\cdot)}{\partial \alpha_s}        -         \frac{\partial
\Pi(\alpha,\cdot)}{\partial \alpha_s}),
\l{w24a}
$$
$\hat{\Gamma} = \hat{\Gamma}(\Phi(\alpha,\cdot), \Pi(\alpha,\cdot))$,
$\hat{\cal R}= \hat{\cal R}(\Phi(\alpha,\cdot), \Pi(\alpha,\cdot))$.
If the isotropic manifold $(\Phi(\alpha,\cdot), \Pi(\alpha,\cdot))$ is
non-degenerate, the  functions  $B_s(\alpha,{\bf  x})$  are   linearly
independent.

The Poincare transformation of the composed semiclassical state
$
\left(
\begin{array}{c}
\{ X(\alpha) \} \\ Y(\cdot)
\end{array}
\right)
$
is
$
\left(
\begin{array}{c}
\{ u_{a,\Lambda}X(\alpha) \} \\
U_{a,\Lambda} (u_{a,\Lambda}X(\alpha) \gets X(\alpha))
Y(\cdot)
\end{array}
\right)
$
Let us investigate some properties
of the inner product \r{w17} in order to check  its  invariance  under
Poincare transformations.

\subsubsection{Constrained Fock space}

The purpose of this subsection is to investigate the properties of the
inner product
$$
<Y_1,Y_2> =
\int d\beta
(Y_1, exp  [\sum_{s=1}^k
\beta_s \int d{\bf x} (B_s({\bf x}) A^+({\bf
x}) - B_s^*({\bf x}) A_s^-({\bf x}))] Y_2)
\l{w18}
$$
for the fock vectors $Y_1$, $Y_2$. Suppose the functions $B_1,...,B_k$
to be linearly independent.  Since the inner product \r{w18} resembles
the inner  products for constrained systems \c{Hen},  we will call the
space under construction as a constrained Fock space.

First of all,  investigate the problem of convergence of the  integral
\r{w18}. Note that the operator
$$
U[B] =  \exp  [\int  d{\bf  x} (B({\bf x}) A^+({\bf x}) - B^*({\bf x})
A^-({\bf x}))]
$$
is a well-defined unitary  operator  \c{Ber},  provided  that  $B  \in
L^2({\bf R}^d)$, and obey the relations
$$
\begin{array}{c}
A^-({\bf x}) U[B] = U[B] (A^-({\bf x}) + B({\bf x}));\\
A^+({\bf x}) U[B] = U[B] (A^+({\bf x}) + B^*({\bf x})).
\end{array}
$$

{\bf Lemma 5.1.} (cf. \c{MS3}). {\it The following estimation is satisfied:
$$
||B||^m |(Y_1,   U[B]  Y_2)|  \le  \sum_{k=0}^m  \frac{m!}{k!  (m-k)!}
||Y_1||_{k/2} ||Y_2||_{(m-k)/2}.
\l{w19}
$$
}

{\bf Proof.} One has
$$
[\int d{\bf x} B^*({\bf x}) A^-({\bf x}); U[B]] = ||B||^2 U[B],
$$
so that
$$
||B||^2 (Y_1,  U[B] Y_2) = (\int d{\bf x} B({\bf x}) A^+({\bf x}) Y_1,
U[B] Y_2) - (Y_1, U[B] \int d{\bf x} B^*({\bf x}) A^-({\bf x}) Y_2).
$$
Applying this identity $m$ times, we obtain:
$$
\begin{array}{c}
||B||^{2m} (Y_1,U[B]Y_2)  =   \sum_{k=0}^m   (-1)^{m-k}   \frac{m!}{k!
(m-k)!} ((\int  d{\bf  x}  B({\bf x}) A^+({\bf x}))^k Y_1,
\\
U[B] (\int
d{\bf x} B^*({\bf x}) A^-({\bf x}))^{m-k} Y_2).
\end{array}
$$
Making use of the result of lemma B.3,
$$
||\int d{\bf  x}  B({\bf  x})  A^{\pm}({\bf  x})   Y||_l   \le   ||B||
||Y||_{l+1/2},
$$
we find:
$$
||B||^{2m} |(Y_1,  U[B]  Y_2)|  \le  \sum_{s=0}^m  \frac{m!}{s!(m-s)!}
||B||^m ||Y_1||_{s/2} ||Y_2||_{(m-s)/2}.
$$
Lemma 5.1 is proved.

{\bf Corollary 1.} {\it  Let  $B_1,...,B_k$  be  linearly  independent
functions. Then  for some constant $C_1>0$ the following estimation is
satisfied:
$$
|\beta|^m |(Y_1,  U[\sum_s \beta_s B_s] Y_2)| \le C_1^m  ||Y_1||_{m/2}
||Y_2||_{m/2}.
$$
}

{\bf Proof.}  It is sufficient to notice that for linearly independent
$B_1,...,B_k$ the  matrix  $(B_m,B_s)$  is  not  degenerate,  so  that
$||\frac{1}{2} \sum_m \beta_m B_m||^2 \ge C_1^{-1} |\beta|^2$
for some $C_1$.  Applying the property $||Y||_{s/2}  \le  ||Y||_{m/2}$
for $s\le m$, making use of eq.\r{w19}, we prove corollary 1.

{\bf Corollary 2.} {\it Let $||Y_1||_{m/2} <\infty$,
$||Y_2||_{m/2} <\infty$ for some $m>k$, Then the integrand entering to
eq.\r{w18} obeys the relation
$$
|(Y_1, U[\sum_s\beta_s B_s] Y_2)| \le \frac{const}{(|\beta|+1)^m}
\l{w20}
$$
and integral \r{w18} converges.
}

{\bf Corollary 3.} {\it Let $Y_{1,n}$,  $Y_{2,n}$ be such sequences of
Fock vectors    that     $||Y_1,n||_{m/2}     \to_{n\to\infty}     0$,
$||Y_2,n||_{m/2} \le    C$    for    some   $m>k$.   Then   $<Y_1,Y_2>
\to_{n\to\infty} 0$. }

Let us investigate the property of  nonnegative  definiteness  of  the
inner product \r{w18}.

{\bf Lemma  5.2.}  {\it  Let  $||Y||_m <\infty$ for some $m>k$ and $Im
(B_s,B_l)=0$. Then $<Y,Y> \ge 0$.
}

{\bf Proof.} Introduce the following "regularized" inner  product
$$
<Y,Y>_{\varepsilon} =  \int  d\beta  e^{-{\varepsilon}  |\beta|^2} (Y,
U[\sum_s \beta_s B_s] Y).
$$
It follows form estimation \r{w20} and  the  Lesbegue  theorem  \c{KF}
that
$$
<Y,Y>_{\varepsilon} \to_{{\varepsilon}\to 0} 0.
$$
It is  sufficient then to prove that $<Y,Y>_{\varepsilon} \ge 0$.  One
has:
$$
e^{-{\varepsilon}|\beta|^2} = (4\pi{\varepsilon})^{k/2}  \int  d\beta'
e^{-2{\varepsilon}|\beta-\beta'|^2 - 2{\varepsilon}|\beta'|^2}
$$
Therefore,
$$
<Y,Y>_{\varepsilon} =  \int d\beta' d\beta'' (4\pi{\varepsilon})^{k/2}
e^{-2{\varepsilon} (|\beta'|^2 + |\beta''|^2} (U[\sum_s \beta_s'' B_s]
Y, U[\sum_s \beta_s' B_s] Y),
\l{w21}
$$
here the shift of variable $\beta = \beta' - \beta''$ is made. We have
also taken into account that
$$
U[\sum_s\beta_s' B_s] U[-\sum_s\beta_s'' B_s]
= U[\sum_s(\beta_s'-\beta_s'') B_s],
$$
provided that the operators
$$
\int d{\bf x} (B_s({\bf x})A^+({\bf x}) - B_s^*({\bf x}) A^-({\bf x})]
$$
commute (i.e. $Im(B_s,B_l)=0$).
Formula \r{w21} is taken to the form
$$
<Y,Y>_{\varepsilon} =    ||\int    d\beta    (4\pi{\varepsilon})^{k/4}
e^{-2{\varepsilon}|\beta|^2} U[\sum_s \beta_s B_s] Y||^2 \ge 0.
$$
Lemma 5.2 is proved.

The expression  \r{w18}  depends  on  $k$   functions   $B_1,...,B_k$.
However, one may perform linear substitutions of variables $\beta$, so
that only the subspace $span \{B_1, ..., B_k\}$ is essential.

{\bf Definition  5.1.}  {\it  A  $k$-dimensional  subspace  $L_k   \in
L^2({\bf R}^d)$  is  called as a $k$-dimensional isotropic plane if $Im
(B',B'') = 0$ for all $B',B'' \in L_k$.}

Let $L_k$ be a $k$-dimensional isotropic plane with an invariant under
shifts measure $d\sigma$.  Let $B_1,...,B_k$ be a basis on $L_k$.  One
can assign then coordinates $\beta_1,...,\beta_n$ to any element $B\in
L_k$ according  to  the  formula $B= \sum_s \beta_s B_s$.  The measure
$d\sigma$ is presented as $d\sigma = a d\beta_1 ... d\beta_k$ for some
constant $a$. Consider the inner product
$$
<Y_1,Y_2>_{L_k} =  a  \int  d\beta (Y_1,  U[\sum_s \beta_s B_s] Y_2) =
\int d\sigma (Y_1, U[B] Y_2),
\l{w22}
$$
$||Y_{1,2}||_{[k/2+1]} \le \infty$. This definition is invariant under
change of basis.

By ${\cal F}_{[k/2+1]}$ we denote space of such Fock vectors $Y$  that
$||Y||_{[k/2+1]} <  \infty$.  We say that $Y\stackrel{L_k}{\sim} 0$ if
$<Y,Y>_{L_k} = 0$.  Thus,  the space ${\cal F}_{[k/2+1]}$  is  divided
into equivalence classes. Introduce the following inner product on the
factor-space ${\cal F}_{[k/2+1]}/\sim$:
$$
<[Y_1], [Y_2]>^{L_k} = <Y_1,Y_2>_{L_k}
\l{w23}
$$
for all $Y_1 \in [Y_1]$,  $Y_2 \in [Y_2]$.  This definition is correct
because of the following statement.

{\bf Lemma 5.3.} {\it Let $<Y,Y>_{L_k} = 0$.  Then $<Y,Y'>_{L_k} =  0$
for all $Y'$.}

The proof is standard (cf,, forexample, \c{KF}). One has
$$
0 \le  <Y'  + \sigma Y,  Y'+\sigma Y>_{L_k} = <Y',Y'>_{L_k} + \sigma^*
<Y,Y'>_{L_k} + \sigma <Y',Y>_{L_k}
$$
for all $\sigma \in {\bf C}$, so that $<Y,Y'>_{L_k} = 0$.

{\bf Definition 5.2.} {\it A constrained  Fock  space  ${\cal  F}(L_k,
d\sigma)$ is the completeness of the factor-space
${\cal F}_{[k/2+1]}/\sim$ with respect to the inner product \r{w23},
$$
{\cal F}(L_k) = \overline{ {\cal F}_{[k/2+1]}/\sim}.
$$
}

\subsubsection{Transformations of constrained Fock vectors}

Let us investigate evolution of constrained Fock vectors. Consider the
Cauchy problem for eq.\r{bb3} (Appendix B):
$$
i\dot{\Psi}_t = \left[
\frac{1}{2} A^+ {\cal H}_t^{++} A^+
A^+ {\cal H}_t^{+-} A^- +
\frac{1}{2} A^- {\cal H}_t^{--} A^-
+ \overline{H}_t
\right]\Psi_t.
\l{w22a}
$$

{\bf Lemma 5.4.} {\it Let $\Psi_0 \in {\cal F}_m$.  Then  $\Psi_t  \in
{\cal F}_m$.}

{\bf Proof.} Analogously to proof of lemma B.11, one has
$$
\begin{array}{c}
||U_t\Psi_0|| = ||U_t^{-1} (A^+A^-+1)^m U_t \Psi_0|| = ||(1+ (A^+G_t^T +
A^-F_t^+)(F_t A^+ + G_t^* A^-))^m \Psi_0||
\end{array}
$$
It follows from Lemmas B.2, B.3 that
$$
\begin{array}{c}
||(1+ (A^+G_t^T +
A^-F_t^+)(F_t A^+ + G_t^* A^-)) \Psi||_l \le \\
||\Psi||_l +   ||F_t||_2^2   ||\Psi||_l  +  \sqrt{2}  ||G^T_t  F_t||_2
||\Psi||_{l+1} +  ||F^T_t  F_t^*  +  G^T_t  G^*_t||  ||\Psi||_{l+1}  +
||F^+_t G^*_t||_2 ||\Psi||_{l+1}\\
\le C ||\Psi||_{l+1}
\end{array}
$$
with
$$
C = 1 +   ||F_t||_2^2  +  \sqrt{2}  ||G^T_t  F_t||_2
 +  ||F^T_t  F_t^*  +  G^T_t  G^*_t||  +
||F^+_t G^*_t||_2.
$$
Applying this estimation, we obtain by induction:
$$
||U_t\Psi_0||_m \le C^m ||\Psi_0||_m.
$$
Lemma is proved.

Let $L_k$ be a $k$-dimensional isotropic plane with invariant  measure
$d\sigma$. Define its evolution transformation $L_k^t$ as follows. Let
$(B_1,...,B_k)$ be a basis on $L_k$.  Let $B_s^t$ be solutions to  the
Cauchy problems
$$
\begin{array}{c}
i \dot{B}_s^t = {\cal H}_t^{+-} B_s^t + {\cal H}_t^{++} (B_s^t)^*;
\\
- i \dot{B}_s^{t*}  =  {\cal  H}_t^{-+}  B_s^{t*}  +  {\cal  H}_t^{--}
B_s^t;\\
B_s^0 = B_s; B_s^{0*} = B_s^*.
\end{array}
\l{w23a}
$$
they can be expressed as
$$
\begin{array}{c}
B_s^t = F_t B_s^* + G_t^* B_s; \\
B_s^{t*} = F_t^* B_s + G_t B_s^*.
\end{array}
\l{w23b}
$$

{\bf Lemma  5.5.} {\it Let $Im (B_i,B_j)=0$.  Then $Im (B_i^t,B_j^t) =
0$.}

{\bf Proof.} One has:
$$
\begin{array}{c}
2i Im(B_i^t,B_j^t) = (B_i^t,B_j^t) - (B_j^t,B_i^t) = \\
(F_t B_i^* + G_t B_i, F_t B_j^* + G_t B_j) -
(F_t B_j^* + G_t B_j, F_t B_i^* + G_t B_i) = \\
(B_i,B_j) - (B_j,B_i) = 0.
\end{array}
$$
because of relations \r{bb16} of Appendix B.

Therefore, $L_k^t$  is  also  an  isotropic plane.  Define the measure
$d\sigma^t$ on $L_k^t$ as  follows.  For  the  choice  of  coordinates
$\beta_1,..,\beta_k$ on  $L_k^t$  according to the formula $B = \sum_s
\beta_s B_s^t$,  set $d\sigma = a d\beta_1 ...  d\beta_k$,  where  $a$
does not depend on $t$.

{\bf Lemma  5.6.}  {\it  The  inner product $<\cdot,\cdot>_{L_k^t}$ is
invariant under time evolution:
$$
<\Psi_t,\Psi_t>_{L_k^t} = <\Psi_0,\Psi_0>_{L_k}.
$$
}

{\bf Proof.} By definition, one has
$$
<\Psi_t,\Psi_t>_{L_k^t} = a  \int  d\beta  (\Psi_t,  U[\sum_s  \beta_s
B_s^t] \Psi_t) = a \int d\beta (\Psi_0,  U_t^+ U[\sum_s \beta_s B_s^T]
U_t\Psi_0).
$$
Eq.\r{w23b} implies that
$$
U[\sum_s \beta_s B_s^t] = \exp \sum_s \beta_s \it d{\bf x} (A_t^+({\bf
x}) B_s({\bf x}) - A_t^-({\bf x}) B_s^*({\bf x})).
$$
Making use of the relation
$$
U_t^+ A_t^{\pm} ({\bf x}) U_t = A^{\pm}({\bf x}),
$$
we obtain statement of lemma 5.6.

It follows  from lemma 5.6 that operator $U_t$ takes equivalent states
to equivalent. Therefore, it can be reduced to the factorspace
${\cal F}_{[k/2+1]}/\sim$.  Since it is unitary, it can be extended to
${\cal F}(L_k)$.

\subsubsection{Definition of  composed  semiclassical  state  and  its
Poincare transformation}

Let us formulate a definition of a composed semiclassical state.

Let $\Lambda^k$     be     a     smooth    $k$-dimensional    manifold
$(S(\alpha)<\Pi(\alpha,\cdot),\Phi(\alpha,\cdot))$ in   the   extended
phase space  with  measure $d\Sigma$ such that the property \r{w14} is
satisfied. Such manifolds are called isotropic.

Specify an isotropic plane $L_k(\alpha) \equiv L_k(\alpha: \Lambda^k)$
as follows.  Let  $B_s$  have  the   form   \r{w24a}.   The   subspace
$span\{B_1,...,B_k\}$ does   not   depend   on  particular  choice  of
coordinates $\alpha_1,...,\alpha_k$. Consider the quantity
$$
\begin{array}{c}
(B_s,B_l) -  (B_l,B_s)  =  (\frac{\partial\Phi}{\partial   \alpha_s}),
(\hat{\cal R}^*   -   \hat{\cal  R})  \hat{\Gamma}^{-1}
\frac{\partial\Pi}{\partial
\alpha_l}) - ( \frac{\partial\Phi}{\partial \alpha_l},  (\hat{\cal R}^*  -
\hat{\cal R}) \hat{\Gamma}^{-1} \frac{\partial\Pi}{\partial \alpha_s})
= \\
i \int d{\bf x} \left(
\frac{\partial\Phi(\alpha,{\bf x})}{\partial\alpha_l}
\frac{\partial\Pi(\alpha,{\bf x})}{\partial\alpha_s} -
\frac{\partial\Phi(\alpha,{\bf x})}{\partial\alpha_s}
\frac{\partial\Pi(\alpha,{\bf x})}{\partial\alpha_l}
\right).
\end{array}
\l{w24}
$$
Differentiating \r{w14}  with  respect  to $\alpha_l$,  we obtain that
quantity \r{w24} vanishes. Thus, plane $L_k(\alpha)$ is isotropic.

Introduce the following measure $d\sigma(\alpha)$ on $L_k(\alpha)$:
$$
d\sigma(\alpha) = \frac{D\Sigma(\alpha)}{D\alpha}(\alpha) d\beta_1 ...
d\beta_k,
\l{w25}
$$
where $\beta_1$, ..., $\beta_k$ are coordinates on $L_k(\alpha)$ which
are determined as $B=\sum_s \beta_s B_s$.

Definition \r{w25}  is invariant under change of coordinates.  Namely,
let $(\alpha_1',...,\alpha_k')$ be another set  of  local  coordinates
chosen instead of $(\alpha_1,...,\alpha_k)$. Then
$$
B_l' = \sum_{s=1}^k \frac{\partial\alpha_s}{\partial\alpha_l'} B_s,
$$
so that  property  $\sum_l \beta_l' B_l' = \sum_s \beta_s B_s$ implies
that coordinate sets $\beta$ and $\beta'$ should be related as follows:
$$
\beta_s = \sum_l \frac{\partial\alpha_s}{\partial\alpha_l'} \beta_l'.
$$
Therefore, for the choice of coordinates $\alpha'$ one has
$$
d\sigma' = \frac{D\Sigma}{D\alpha'} d\beta_1'... d\beta_k' =
\frac{D\Sigma}{D\alpha'} \frac{D\alpha'}{D\alpha}     d\beta_1     ...
d\beta_k = d\sigma.
$$
The invariance property is checked.

Introduce the vector (Hilbert) bundle  $\pi_{\Lambda^k}$  as  follows.
The base  of  the  bundle  is the isotropic manifold $\Lambda^k$.  The
fibre that corresponds to the point $\alpha\in  \Lambda^k$  is  ${\cal
H}_{\tau} = {\cal F}(L_k(\alpha))$.  Composed semiclassical states are
introduced as sections of bundle $\pi_{\Lambda^k}$.

{\bf Definition 5.2.} {\it A composed semiclassical state is a set  of
isotropic manifold   $\Lambda^k$   and   section  $Z$  of  the  bundle
$\pi_{\Lambda^k}$, such that the inner product
$$
<(\Lambda^k,Z), (\Lambda^k,Z)> = \int_{\Lambda^k} d\Sigma  (Z(\alpha),
Z(\alpha))_{{\cal F}(L_k(\alpha))}
$$
converges.
}

Poincare transformation   of   isotropic   manifold    $\Lambda^k    =
\{X(\alpha)\}$ is determined as
$$
u_{a,\Lambda} \{X(\alpha)\} = \{ u_{a,\Lambda} X(\alpha)\}.
$$
Section $\{Z(\alpha)\}$  is  transformed as follows.  Let $Z(\alpha) =
[Y(\alpha)]$. define   $U_{a,\Lambda}   Z(\alpha)   =   [U_{a,\Lambda}
Y(\alpha)]$. This  definition  is  correct  because  of the resulta of
previous subsubsection, provided that
$$
L_k(\alpha: u_{a,\Lambda}       \Lambda^k)       =       U_{a,\Lambda}
L_k(\alpha:\Lambda^k).
\l{w26}
$$
It is sufficient to prove property \r{w26} for partial cases:  spatial
translations and rotations, time evolution and $x^1$-boost.

{\bf 1. Spatial translations.}

The functions  $\Phi,\Pi$ obey eqs.\r{w27},  while eq.\r{w22a} has the
form \r{w28}. Eqs.\r{w23a} for $B_s^t$ take the form
$$
\dot{B}^{\tau}_s = - a^k \partial_k B_s^{\tau}.
$$
This equation  is  satisfied  because  of  eqs.\r{w27},  \r{w24a}  and
property \r{f7}.

{\bf 2. Spatial rotations.}

The functions  $\Phi,\Pi$  obey eqs.\r{e9},  while eq.\r{w22a} has the
forem \r{w20}. Eqs.\r{w23a} take the form
$$
\dot{B}_i^{\tau} =    \frac{1}{2}   \theta_{sm}   (x^s\partial_m   -
x^m\partial_s) B_i^{\tau}.
$$
This equation   is  satisfied  because  of  eqs.\r{e9},  \r{w24a}  and
property \r{f7}.

{\bf 3. Time evolution.}

The functions $\Phi,\Pi$ obey eq.\r{3}, while eq.\r{w22a} has the form
\r{w31}. Eq.\r{w23a} takes teh form:
$$
i\dot{B}_j^t = ({\cal H}+\hat{\omega})B_j^t + {\cal H}^{++} B_j^{t*},
$$
where $\cal  H$  and  ${\cal  H}^{++}$  have  the form \r{w31}.  It is
satisfied because of eqs.\r{3} and \r{w24a}.

{\bf 4. The $x^1$-boost.}

The functions $\Phi,\Pi$ obey eq.\r{w12} with ${\bf n}=  (1,0,...,0)$,
while eq.\r{w22a} has the form \r{w32}. Eq.\r{w23a} has the form
$$
i\dot{B}_j^t = (L_k + {\cal B}^k) B_j^t + {\cal B}^{k++} B_j^{t*},
$$
where ${\cal  B}^k$ and ${\cal B}^{k++}$ have the form \r{b28}.  It is
satisfied because of eqs.\r{e12} and \r{w24a}.

Thus, the   composed   semiclassical   states   and   their   Poincare
transformations are introduced.

\section{Conclusions}

In this  paper  a  notion  of  a  semiclassical  state  is introduced.
"Elementary" semiclassical state are specified by a set $(X,\Psi)$  of
classical  field  configuration $X$ (point on the infinite-dimensional
manifold $\cal X$,  see section 2  and  subsection  3.2)  and  element
$\Psi$  of  the space $\cal F$.  Set of all "elementary" semiclassical
states may be viewed as a semiclassical bundle.

The physical  meaning  of classical field $X$ is evident.  Discuss the
role of $\Psi$. In the soliton quantization language \c{DHN,R} $\Psi$
specifies whether  the  quantum  soliton  is  in the ground or excited
state. For the Gaussian approach \c{G1,G2,G3,G4}, $\Psi$ specifies the
form of the Gaussian functional,  while for QFT in the strong external
classical field \c{GMM,BD} $\Psi$ is a state of a quantum field in the
classical background.

The "composed" semiclassical states have been also introduced (section
5). They can be viewed as superpositions of "elementary" semiclassical
states and  are  specified  by  the  functions  $(X(\tau),\Psi(\tau))$
defined on some domain of ${\bf R}^k$ with values on the semiclassical
bundle.

Not arbitrary  superposition  of  elementary  semiclassical  states is
nontrivial. The  isotropic  condition  \r{w14}  should  be  satisfied.
Moreover, the  inner  product  of  the "composed" semiclassical states
(eq.\r{w17}) is degenerate, so that there is a "gauge freedom" \r{w16}
in specifying composed semiclassical states.

The composed   semiclassical   states  are  used  \c{MS3}  in  soliton
quantization, since there are translation zero modes and solitons  can
be shifted. They are useful if there are conserved integrals of motion
like charges.  The  correspondence  between  composed  and  elementary
semiclassical states in QFT resembles the relationship between WKB and
wave packet approximations in quantum mechanics.

An important feature of QFT is the property of Poincare invariance. In
this paper  an  explicit  check  of  this  property  is  presented for
semiclassical QFT.  The Poincare  transformations  of  elementary  and
composed semiclassical states have been constructed as follows. First,
the simplest Poincare transformations like  spatial  translations  and
rotations,  evolution  and  boost  are  considered.  The infinitesimal
transformations  are  investigated,  the  Lie  algebraic   commutation
relations  have  been  checked  and  the  group  properties  have been
justified.

For the  "composed"  states,  conservation  of  the  degenerate  inner
product and  isotropic  condition  under  Poincare transformation have
been checked.

An important feature of QFT is a notion of field.  In this paper  this
notion is  introduced for semiclassical QFT.  The property of Poincare
invariance of semiclassical field is checked.

This work  was supported by the Russian Foundation for Basic Research,
project 99-01-01198.

\appendix

\section{Symmetries of the semiclassical theory under Lie
groups}

Let the semiclassical  field  theory  be  symmetric  under  Lie  group
transformations. Let $\cal G$ be a Lie group.
Let $\cal  X$  be  a  smooth  (maybe,  infinite-dimensional) manifold,
${\cal H}$ be a Hilbert space.
Suppose that\\
( ) a smooth mapping $u:  {\cal G} \times {\cal  X}  \to  {\cal
X}$ is specified; the  smooth mappings
$u_g: {\cal  X}  \to  {\cal  X}$  associated  with the mapping $u$
of the form $u_gX = u(g,X)$ satisfy
the group property $u_{g_1}u_{g_2} = u_{g_1g_2}$;\\
(¡) for each  $X\in{\cal  X}$  unitary operators
$U_g[X] \equiv U_g(u_gX\gets X): {\cal F} \to {\cal F}$ are specified;
the group property
$$
U_{g_1} (u_{g_1g_2}X \gets u_{g_2}X)  U_{g_2}  (u_{g_2}X  \gets  X)  =
U_{g_1g_2} (u_{g_1g_2}X \gets X)
\l{x0}
$$
are satisfied.

Let us investigate the properties of infinitesimal transformations.

By $T_e{\cal G}$ we denote the tangent space to the
Lie group  $\cal G$ at $g=e$.  Let $A\in T_e{\cal G}$,  $g(\tau)$ be a
smooth curve on the group $\cal G$ with the tangent vector
$A$ at the point $g(0)=e$.
Introduce the differenial operator
$\delta[A]$ on the space of differentiable functionals
$F$  on $\cal   X$:
$$
(\delta[A]F)(X) = \frac{d}{d\tau}|_{\tau=0} F(u_{g(\tau)}X).
\l{x1}
$$

{\bf Lemma A.1} {\it 1.  The quantity \r{x1} does not  depend  on  the
choice of the curve $g(\tau)$ with the tangent vector $A$.\\
2. The following property
$$
\delta[A_1+\alpha A_2] =  \delta[A_1]  +  \alpha  \delta[A_2],   \alpha
\in{\bf R}, A_1,A_2 \in T_e{\cal G}
$$
is satisfied.
}

{\bf Proof.}  Let  $g_1(\tau)$ and $g_2(\tau)$ be smooth curves on the
Lie group ${\cal G}$ such that $g_1(0)=e$, $g_2(0)=e$. One has
$$
\frac{d}{d\tau}|_{\tau=0} F(u_{g_1(\tau)g_2(\tau)} X) =
\frac{d}{d\tau}|_{\tau=0} F(u_{g_1(\tau)} X)
\frac{d}{d\tau}|_{\tau=0} F(u_{g_2(\tau)} X).
\l{x1a}
$$
since
$$
\frac{F(u_{g_1(\tau)g_2(\tau)} X}{d\tau}   |_{\tau=0}   =   \int  d\xi
\frac{\partial}{\partial \tilde{\tau}}         F(u_{g_1(\tilde{\tau})}
u_{g_2({\tau})} X)|_{\tilde{\tau}         =         \xi\tau}         +
\frac{F(u_{g_2(\tau)}X)- F(X) }{\tau}.
$$
Let $g(\tau)$ and $\tilde{g}(\tau)$ be curves on  $\cal  G$  with  the
tangent vector   $A$.  Choose  $g_1(\tau)  =  g(\tau)$,  $g_2(\tau)  =
g^{-1}(\tau) \tilde{g}(\tau)$.  Since $g_2(\tau)  -  e  =  O(\tau^2)$,
$\frac{d}{d\tau}|_{\tau=0} F(u_{g_2(\tau)}X)  =  0$.  It  follows from
eq\r{x1a} that
$\frac{d}{d\tau}|_{\tau=0} F(u_{g(\tau)}X)  =
\frac{d}{d\tau}|_{\tau=0} F(u_{\tilde{g}(\tau)}X)$.
Thus, definition \r{x1} is correct.

Let $g_1(\tau)$,  $g_2(\tau)$  be  curves with tangent vectors $A$ and
$B$ correspondingly.  Then $g_1(\tau)g_2(\tau)$ is a  curve  with  the
tangent vector $A+B$. Eq.\r{x1a} implies that
$$
\delta[A+B] = \delta[A] + \delta[B].
$$
Finally, consider  a  curve  $g(\tau)$ with the tangent vector $A$ and
teh curve $\tilde{g}(\tau)= g(\alpha \tau)$ with  the  tangent  vector
$\alpha A$. One has
$$
\frac{d}{d\tau}|_{\tau=0} F(u_{\tilde{g}(\tau)}     X)     =    \alpha
\frac{d}{d\tau}|_{\tau=0} F(u_{g(\tau)}X).
$$
Thus, $\delta[\alpha A] =\alpha \delta[A]$. Lemma A.1 is proved.

Let $g\in{\cal G}$, $A\in T_e{\cal G}$, $h(\tau)$ be a curve on ${\cal
G}$  with tangent vector $B$ at $h(0)=e$.  Then the tangent vector for
the curve $g(\tau)h(\tau)g^{-1}(\tau)$ at $h(0)=e$ does not depend  on
the choice of the curve $h(\tau)$. Denote it by $gBg^{-1}$.

Define the operator $W_g$ on the space of functionals $F$ as
$W_gF[X] = F[u_{g^{-1}}X]$.

{\bf Lemma A.2.} {\it The following property is satisfied:
$$
W_g \delta[B] W_{g^{-1}} F = \delta[gBg^{-1}]F.
\l{x3}
$$
}

The proof is straightforward.

Let $g=g(\tau)$ be a curve with the tangent vector
$A$ at $g(0)=e$.  Differentiating expression
\r{x3} by $\tau$ at $\tau=0$, we obtain:

{\bf Lemma A.3.} {\it The following relation is satisfied:
$$
([\delta[A],\delta[B]] + \delta([A,B]))F = 0.
\l{x4}
$$
Here $[A,B]$ is the Lie-algrbra commutator for the group $\cal G$.
}

{\bf Proof.} Let $g(\tau)$ be a smooth curve on the Lie group $\cal G$
with tangent vector $A$ at $g(0)=e$. Make use of the property \r{x3}:
$$
W_{g(\tau)} \delta[B]   W_{g^{-1}(\tau)}   =   \delta    [g(\tau)    B
g^{-1}(\tau)] F,
\l{x3a}
$$
rewrite definition \r{x1} as
$$
\delta [A] = \frac{d}{d\tau}|_{\tau = 0} W_{g^{-1}(\tau)},
$$
remember that the Lie commutator can be defined as
$$
[A;B] \equiv \frac{d}{d\tau}|_{\tau=0} g(\tau) B g^{-1}(\tau).
$$
Consider the derivatives of sides of eq.\r{x3a}  at  $\tau  =  0$.  We
obtain property \r{x4}. Lemma A.3 is proved.

Consider now the infinitesimal properties of the transformation
$U$. Suppose that on some dense subset $D$ of
$\cal F$ the vector functions  $U_g[X]\Psi$ ($\Psi\in D$)
are strongly continously differentiable  with respect to
$g$ and smooth with respect to $X$. Define operators
$$
H(A:X) \Psi = i\frac{d}{d\tau}|_{\tau=0} U_{g(\tau)}[X] \Psi,
\l{x4a}
$$
where $g(\tau)$  is  a  curve  on  the group $\cal G$ with the tangent
vector $A$ at $g(0)=e$.

{\bf Lemma A.4.} {\it
1. The operator  $H(A:X)$  does not depend on
the choice of the curve  $g(\tau)$   with the tangent vector $A$.\\
2. The following property is satisfied:
$$
H(A_1+\alpha A_2:X) = H(A_1:X) + \alpha H(A_2:X).
$$
}

The proof is analogous to lemma A.1.

Let $h(\tau)$ be a curve with the tangent vector $B$ at $h(0)=e$.
Eq.\r{x0} implies:
$$
U_g[u_{h(\tau)}X] U_{h(\tau)}[X]U_g^{-1}[X]\Psi  =
U_{gh(\tau)g^{-1}}[u_gX]\Psi, \qquad \Psi \in D
$$
Differentiating this identity by $\tau$ at $\tau=0$, we obtain:

{\bf Lemma A.5.} {\it For $\Psi\in D$
$$
-iU_g[X] H(B:X) U_g^{-1}[X]\Psi + (\delta[B]U_g)[X] U_g^{-1}[X] \Psi =
-i H(gBg^{-1}; u_gX)\Psi.
\l{x5}
$$
}

Let  $g=g(t)$  be a curve with the tangent vector
$A$ at $g(0)=e$. Differentiating eq.\r{x5} by $t$ in a weak sense,
we obtain:

{\bf Lemma A.6.}
{\it On the subset $\cal D$ the following bilinear form vanishes:
$$
-[H(A;X): H(B;X)]   -  i\delta[B] H(A:X) + i\delta[A] H(B:X)
+ i H([A;B]: X) = 0.
\l{x6}
$$
}

{\bf Renark.} Introduce the operator
$$
\breve{H}(A:X) = H(A:X) - i\delta[A]
\l{x7}
$$
on the space of smooth sections of the bundle
${\cal  X} \times {\cal D} \to {\cal X}$. The property \r{x6} can be rewritten as
$$
[\breve{H}(A;X); \breve{H}(B;X)] = i \breve{H}([A,B];X).
\l{x7*}
$$
Note that the operator
\r{x7} is an analog of the covariant differentiation operator
in the theory of bundles (see, for example, \c{Post}).

Thus, the group property
\r{x0} is reformulated in terms of Lie algebras.

Investigate now the problem of reconstructing the group representation
if the algebra representation is known.
Our purpose  is  to prove some lemmas which are useful in constructing
the representation of the Poincare group.

Impose the following conditions on the operators
$H(A:X)$,  $A  \in
T_e{\cal G}$, $X\in {\cal X}$.

H1. {\it Hermitian operators $H(A:X)$ are defined on a common
domain $\cal D$ which is dense in $\cal F$.}

H2. {\it For each smooth curve $h(\alpha)$ on $\cal G$  and each  $\Psi
\in {\cal  D}$  the  vector  function   $H(A:u_{h(\alpha)}X)\Psi$   is
strongly continously differentiable with respect to
$\alpha$.}

H3. {\it The bilinear form \r{x6} vanshes on $\cal D$.}

Let ${\cal  Z} \in T_e {\cal G}$ be a subset of the Lie algebra of the
group $\cal G$. Let $B \in {\cal Z}$,  while
$g_B(t)$  is a one-parametric subgroup of the Lie group
$\cal  G$ with the tangent vector
$B$  at  $t=0$,  $g_B(0)=e$.
Denote by $U_B^t(X)$ the operator taking the initial condition
$\Psi_0 \in {\cal D}$ of the Cauchy problem for the equation
$$
i \frac{\partial\Psi_t}{\partial t} = H(B:u_{g_B(t)}X) \Psi_t
\l{x8}
$$
($\partial\Psi_t/\partial t$  is a strong derivative)
to the solution  $\Psi_t
\in {\cal  D}$ of the Cauchy problem,
$\Psi_t =  U_B^t(X)\Psi_0$.  This  definition  is  correct  under  the
following condition.

H4. {\it  Let $B \in {\cal Z}$.
If  $\Psi_0\in  {\cal D}$, there exists a solution
of the Cauchy problem for eq.\r{x8}. }

Uniqueness of the solution is a corollary of the property
$||\Psi_t||  =
||\Psi_0||$ which is checked directly by differentiation.
The isometric operator  $U_B^t(X)$ can be  expanded
then from  ${\cal D}$ to $\cal F$. It satisfies the
property
$$
U_B^{t_1}(u_{g_B(t_2)}X) U_B^{t_2}(X) = U_B^{t_1+t_2} (X).
$$
Therefore, it is invertible and unitary.

Impose also the following conditions.

H5. {\it Let $B\in {\cal Z}$.
For each smooth curve $h(\alpha)$ on  $\cal  G$  and each
$\Psi_0\in {\cal   D}$ the quantity $||\frac{\partial}{\partial\alpha}
H(A:u_{h(\alpha)}X) \Psi_t||$ is bounded uniformly
with respect to  $\alpha,t  \in
[\alpha_1,\alpha_2] \times [t_1,t_2]$ for any finite
$\alpha_1,\alpha_2,t_1,t_2$.
}

H6. {\it For $\psi \in {\cal D}$, $B\in {\cal Z}$, $A\in T_e{\cal G}$,
the following property is satisfied:
$$
|| H(g_B(\tau)Ag_B^{-1}(\tau): u_{g_B(\tau)} X) [U_B^{\tau}(X)
\Psi - \Psi]
|| \to_{\tau\to 0} 0.
$$
}

Under these conditions, we obtain:

{\bf Lemma A.7.} {\it Let $B\in {\cal Z}$, $A\in T_e{\cal G}$,
The following property is satisfied on the domain
$\cal D$:
$$
H(A:X) +   i  (U^t_B(X))^{-1}  (\delta[A]U_B^t)(X)  -  (U^t_B(X))^{-1}
H(g_B(t)A g_B(t)^{-1}: u_{g_B(t)} X) U_B^t(X) = 0.
\l{x9}
$$
}

{\bf Proof.} Let us check that under  these  conditions  the  operator
$(\delta[A]U_B^t)(X)$ is correctly defined, i.e. the strong derivative
$$
(\delta[A] U_B^t)(X)  \Psi  =  \frac{d}{d\alpha}|_{\alpha  =  0} U_B^t
(u_{h(\alpha)} X) \Psi
\l{x9+}
$$
exists for  all  $\Psi \in {\cal D}$,  where $h(\alpha)$ is a curve on
$\cal G$ with tangent vector $A$.

Denote
$$
\Psi_{\alpha, t} = U_B^t(u_{h(\alpha)}X) \Psi.
$$
This vector obeys the equation
$$
i \frac{\partial}{\partial     t}     \Psi_{\alpha,t}      =      H(B:
u_{g_B(t)h(\alpha)} X) \Psi_{\alpha,t},
$$
so that
$$
\begin{array}{c}
i \frac{\partial}{\partial  t}
(\Psi_{\alpha+  \delta  \alpha,  t } - \Psi_{\alpha,t})
= H(B:u_{g_B(t) h(\alpha+\delta\alpha)} X)
(\Psi_{\alpha+  \delta  \alpha,  t } - \Psi_{\alpha,t})
\\
+ (H(B:  u_{g_B(t)  h(\alpha+\delta  \alpha)}X)  - H(B:  u_{g_B(t)}X))
\Psi_{\alpha,t}.
\end{array}
$$
Since $\Psi_{\alpha,0} = \Psi_{\alpha+\delta\alpha,0} = \Psi$, we have
$$
\Psi_{\alpha+\delta \alpha,  t} - \Psi_{\alpha,t} = - i \int_0^t d\tau
U_B^{t-\tau} (u_{g_B(\tau) h(\alpha+\delta \alpha)} X)
(H(B: u_{g_B(\tau) h(\alpha+\delta \alpha)}X) - H(B:  u_{g_B(\tau)}X))
\Psi_{\alpha,\tau}.
$$
Because of   unitarity   of   the  operators  $U_B^t$,  the  following
estimation takes place:
$$
||\Psi_{\alpha+\delta \alpha,  t} - \Psi_{\alpha,t}||  \le
\int_0^t d\tau
|| (H(B: u_{g_B(\tau) h(\alpha+\delta \alpha)}X) - H(B:  u_{g_B(\tau)}X))
\Psi_{\alpha,\tau}||.
$$
Making use  of  the Lesbegue theorem \c{KF} and condition H5,  we find
that $||\Psi_{\alpha+\delta \alpha,t} - \Psi_{\alpha,t}||  \to_{\delta
\alpha =  0}  0$,  so  that  the  operator  $U_B^t(u_{h(\alpha)}X)$ is
strongly continous with respect to $\alpha$.

Furthermore,
$$
\frac{\Psi_{\alpha +  \delta   \alpha,t}   -   \Psi_{\alpha,t}}{\delta
\alpha} =  - i\int_0^{\tau} \int_0^1 d\gamma U_B^{t-\tau}(u_{g_B(\tau)
h(\alpha+ \delta \alpha)} X ) \left(
\frac{\partial}{\partial \alpha}  H(B:  u_{g_B(\tau) h(\alpha + \gamma
\delta \alpha)}X)
\right) \Psi_{\alpha, \tau}
$$
Denote
$$
\frac{\partial \Psi_{\alpha,t}}{\partial  \alpha}  \equiv - i \int_0^t
d\tau U_B^{t-\tau}(u_{g_B(\tau)
h(\alpha+ \delta \alpha)} X ) \left(
\frac{\partial}{\partial \alpha}  H(B:  u_{g_B(\tau) h(\alpha)}X)
\right) \Psi_{\alpha, \tau}.
\l{x9b}
$$
The following estimation takes place:
$$
\begin{array}{c}
|| \frac{\Psi_{\alpha  + \delta \alpha,  t} - \Psi_{\alpha,t} }{\delta
\alpha} -  \frac{\partial  \Psi_{\alpha,t}}{\partial  \alpha}  ||  \le
\int_0^t d\tau \int_0^1 d\gamma
\left(
||U_B^{t-\tau} (u_{g_B(\tau) h(\alpha+ \delta \alpha)}X)||
\right. \\ \times
||\left[
\frac{\partial H}{\partial   \alpha}  (B:  u_{g_B(\tau)h(\alpha+\gamma
\delta \alpha)}X) -
\frac{\partial H}{\partial   \alpha}  (B:  u_{g_B(\tau)h(\alpha)}X)
\right]
\Psi_{\alpha,\tau}||
+
 \\ \left.
||(U_B^{t-\tau} (u_{g_B(\tau) h(\alpha+ \delta \alpha)}X)
- U_B^{t-\tau} (u_{g_B(\tau) h(\alpha)}X))
\frac{\partial H}{\partial   \alpha}  (B:  u_{g_B(\tau)h(\alpha)}X)
\Psi_{\alpha,\tau}
\right)
\end{array}
\l{x9a}
$$
Making use of the Lesbegue theorem, conditions H2,H5, we find that the
quantity \r{x9a} tends to zero as $\delta \alpha  \to  0$.  Thus,  the
vector \r{x9+} is correctly defined.

It follows from the expression \r{x9b} that
$$
\frac{\partial}{\partial \tau}    \frac{\partial}{\partial     \alpha}
U_B^t(u_{h(\alpha)} X) - i H(B: u_{g_B(t) h(\alpha)}X)
\frac{\partial U_B^t (u_{h(\alpha)}X)}{\partial \alpha}
$$
in a strong sense.

Let us prove now property \r{x9}.
At $t=0$ the property \r{x9} is satisfied. The derivative with
respect to $t$ of any matrix element of the
operator \r{x9} under conditions H1-H6 vanishes.
Therefore, equality  \r{x9}  viewed  in  terms  of  bilinear  forms is
satisfied on $\cal D$. Since the left-hand side of
eq.\r{x9} is defined on $\cal D$, it also vanshes on $\cal D$.

To construct  the representation of the Poincare group,  the following
statement is used in this paper.

Let the property
$$
g_{B_n}(t_n(\alpha)) ... g_{B_1} (t_1(\alpha)) = e,
\l{x11}
$$
be satisfy for $\alpha \in [0,\alpha_0]$ and
$B_1, ..., B_n\in {\cal Z}$. Here
$t_k(\alpha)$ are smooth functions. Denote $h_k(\alpha) =
g_{B_k}(t_k(\alpha))$, $s_k(\alpha) = h_k(\alpha) ... h_1(\alpha)$.

{\bf Lemma A.8.} {\it Under condition \r{x11} the operator
$$
U_{B_n}^{t_n(\alpha)}(u_{s_{n-1}(\alpha)}X)
...
U_{B_1}^{t_1(\alpha)}(X)
\l{x12}
$$
is $\alpha$-independent.
}

To prove lemma, denote
$U_k \equiv U_k(u_{s_{k-1}(\alpha)}X) \equiv
U_{B_k}^{t_k(\alpha)}(u_{s_{k-1}(\alpha)}X)$.
Let us use the following lemma.

{\bf Lemma A.9.} {\it Let $s(\alpha)$ be a smooth curve on the group
$\cal G$,  $t(\alpha)$ is a smooth real function,  $B\in
T_e{\cal G}$. Then the operator function
$U_B^{t(\alpha)} (u_{s(\alpha)}X)$    is    strongly    differentiable
with respect to $\alpha$ on $\cal D$ and
$$
\frac{\partial}{\partial\alpha} U_B^{t(\alpha)} (u_{s(\alpha)}X) = - i
\frac{dt}{d\alpha} H(B:u_{g_B(t(\alpha)) s(\alpha)}X)
U_B^{t(\alpha)}
(u_{s(\alpha)}X) + (\delta[\frac{ds}{d\alpha} s^{-1}]
U_B^{t(\alpha)}) (u_{s(\alpha)}X),
\l{x10}
$$
where
$\frac{ds}{d\alpha} s^{-1}$ is a
tangent vector to the curve
$s(\alpha+\tau)s^{-1}(\alpha)$ at $\tau=0$.
}

{\bf Proof.}
Let $\Psi \in {\cal D}$. One has
$$
\begin{array}{c}
\frac{1}{\delta \alpha}
(U_B^{t(\alpha+\delta
\alpha)}(u_{s(\alpha+\delta \alpha)}X)
- U_B^{t(\alpha)}(u_{s(\alpha)}X) \Psi = \\
\frac{1}{\delta \alpha}
(U_B^{t(\alpha+\delta
\alpha)}(u_{s(\alpha)}X)
- U_B^{t(\alpha)}(u_{s(\alpha)}X) \Psi
+
\frac{1}{\delta \alpha}
(U_B^{t(\alpha+\delta
\alpha)}(u_{s(\alpha+\delta \alpha)}X)
- U_B^{t(\alpha+ \delta \alpha)}(u_{s(\alpha)}X) \Psi.
\end{array}
\l{x10a}
$$
The first term in the right-hand side of eq.\r{x10a} tends to
$$
-i \frac{dt}{d\alpha} H(B:u_{g_B(t(\alpha)) s(\alpha)}X) \Psi
$$
by definition of the operator $U_B^t(X)$. Consider the second term. It
can be represented as
$$
\int_0^1 d\gamma           \frac{\partial}{\partial            \alpha}
U_B^{t(\overline{\alpha})} (u_{s(\alpha      +      \gamma      \delta
\alpha)}X)|_{\overline{\alpha} = \alpha + \delta \alpha} \Psi.
$$
Making use of eq. \r{x9b}, we take this term to the form
$$
-i \int_0^1 d\gamma \int_0^{t(\alpha+\delta \alpha)} d\tau
U_B^{t(\alpha+\delta\alpha)-\tau} (u_{g_B(\tau)    s(\alpha+    \gamma
\delta \alpha)}           X)           \frac{\partial}{\partial\alpha}
H(B:u_{g_B(\tau)s(\alpha+ \gamma\delta       \alpha}X)      U_B^{\tau}
(u_{s(\alpha + \gamma \delta \alpha)}X) \Psi.
\l{x10b}
$$
Making use of the Lesbegue theorem and property H6,  we see  that  the
vector \r{x10b} is strongly continous as $\delta\alpha \to 0$, so that
it is equal to
$$
\frac{\partial}{\partial \alpha}            U_B^{t(\overline{\alpha})}
(u_{s(\alpha)}X)|_{\overline{\alpha}= \alpha} \Psi =
((\delta[\frac{ds}{d\alpha}s^{-1}] U_B^{t(\alpha)})  (u_{s(\alpha)} X)
\Psi.
$$
We obtain formula \r{x10}.

Let us return to proof of lemma A.8.
To check formula \r{x12}, let us obtain that
$$
\frac{d}{d\alpha} (U_n...U_1) (U_n...U_1)^{-1} = 0
\l{x13}
$$
on $\cal  D$  (the  derivative  is  viewed  in the strong sense).  The
property \r{x13} is equivalent to the folowing relation:
$$
\sum_{k=1}^n U_n...U_{k+1}    \frac{\partial   U_k}{\partial   \alpha}
U_k^{-1}...U_n^{-1} = 0.
\l{x14}
$$
Making use of eq.\r{x10}, we take eq.\r{x14} to the form
$$
\begin{array}{c}
\sum_{k=1}^n U_n...U_{k+1}  H(-i\frac{dt_k}{d\alpha}  B_k :  u_{s_k}X)
U_{k+1}^{-1} ...   U_n^{-1}   +  \\ \sum_{k=1}^n   U_n    ...    U_{k+1}
(\delta[\frac{ds_{k-1}}{d\alpha}       s_{k-1}^{-1}]U_k)(u_{s_{k-1}}X)
U_k^{-1} ... U_n^{-1}=0.
\end{array}
\l{x15}
$$
Applying properties \r{x9} $n-k$ times, we obtain
$$
\begin{array}{c}
-iH(\sum_{k=1}^n \frac{dt_k}{d\alpha} h_n ... h_{k+1} B_k h_{k+1}^{-1}
... h_n^{-1}: X) + \sum_{l=1}^n U_n ... U_{l+1}
(\delta[\frac{ds_{l-1}}{d\alpha}       s_{l-1}^{-1}]U_l)
(u_{s_{l-1}}X)
U_l^{-1} ... U_n^{-1}
\\
- \sum_{k=1}^n  \sum_{l=k+1}^n  U_n  ...  U_{l+1} \frac{dt_k}{d\alpha}
\delta[h_{l-1}... h_{k+1} B_k h_{k+1}^{-1} ... h_{l-1}^{-1}]U_l)
(u_{s_{l-1}}X)
U_l^{-1} ... U_n^{-1}.
\end{array}
\l{x16}
$$
Eq.\r{x11} implies that
$$
\begin{array}{c}
\sum_{k=1}^n \frac{dt_k}{d\alpha} h_n ... h_{k+1} B_k h_{k+1}^{-1}
... h_n^{-1} = 0; \\
\frac{ds_{l-1}}{d\alpha} s_{l-1}^{-1} - \sum_{k=1}^{l-1}
h_{l-1}... h_{k+1} B_k h_{k+1}^{-1} ... h_{l-1}^{-1}
= 0.
\end{array}
$$
Lemma A.8 is proved.

{\bf Corollary.} {\it Let $t_k(0)=e$. Then
$$
U_{B_n}^{t_n(\alpha)}(u_{s_{n-1}(\alpha)}X)
...
U_{B_1}^{t_1(\alpha)}(X) = 1
$$
under conditions of lemma A.8.
}

\section{Some properties of quadratic Hamiltonians in the
Fock space}

The purpose  of this appendix is to introduce some notations and check
some properties of operators in Fock space.

Remind that the Fock space ${\cal F}(L^2({\bf R}^d))$ is defined  as  a
space of sets
$$
\Psi= (\Psi_0,  \Psi_1({\bf  x}_1),  ...  ,\Psi_n({\bf   x}_1,...,{\bf
x}_n),...)
$$
of symmetric  with  respect to $x_1$,  ...,  $x_n$ symmetric functions
$\Psi_n$ such that $||\Psi||^2 =  \sum_{n=0}^{\infty}  |\Psi_n||^2  <
\infty$. By $A^{\pm}({\bf x})$ we denote,  as usual,  the creation and
annihilation operator distributions:
$$
\begin{array}{c}
(\int d{\bf  x}  A^+({\bf  x})  f({\bf x}) \psi)_n ({\bf x}_1,...,{\bf
x}_n) = \frac{1}{\sqrt{n}} \sum_{l=1}^n f({\bf x}_j) \psi_{n-1}  ({\bf
x}_1,..., {\bf x}_{j-1}, {\bf x}_{j+1}, ..., {\bf x}_n); \\
(\int d{\bf x} A^-({\bf x})  f^*({\bf x}) \psi)_{n-1}
({\bf x}_1,...,{\bf  x}_{n-1})  =  \sqrt{n} \int d{\bf x} f^*({\bf x})
\psi_n({\bf x},{\bf x}_1,...,{\bf  x}_{n-1}).
\end{array}
$$
By $|0>$   we  denote,  as  usual,  the  vacuum  vector  of  the  form
$(1,0,0,...)$. Arbitrary vector of the Fock space can be presented via
the creation operators and vacuum vector as follows
\c{Ber}
$$
\Psi = \sum_{n=0}^{\infty} \frac{1}{\sqrt{n!}}
\int d{\bf x}_1 ...  d{\bf x}_n \Psi_n({\bf
x}_1,...,{\bf x}_n) A^+({\bf x}_1) ... A^+({\bf x}_n)|0>.
$$
Introduce the  operator  of  number of particles $\hat{n}$ as $(\hat{n}
\psi)_n = n \psi_n$.  Let $T$ be a nonbounded self-adjoint operator in
$L^2({\bf R}^d)$  such  that  $T-1\ge  0$.  By $A^+TA^-$ we denote the
operator of the form
$$
(A^+TA^- \psi)_n = \sum_{j=1}^n  1^{\otimes  j-1}  \otimes  T  \otimes
1^{\otimes n-j} \psi_n.
$$
Introduce also the following norms in the Fock space:
$$
||\psi||_m = ||(\hat{n}+1)^m \psi||, \qquad
||\psi||^T_m = || (A^+TA^-+1)^m \psi||.
$$

{\bf Lemma B.1.} {\it Let $||\psi||^T_m <\infty$.  Then $||\psi||_m \le
||\psi||_m^T$.}

{\bf Proof.} It is sufficient to check that
$$
(\psi_n, (\sum_{j=1}^n 1^{\otimes j-1} \otimes  T  \otimes  1^{\otimes
n-j} +1)^{2m} \psi_n ) \ge (\psi_n, (\sum_{j=1}^n 1 + 1)^{2m} \psi_n).
$$
This relation is a corollary of the formula
$$
(\psi_n, T^{2l_1}   \otimes   ...   \otimes   T^{2l_n}   \psi_n)   \ge
(\psi_n,\psi_n)
$$
for all $l_1,...,l_n\ge 0$.  The latter formula is obtained  from  the
relation $||T^{-l_1} \otimes ...  \otimes T^{-l_n} || \le 1$. Lemma B.1
is proved.

Let ${\cal H}^{+-}$ be a nonbounded operator in $L^2({\bf R}^d)$  such
that operators $T^{-1/2}{\cal H}^{+-} T^{-1/2}$ and
${\cal H}^{+-}T^{-1}$ are bounded.

{\bf Lemma B.2.} {\it The following estimation is satisfied:
$$
||A^+{\cal H}^{+-}A^- \psi|| \le C ||\psi||^T_1
$$
with $C=max(||T^{-1/2}{\cal         H}^{+-}         T^{-1/2}||,||{\cal
H}^{+-}T^{-1}||)$.
}

{\bf Proof.} One should check
$$
(\psi_n,{\cal H}^{+-}_i{\cal H}^{+-}_j \psi_n) \le C
(\psi_n, T_iT_j \psi_n)
\l{bb1}
$$
with ${\cal H}^{+-}_i = 1^{i-1} \otimes {\cal H}^{+-} \otimes 1^{n-i}$,
$T_i =     1^{i-1}     \otimes     T    \otimes    1^{n-i}$.    Denote
$T_i^{1/2}T_j^{1/2}\psi_n =\phi_n$. Inequality \r{bb1} takes the form
$$
(\phi_n, T_i^{-1/2}  T_j^{-1/2}   {\cal   H}^{+-}_i   {\cal   H}^{+-}_j
T_i^{-1/2} T_j^{-1/2} \phi_n) \le C^2 (\phi_n,\phi_n).
\l{bb2}
$$
For $i\ne j$, property \r{bb2} is satisfied if $C = ||T^{-1/2}{\cal
H}^{+-} T^{-1/2}||$  as a corollary of the Cauchy-Bunyakovski-Schwartz
inequality. For $i=j$,  property \r{bb2} is  satisfied  if  $C=||{\cal
H}^{\pm}T^{-1}||$. Lemma B.2 is proved.

{\bf Lemma B.3.} {\it Consider the operator
$$
\hat{\varphi} =  \int d{\bf x}_1 ...  d{\bf x}_m d{\bf y}_1 ...  d{\bf
y}_k \varphi  ({\bf  x}_1,...,{\bf  x}_n,  {\bf  y}_1,...,{\bf   y}_k)
A^+({\bf x}_1) ... A^+({\bf x}_m) A^-({\bf y}_1) ... A^-({\bf y}_k)
$$
with $\varphi \in L^2({\bf R}^{d(m+k)})$.  Let $\Psi \in {\cal F}$ and
$||\Psi||_{l+\frac{k+m}{2}} < \infty$. Then
$$
||\hat{\varphi} \Psi||_l  \le  C   ||\varphi||_{L^2}   ||\Psi||_{l   +
\frac{k+m}{2}},
$$
where $C^2 = \max\{1,(m-k)!(m-k)^{2l}\}$.
}

{\bf Proof.} One has
$$
\begin{array}{c}
(\hat{\varphi}\Psi)_n({\bf z}_1,...,{\bf z}_n) =
\sqrt{\frac{(n-m+k)!}{(n-m)!}} \sqrt{\frac{n!}{(n-m)!}}
Sym \int d{\bf y}_1 ... d{\bf y}_k \varphi({\bf z}_1,...,{\bf z}_m,
{\bf y}_1,...,{\bf  y}_k) \\ \times
\Psi_{n-m+k}({\bf y}_1,...,{\bf y}_k,  {\bf
z}_{m+1},...,{\bf z}_n)
\end{array}
$$
where $Sym$ is a symmetrization operator.  Since $||Sym  \Phi_n||  \le
||\Phi_n||$ and
$$
||\int dy   \varphi({\bf   z},{\bf  y})  \Psi({\bf  y},{\bf  z}')||  \le
||\varphi|| ||\Psi||,
$$
one has
$$
||(\hat{\varphi}\Psi)_n|| \le
\sqrt{\frac{(n-m+k)!}{(n-m)!}} \sqrt{\frac{n!}{(n-m)!}}
||\varphi||
||\Psi_{n-m+k}||.
$$
Therefore,
$$
\begin{array}{c}
||\hat{\varphi}\Psi||_l^2 =       \sum_{n=0}^{\infty}       (n+1)^{2l}
\frac{(n-m+k)!}{(n-m)!} \frac{n!}{(n-m)!}                ||\varphi||^2
||\Psi_{n-m+k}||^2 \\
= \sum_{s=0}^{\infty}  \frac{(s+m-k)^{2l}  (s-k+1)  ...  s (s-k+1) ...
(s-k+m)}{(s+1)^{2l+k+m}} ||\varphi|| ||\Psi_s||^2 (s+1)^{2l+k+m} \\
\le C^2 |\varphi||^2 ||\Psi||^2_{l+\frac{k+m}{2}},
\end{array}
$$
where $s=m-n+k$. Lemma is proved.

{\bf Corollary.}
$$
||\frac{1}{2} A^{\pm}    {\cal    H}^{\pm    \pm}    A^{\pm} \Psi||
\le
\frac{1}{\sqrt{2}} ||{\cal H}^{\pm \pm}||_2 ||\Psi||_1,
$$

Here $||\cdot||_2$  is  a  Hilbert-Schmidt  norm  $||A||_2  = \sqrt{Tr
A^+A}$.

Consider the quadratic Hamiltonian
$$
H = \frac{1}{2} A^+{\cal  H}^{++}  A^-  +  A^+  {\cal  H}^{+-}  A^-  +
\frac{1}{2} A^- {\cal H}^{--} A^- + \overline{\cal H}.
$$
Let $H$, ${\cal H}^{++}$,
${\cal H}^{+-}$, ${\cal H}^{--}$,
$\overline{\cal H}^{++}$  be $\alpha$-dependent,  $\alpha\in {\bf R}$,
and
$({\cal H}^{++})^+ = {\cal H}^{--}$,
$({\cal H}^{+-})^+ = {\cal H}^{+-}$.

{\bf Lemma  B.4.}  {\it  Let   $\overline{\cal   H}_{\alpha}$   be   a
continuously differentiable function, ${\cal H}_{\alpha}^{++}$ be a
continously differentiable   Hilbert-Schmidt   operator  in  the  norm
$||\cdot||_2$, while $T^{-1/2}{\cal H}^{+-}_{\alpha}T^{-1/2}$ and
${\cal H}^{+-}_{\alpha}T^{-1}$ are operator functions being
continuously differentiable  in  the operator norm.  Let $\psi\in{\cal
F}$, $||\psi||_1^T <\infty$. Then the vector function $H_{\alpha}\psi$
is continously differentiable in the strong topology.
}

{\bf Proof.} One has from lemmas B.2 and B.3 that
$$
\begin{array}{c}
||(\frac{H_{\alpha+\delta\alpha} -     H_{\alpha}}{\delta\alpha}     -
\frac{dH}{d\alpha})\psi|| \le
\sqrt{2} ||
\frac{{\cal H}^{++}_{\alpha + \delta \alpha} - {\cal H}^{++}_{\alpha}}
{\delta \alpha}    -    \frac{d}{d\alpha}{\cal    H}^{++}_{\alpha}||_2
||\psi||_1 + \\
max\left(
|| T^{-1/2} \left[
\frac{{\cal H}^{++}_{\alpha + \delta \alpha} - {\cal H}^{++}_{\alpha}}
{\delta \alpha} - \frac{d}{d\alpha}{\cal H}^{++}_{\alpha} \right]
T^{-1/2}||,
|| \left[
\frac{{\cal H}^{++}_{\alpha + \delta \alpha} - {\cal H}^{++}_{\alpha}}
{\delta \alpha} - \frac{d}{d\alpha}{\cal H}^{++}_{\alpha} \right]
T^{-1}||
\right) ||\psi||_1^T + \\
\left|
\frac{\overline{\cal H}_{\alpha + \delta \alpha} -
\overline{\cal H}_{\alpha}}
{\delta \alpha} - \frac{d}{d\alpha}\overline{\cal H}_{\alpha} \right|
 \to_{\delta\alpha\to 0} 0.
\end{array}
$$
The fact   that  $\frac{d}{d\alpha}H_{\alpha}\psi$  is  continuous  is
checked analogously. Lemma B.4 is proved.

Consider now the Cauchy problem for the equation
$$
\begin{array}{c}
i \frac{d\Psi_t}{dt} = H_t\Psi_t,
\\
H_t = \frac{1}{2} A^+{\cal  H}^{++}_t  A^-  +  A^+  {\cal  H}^{+-}_t  A^-  +
\frac{1}{2} A^- {\cal H}^{--}_t A^- + \overline{\cal H}_t.
\end{array}
\l{bb3}
$$
on the Fock vector $\Psi_t$; the strong derivative enters to eq.\r{bb3}.

Formally, the solution for the initial condition
$$
\Psi_0 = \sum_{n=0}^{\infty} \frac{1}{\sqrt{n!}} \int d{\bf  x}_1  ...
d{\bf x}_n   A^+({\bf   x}_1)  ...  A^+({\bf  x}_n)  \Psi_{0,n}  ({\bf
x}_1,...,{\bf x}_n)|0>
\l{bb4}
$$
is looked for in the following form \c{Ber,MS-RJMP}
$$
\Psi_t = \sum_{n=0}^{\infty} \frac{1}{\sqrt{n!}} \int d{\bf  x}_1  ...
d{\bf x}_n   A_t^+({\bf   x}_1)  ...  A_t^+({\bf  x}_n)  \Psi_{0,n}  ({\bf
x}_1,...,{\bf x}_n)|0>_t
\l{bb5}
$$
with
$$
|0>_t = c^t \exp [\frac{1}{2}\int d{\bf x}d{\bf  y}  M^t({\bf  x},{\bf
y}) A^+({\bf x}) A^+({\bf y})]|0>.
\l{bb6}
$$
while operators $A^{+}_t({\bf x})$ are chosen to be
$$
A_t^{+}({\bf x}) = \int d{\bf y} [A^+({\bf y}) G_t^*({\bf y},{\bf  x})
- A^-({\bf y}) A_t^*({\bf y},{\bf x})].
$$
Namely, the Gaussian ansatz \r{bb6} formally satisfies eq.\r{bb3} if
$$
\begin{array}{c}
i \frac{dc^t}{dt}  =   \frac{1}{2}   Tr   {\cal   H}^{--}_tM^t   c^t   +
\overline{\cal H}_t c^t,\\
i \frac{dM^t}{dt} = {\cal H}^{++}_t + {\cal H}_t^{+-} M_t + M_t  {\cal
H}_t^{-+} + M_t {\cal H}_t^{--} M_t.
\end{array}
\l{bb7}
$$
Here $M_t$ is the operator with kernel $M^t({\bf x},{\bf y})$,  ${\cal
H}_t^{-+} =  ({\cal  H}_t^{+-})^*$.  The  operators  $A_t^+({\bf  x})$
commute with $i\frac{d}{dt}-H_t$ if
$$
\begin{array}{c}
i \frac{dF^t}{dt} = {\cal H}_t^{+-} F_t + {\cal H}_t^{++} G_t,
\qquad
-i \frac{dG^t}{dt} = {\cal H}_t^{-+} G_t + {\cal H}_t^{--} F_t.
\end{array}
\l{bb8}
$$
Here $F_t$,  $G_t$  are  operators with kernels $F_t({\bf x},{\bf y})$
and $G_t({\bf x},{\bf y})$. Note that the operator $M_t = F_tG_t^{-1}$
formally satisfies   eq.\r{bb7}.   Initial   conditions   \r{bb4}  are
satisfied if $F_0=0$, $G_0=1$.

Let us check that eq.\r{bb3} is satisfied in a strong sense.

First of all, let us present some auxiliary lemmas.

{\bf Lemma B.5.} {\it  Let  $M$  be  a  Hilbert-Schmidt  operator  and
$||M||<1$. Then
$$
\exp[\frac{1}{2} A^+MA^+]|0>
\l{bb8*}
$$
}

The proof is presented in \c{Ber}.

{\bf Corollary.}  {\it  For  the  state   \r{bb8*},   the   estimation
$$
||\psi_n|| \le  A  e^{- \alpha  n}
\l{bb9}
$$
is  satisfied under conditions of
lemma B.5 for some $A$ and $0< \alpha \le - \frac{1}{2}log ||M||$. }

{\bf Proof.} Since $||M||<1$, $||e^{2\alpha}M|| <1$. Since expression
$\tilde{\psi} =  exp[\frac{1}{2}  e^{2\alpha} A^+ MA^+]|0>$ specifies a
Fock space   vector,   $||\tilde{\psi}_{2n}||   =   ||e^{2\alpha   n}
\psi_{2n}|| \le A$. Corollary is proved.

{\bf Lemma   B.6.}   {\it  Let  $M$,  $\delta  M$  be  Hilbert-Schmidt
operators, $||M||\le 1$, $||M+\delta M||\le 1$ and
$$
||\delta M||_2 \le \frac{1}{4} \log||M||^{-1} ||M||^{-3/8}.
$$
Then
$$
\exp[\frac{1}{2}A^+(M+\delta M)A^+]|0>      =      \sum_{k=0}^{\infty}
\frac{1}{k!} [\frac{1}{2}A^+\delta M A^+]^k \exp[\frac{1}{2}A^+MA^+]|0>
$$
}

{\bf Proof.} One should check that
$$
\begin{array}{c}
s-\lim_{N\to\infty} \sum_{k,l,  k+l\le  N}  \frac{1}{2^kk!} (A^+\delta M
A^+)^k \frac{1}{2^ll!}    (A^+MA^+)^l    |0>    = \\
s-\lim_{N\to\infty}
\sum_{k=0}^N \frac{1}{2^kk!}         (A^+\delta        M        A^+)^k
e^{\frac{1}{2}A^+MA^+}|0>
\end{array}
\l{bb10*}
$$
Since the strong limit in the left-hand side of equality \r{bb10*}  exists,
eq.\r{bb10*} can be presemted as
$$
\sum_{k=0}^N \sum_{l=N-k+1}^{\infty} \psi_{k,l} \to_{N\to\infty} 0
\l{bb10}
$$
with
$$
\psi_{k,l} =   \frac{1}{2^kk!}   (A^+\delta   MA^+)^k  \frac{1}{2^ll!}
(A^+MA^+)^l |0>.
$$
Since
$$
([A^+\delta M A^+]\psi)_n({\bf x}_1,...,{\bf x}_n) = Sym \sqrt{n(n-1)}
\delta M({\bf x}_1,{\bf x}_2) \psi_{n-2} ({\bf x}_3,...,{\bf x}_n),
$$
one has
$$
||([A^+\delta MA^+]\psi)_n||    \le   \sqrt{n(n-1)}   ||\delta   M||_2
||\psi||_{n-2}.
$$
By induction, one obtains:
$$
||[A^+\delta MA^+]^k   \psi_{n-2k}||   \le   \sqrt{\frac{n!}{(n-2k)!}}
||\delta M||_2^k ||\psi_{n-2k}||.
$$
It follows from the extimation \r{bb9} that
$$
\begin{array}{c}
||\psi_{k,l}|| \le \sqrt{\frac{(l+2k)!}{k!}} \frac{||\delta M||_2^k}
{2^kk!} A  e^{-\alpha  l/2}  e^{-\alpha  l/2} \\
\le   \max_l   (l+2k)^k
e^{-\alpha (l+2k)/2}   A   e^{-\alpha   l/2}   \frac{(||\delta   M||_2
e^{\alpha})^k}{2^kk!} =  A    e^{-\alpha    l/2}    \frac{k^k}{k!e^k}
\left(\frac{||\delta M||_2e^{\alpha}}{\alpha}\right)^k
\end{array}
$$
Since $k!   \sim   (k/e)^k\sqrt{2\pi  k}$  as  $k\to\infty$,  one  has
$e^{-k}k^k/k!\le A_1$. Therefore,
$$
||\psi_{k,l}|| \le AA_1 e^{-\alpha l/2} b^k
\l{bb10x}
$$
with $b = ||\delta M||_2e^{\alpha}/\alpha$. Therefore,
$$
\sum_{k=0}^N \sum_{l=N-k+1}^{\infty}   ||\psi_{k,l}||  =  \sum_{k=0}^N
AA_1 b^k e^{-\frac{\alpha}{2}(N-k+1)} \frac{1}{1-e^{-\alpha/2}}
\le AA_1 \frac{e^{-\alpha(N+1)/2}}{(1-e^{-\alpha/2})(1-be^{-\alpha/2})}.
$$
Therefore, for  $||\delta  M||_2  e^{3\alpha/2}  \le  \alpha$ property
\r{bb10} is satisfied. Choosing $\alpha = - \frac{1}{4} log ||M||$, we
obtain statement of lemma.

{\bf Lemma B.7.} {\it Let $M_t$,  $t\in [t_1,t_2]$ be a differentiable
operator function, $||M_t||_2 < \infty$,
$$
|| \frac{M_{t+\delta   t}-M_t}{\delta   t}    -    \frac{dM_t}{dt}||_2
\to_{\delta t \to 0} 0.
\l{bb10a}
$$
Then
$$
|| \frac{e^{\frac{1}{2}A^+M_{t+\delta t}A^+}|0>
- e^{\frac{1}{2}A^+M_{t}A^+}|0>}{\delta  t}  -  \frac{1}{2} A^+
\frac{dM_t}{dt} A^+e^{\frac{1}{2}A^+M_{t}A^+}|0>||_m
\to_{\delta t \to 0} 0.
\l{bb11}
$$
}

{\bf Proof.}  Denote  $\delta  M  \equiv  \delta  M_{t,\delta   t}   =
M_{t+\delta t} - M_t$. It is sufficient to check the following formulas:
$$
||\frac{e^{\frac{1}{2}A^+M_{t+\delta t}A^+}|0> - 1 -
{\frac{1}{2}A^+M_{t}A^+}|0>}{\delta t}
e^{\frac{1}{2}A^+M_{t}A^+}|0>||_m \to_{\delta t \to 0} 0;
\l{bb12}
$$
$$
||A^+\frac{\delta M}{\delta t}- \frac{dM}{dt})A^+
e^{\frac{1}{2}A^+M_{t}A^+}|0>||_m \to_{\delta t \to 0} 0.
\l{bb13}
$$
The latter formula is a direct corollary of lemma B.3, property
$||e^{\frac{1}{2}A^+M_{t+\delta t}A^+}|0>||_{m+1} <\infty$  following
from formula \r{bb9} and relation
$||\frac{\delta M}{\delta t} - \frac{dM}{dt}||_2 \to_{\delta t \to  0}
0$. Formula \r{b12} is a corollary of the relation
$$
\sum_{k=2}^{\infty} \sum_{l=0}^{\infty}  \frac{1}{\delta t} (2k+1+l)^m
||\psi_{k,l}|| \to_{\delta t \to 0} 0.
\l{bb14}
$$
Makibg use   of   the   estimation  \r{bb10x}  and  formula  $||\delta
M||_2^2/\delta t \to_{\delta t\to 0} 0$,  we prove relation  \r{bb14}.
Lemma B.7 is proved.

{\bf Lemma  B.8.}  {\it  Let  $T$ be such nonbounded self-adjoint
operator in $L^2({\bf R}^d)$
that $T-1 >0$, $D(T)\subset D({\cal H}^{+-})$, ${\cal H}_t^{+-}T^{-1}$
be uniformly bpinded operator.
Let the initial condition for eq.\r{bb3} be of
the form \r{bb4}, where $\Psi_{0,n}=0$ as $n\ge N_0$,
$$
\Psi_{0,n}({\bf x}_1,...,{\bf  x}_n)  =  \sum_{j=1}^{J_0}   f_j^1({\bf
x}_1) ... f_j^n({\bf x}_n), \qquad f_j^s \in D(T).
\l{bb14a}
$$
Let Hilbert-Schmidt operator $M_t$ satisfy eq.\r{bb7}  (the  derivative
is defined   in   the  Hilbert-Schmidt  sense  \r{bb10a})  and  initial
condition $M_0=0$,  ,  $c_t$  obey  eq.\r{bb7},  $F_t$  and  $G_t$  be
uniformly bounded operators $F_t:D(T)\to D(T)$, $G_t:D(T)\to D(T)$
satisfying eq.\r{bb8} in the strong sense on $D(T)$, $F_0=0$, $G_0=1$.
Then the Fock vector \r{bb5} obeys eq.\r{bb3} in the strong sense.
}

{\bf Proof.} It is sufficient to prove lemma for the initial condition
$$
\Psi_0 = \frac{1}{\sqrt{n!}} A^+[f^1] ... A^+[f^n]|0>
$$
where $A^+[f]= \int d{\bf x} f({\bf x})A^+({\bf x})$. Let us show that
the Fock vector
$$
\Psi_t = \frac{1}{\sqrt{n!}} A_t^+[f^1] ... A_t^+[f^n]|0>_t
$$
with
$$
A_t^+[f] =  \int  d{\bf  y} [A^+({\bf y}) (G_t^*f)({\bf y}) - A^-({\bf
y}) (F_t^*f)({\bf y})]
$$
satisfies eq.\r{bb3}. Let
$$
\dot{\Psi}_t \equiv \frac{1}{\sqrt{n!}}  (\sum_{j=1}^n  A_t^+[f^1]  ...
\dot{A}_t^+[f^j] ... A_t^+[f^n]|0>_t + A_t^+[f^1]  ...
{A}_t^+[f^j] ... A_t^+[f^n]\frac{d}{dt}|0>_t
$$
with
$$
\begin{array}{c}
\dot{A}_t^+[f] =  \int  d{\bf  y} [A^+({\bf y})
\frac{d}{dt}(G_t^*f)({\bf y}) - A^-({\bf
y}) \frac{d}{dt}(F_t^*f)({\bf y})], \\
\frac{d}{dt}|0>_t \equiv \frac{dc^t}{dt} e^{\frac{1}{2} A^+M_tA^+}|0>
+ c^t \frac{1}{2} A^+ \frac{dM_t}{dt} A^+ e^{\frac{1}{2} A^+M_tA^+}|0>.
\end{array}
$$
One has
$$
\begin{array}{c}
\frac{\Psi_{t+\delta t}-\Psi_t}{\delta    t}    -    \dot{\Psi}_t    =
\frac{1}{\sqrt{n!}} A^+_{t+\delta t} [f^1] ... A^+_{t+\delta t} [f^n]
\left(
\frac{|0>_{t+\delta t} - |0>_t}{\delta t} - \frac{d}{dt}|0>_t
\right)
+ \\
\frac{1}{\sqrt{n!}} [ A^+_{t+\delta t} [f^1] ... A^+_{t+\delta t} [f^n]
-  A^+_{t} [f^1] ... A^+_{t} [f^n]] \frac{d}{dt}|0>_t +\\
\sum_{j=1}^n  A^+_{t+\delta t} [f^1] ... A^+_{t+\delta t} [f^{j-1}]
[\frac{A^+_{t+\delta t}[f^j]     -     A^+_t[f^j]}{\delta     t}     -
\dot{A}^+_t[f^j]]  A^+_{t} [f^{j+1}] ... A^+_{t} [f^n]|0>_t +\\
\sum_{j=1}^n \frac{1}{\sqrt{n!}}
(A^+_{t+\delta t}[f^1]...A_{t+\delta t}^+[f^{j-1}] -
 A^+_{t} [f^1] ... A^+_{t} [f^{j-1}]] \dot{A}_t^+[f^j]
 A^+_{t} [f^{j+1}] ... A^+_{t} [f^n]|0>_t.
 \end{array}
$$
It follows from lemmas B.3, B.7 and conditions of lemma B.8 that
$$
||\frac{\Psi_{t+\delta t}-   \Psi_t}{\delta   t}   -    \dot{\Psi}_t||
\to_{\delta t\to 0} 0.
$$
Eqs.\r{bb7}, \r{bb8}  imply  that $\dot{\Psi}_t = -iH_t\Psi_t$.  Lemma
B.8 is proved.

Denote by ${\cal D}_1 \subset {\cal F}$ the set of  all  Fock  vectors
$\Psi \in  {\cal  F}$ such that $\Psi_n$ vanish at $n\ge N_0$ and have
the form \r{bb14a} as $n<N_0$.  Lemma B.8 allows us to  construct  the
mapping $U_t:{\cal  D}_1 \to {\cal F}$ of the form $U_t\Psi_0=\Psi_t$.
Note that the domain ${\cal D}_1$ is dense in ${\cal F}$.

Denote
$$
A_t^-[f] \equiv  (A_t^+[f])^+  \equiv  \int  d{\bf  y}  [A^-({\bf  y})
(G_tf)({\bf y}) - A^+({\bf y}) (F_tf)({\bf y}) ].
$$

{\bf Lemma B.9.} {\it
1. The operators $A_t^{\pm}[f]$ obey the commutation relations
$$
[A^-_t[f], A^+_t[g]] = (f,g), \qquad [A^{\pm}_t[f], A^{\pm}_t[g]] =0.
\l{bb15}
$$
2. The following property is satisfied:
$$
A_t^-[f]|0>_t =0.
\l{bb15a}
$$
3. The operator $U_t$ is isometric.
}

{\bf Proof.} The commutation relations \r{bb15} are rewritten as
$$
\begin{array}{c}
(G_tf,G_tg) - (F_tf,F_tg) = (f,g);\\
(F_t^*f, G_tg) - (G_t^*f,F_tg) =0.
\end{array}
\l{bb16}
$$
They are satisfied at $t=0$.  The time derivatives  of  the  left-hand
sides of  eqs.\r{bb16}  vanish because of eqs.\r{bb8}.  Statement 1 is
proved.

The fact that $U_t$ is an isometric operator is  a  corollary  of  the
property $\frac{d}{dt}(\Psi_t,\Psi_t) =0$.

Analogously to  lemma  B.8,  we find that the vector $\tilde{\Psi}_t =
A^-_t[f]|0>_t$ obeys eq.\r{bb3} in the strong sense.  Since $\Psi_0=0$
and $||\Psi_t||=||\Psi_0||$, one has $\Psi_t=0$. Property \r{bb15a} is
proved. Note that it means that
$$
M_tG_t=F_t.
\l{bb16c}
$$
Lemma B.9 is proved.

Therefore, the operator $U_t$ can be extended to the whole space $\cal
F$, $U_t:{\cal F} \to {\cal F}$.

{\bf Lemma B.10.} {\it Let the operator
$$
\left(
\begin{array}{cc}
G^+ & - F^+ \\ - F^T & G^T
\end{array}
\right)
$$
be invertible.  Then the following relation  is  satisfied  on  ${\cal
D}_1$:
$$
U_t^{-1} A^+TA^- U_t \Psi_0= (A^+G^T_t + A^-F^+)T(FA^++G^*A^-)\Psi_0
\l{bb16a}
$$
}

{\bf Proof.} It follows from lemma B.9 that
$$
\left(
\begin{array}{cc}
G^+ & - F^+ \\ - F^T & G^T
\end{array}
\right)
\left(
\begin{array}{cc}
G &  F^* \\  F & G^*
\end{array}
\right) = 1
$$
Therefore,
$$
\left(
\begin{array}{cc}
G^+ & - F^+ \\ - F^T & G^T
\end{array}
\right)^{-1} =
\left(
\begin{array}{cc}
G &  F^* \\  F & G^*
\end{array}
\right)
$$
and
$$
\begin{array}{c}
A^-({\bf y}) = \int d{\bf z} (F_t({\bf y},{\bf z})A_t^+({\bf z}) +
G_t^*({\bf y},{\bf z}) A_t^-({\bf z})), \\
A^+({\bf y}) = \int d{\bf z} (F^*_t({\bf y},{\bf z})A_t^-({\bf z}) +
G_t({\bf y},{\bf z}) A_t^+({\bf z})).
\end{array}
$$
Identity \r{bb16a} is then a corollary of definition of  the  operator
$U_t$.

By ${\cal  D}  \in {\cal F}$ we denote set of such Fock vectors $\Psi$
that $||\Psi||_1^T <\infty$.

{\bf Lemma B.11.} {\it
Let $\Psi_0 \in {\cal D}$.  Suppose that $TF_t$  and
${\cal H}^{++}$ are
continuous operator   functions   in   the  $||\cdot||_2$-norm,  $G_t$,
$T^{1/2}G_tT^{-1/2}$, $TG_tT^{-1}$, $T^{-1/2}{\cal H}_t^{+-}T^{-1/2}$,
${\cal H}^{+-}T^{-1}$  are  continous  operator   functions   in   the
$||\cdot||$-norm. Then the following statements are satisfied.

1. $\Psi_t \equiv U_t\Psi_0 \in {\cal D}$.

2. $\Psi_t$ obeys eq.\r{bb3} in the strong sense.

3. $$||\Psi_t-\Psi_0||_1^T \to_{t\to 0} 0.
\l{bb17} $$
}

{\bf Proof.} Let $\Psi_0\in {\cal D}_1$.  For $||U_t\Psi_0||_1^T$, one
has the following estimation:
$$
\begin{array}{c}
||U_t\Psi_0||_1^T =  ||U_t^{-1}(\hat{T}+1)U_t\Psi_0|| \le ||\Psi_0|| +
||(A^+G^T+A^-F^+)T(FA^++G^*A^-)\Psi_0|| \le
\\
(1 +  ||F^+||_2 ||TF||_2)
||\Psi_0||  +  (\sqrt{2}||G^TTF||_2  + ||F^+TF|| +
||F^+TG||_2) ||\Psi_0|| \\
+ (||T^{-1/2}G^TTG^*T^{-1/2}||  + ||A^TTA^*T^{-1}||)||\Psi_0||_1^T \le
const ||\Psi||_1^T
\end{array}
$$
at $t\in[0,t_1]$.  Therefore,the  operator  $U_t$  is  bounded in norm
$||\cdot||_1^T$. The extension of the operator $U_t$ to  $\cal  D$  is
then also  a  bounded operator in $||\cdot||_1^T$ norm.  One therefore
has $\Psi_t\in {\cal D}$.

The fact that $||U_t\Psi_0 - \Psi_0||_1^T \to_{t\to 0} 0$  if  $\Psi_0
\in {\cal  D}_1$  is  justified  analogously  to lemma B.8.  Since the
operator $U_t:{\cal D} \to {\cal D}$ is  uniformly  bounded  at  $t\in
[0,t_1]$ in  $||\cdot||_1^T$-norm,  the Banach-Steinhaus theorem (see,
for example, \c{KA}) implies relation \r{bb17}.

To check the second statement, note that lemma B.8 imply that
$$
\frac{U_{t+\delta t}-U_t}{\delta t} \to_{\delta t \to 0} \frac{dU_t}{dt}
\l{bb18}
$$
in the  strong  sense  on  ${\cal  D}_1$.  For  showing  that relation
\r{bb18} is  satisfied  in  the  strong  sense  on  $\cal  D$,  it  is
sufficient to show that the operator
$$
\frac{\delta U}{\delta t} :{\cal D} \to {\cal F}
$$
is uniformly bounded,
$$
||\frac{\delta U}{\delta t} \Psi|| \le C ||\Psi||_1^T.
$$
One has
$$
\begin{array}{c}
||\frac{\delta U}{\delta t} \Psi|| = ||\int_0^1 ds  \dot{U}_{t+s\delta
t}\Psi|| = ||\int_0^1 ds H_{t+s\delta t} U_{t+s\delta t}|| \le
\\
\max_{s\in[0,1]} [\sqrt{2}||{\cal     H}^{++}_{t+s\delta    t}||_2    +
||T^{-1/2}{\cal H}^{+-}_{t_s+ \delta t}T^{-1/2}|| +
||{\cal H}^{+-}_{t_s+\delta t}T^{-1}|| ] ||U_{t+s\delta t}\Psi||_1^T.
\end{array}
$$
Lemma B.11 is proved.

Let us now check properties of operators $F_t$, $F_t$, $M_t$.

First of all, consider the Cauchy problem
$$
\begin{array}{c}
i\dot{f}_t = Y_tf_t + Z_tg_t,\\
-i\dot{g}_t = Z_t^*f_t + Y_t^*g_t,\\
f_0=0, g_0=1,
\end{array}
\l{bb19}
$$
where $g_t$  is   a   bounded   operator   functions,   $f_t$   is   a
Hilbert-Schmidt operator  function.  The  derivatives  in \r{bb19} are
understood as
$$
||(\frac{g_{t+\delta t}-g_t}{\delta    t}    -     \dot{g}_t)\varphi||
\to_{\delta t \to 0} 0, \qquad
||\frac{f_{t+\delta t}-f_t}{\delta    t}    -     \dot{f}_t||_2
\to_{\delta t \to 0} 0.
\l{bb19*}
$$

{\bf Lemma  B.12.}  {\it  Let  $Y_t$  be a strongly continous operator
function, while $||Z_{t+\tau}-Z_t||_2 \to_{\tau\to 0} 0$,  $||TZ_t||_2
\le a_1^t$,  $||TY_tT^{-1}||  \le  a_2^t$,  $||T^{1/2}Y_tT^{-1}||  \le
a_3^t$ for smooth functions $a_k^t$.  Then there exist a  solution  to
the Cauchy problem \r{bb19} such that
$$
||Tf_t||_2 \le a_4^t, \qquad ||T^{1/2}g_tT^{-1/2}|| \le a_5^t, \qquad
||Tg_tT^{-1}|| \le a_6^t, \qquad ||g_t|| \le a_7^t
\l{bb19a}
$$
for smooth functions $a_k^t$.
}

{\bf Proof} (cf.\c{MS-RJMP}). Let us look for the solution to the
Cauchy problem in the following form:
$$
f_t = \sum_{n=0}^{\infty} f_t^n, \qquad
g_t = \sum_{n=0}^{\infty} g_t^n.
\l{bb20}
$$
where $f_t^0=0$, $g_t^0=1$,
$$
\begin{array}{c}
f_t^{n+1} =  -i  \int_0^t   d\tau   (Y_{\tau}f_{\tau}^n   +   Z_{\tau}
g_{\tau}^n), \\
g_t^{n+1} = -i \int_0^t d\tau (Y^*_{\tau}g_{\tau}^n + Z^*_{\tau}
f_{\tau}^n).
\end{array}
\l{bb20*}
$$
By induction we find that $||f_t^n||_2 \le C_1t^n/n!$,
$||g_t||_2 \le C_1t^n/n!$ for $t\in [0,t_1]$. Here $C_1$ is a constant.

Therefore, the series \r{bb20} converge.  $f_t$  is  a  Hilbert-Schmidt
operator, while $g_t$ is a bounded operator. Analogously, we show
$$
||Tf_t^n||_2 \le \frac{C_2 t^n}{n!},
\qquad
||Tg_t^nT^{-1}|| \le \frac{C_2 t^n}{n!},
\qquad
||T^{1/2}g_t^n T^{-1/2}|| \le \frac{C_2 t^n}{n!},
$$
where $t\in [0,t_1]$. Therefore, properties \r{bb19a} are satisfied.

To check relations \r{bb19*}, note that
$$
\begin{array}{c}
f_t = - i \int_0^t d\tau (Y_{\tau}f_{\tau} + Z_{\tau}g_{\tau}),
\\
g_t = i\int_0^t d\tau (Z_{\tau}^*f_{\tau} + Y_{\tau}^*g_{\tau}).
\end{array}
\l{bb20a}
$$
Eqs.\r{bb20a} imply that the  operator  functions  $f_t$,  $g_t$  obey
properties
$$
||T(f_{t+\delta t} - f_t)||_2 \to_{\delta t \to 0} 0,
\qquad
||(g_{t+\delta t} - g_t)|| \to_{\delta t \to 0} 0,
$$
Therefore,
$$
\begin{array}{c}
||i \frac{f_{t+\delta t} - f_t}{\delta t} - Y_t f_t - Z_t g_t||_2 \le
\int_0^1 ds ||Y_{t+s\delta t} f_{t+s\delta t} + Z_{t+s\delta t} g_{t+s
\delta t} - Y_tf_t - Z_tg_t||_2, \\
||(-i \frac{g_{t+\delta  t}  -  g_t}{\delta  t}  -  Z_t^*f_t  -  Y_t^*
g_t)\varphi_t|| \le
\int_0^1 ds  ||(Z_{t+s\delta  t}^* f_{t+s\delta t} + Y^*_{t+s\delta t}
g_{t+s\delta t} - Z_t^* f_t - Y_t^* g_t) \varphi_t||.
\end{array}
$$
Since the  integrands  are  uniformly bounded functions,  the Lesbegue
theorem (see,  for example,  \c{KF}) tells us that it is sufficient to
check that
$$
\begin{array}{c}
||Y_{t+\tau}f_{t+\tau} - Y_tf_t||_2 \to_{\delta t\to 0} 0,
\qquad
||Z_{t+\tau}g_{t+\tau} - Z_tg_t||_2 \to_{\delta t\to 0} 0,\\
s-\lim_{\tau\to 0} Z_{t+\tau}^* f_{t+\tau} = Z_t^*f_t,\\
s-\lim_{\tau\to 0} Y_{t+\tau}^* g_{t+\tau} = Y_t^*g_t.
\end{array}
$$
These relations  are  corollaries  of  conditions  of  lemma  B.12 and
formulas \r{bb20*}.

{\bf Lemma B.13.} {\it Let ${\cal H}_t^{+-} = L + {\cal H}_t$,  ${\cal
H}_t$, $T^{1/2}{\cal   H}T^{-1/2}$,   $T{\cal  H}T^{-1}$  be  strongly
continous operator functions, $||{\cal H}^{++}_{t+\delta t} -
{\cal H}^{++}_{t}||_2   \to_{\delta   t   \to   0}   0$,   $L$   be  a
$t$-independent (maybe nonbounded) self-adjoint  operator,  such  that
$||LT^{-1}||<\infty$, while  $||T^{1/2}  e^{-iLt}  T^{-1/2}||<\infty$,
$||T e^{-iLt} T^{-1}|| <\infty$.  Then there exists a solution to  the
Cauchy problem  for  system \r{bb8} for the initial condition $F_0=0$,
$G_0=1$:
$$
\begin{array}{c}
||i \frac{F_{t+\delta t} - F_t}{\delta t} - {\cal H}_t^{++}
F_t - {\cal H}_t^{+-} G_t||_2 \to_{\delta t\to 0} 0,\\
||(-i \frac{G_{t+\delta t} - G_t}{\delta t} - {\cal H}_t^{--}
G_t - {\cal H}_t^{-+} F_t)\varphi||  \to_{\delta  t\to  0}  0,  \qquad
\varphi \in D(T).
\end{array}
\l{bb22}
$$
Moreover,
$$
||TF_t||_2 \le b(t), \qquad
||TG_tT^{-1}|| \le b(t), \qquad
||T^{1/2} G_t T^{-1/2}|| \le b(t), \qquad
||G_t|| \le b(t)
\l{bb21}
$$
for some smooth function $b(t)$ on $t\in [0,t_1]$.
The properties \r{bb16c} are also satisfied.
}

{\bf Proof.} Consider the operator functions
$$
F_t = e^{-iLt} f_t, \qquad G_t = e^{iL^*t} g_t,
$$
where $(f_t, g_t)$ is a solution to the Cauchy problem \r{bb19} with
$Y_t =  e^{iLt}  {\cal H}_t e^{-iLt}$,  $Z_t = e^{iLt} {\cal H}_t^{++}
e^{iL^*t}$, $f_0=0$,   $g_0=1$.   Check   of   properties   \r{bb21}  is
straightforward. Let us prove relations \r{bb22}. One has
$$
\begin{array}{c}
i \frac{F_{t+\delta  t}-F_t}{\delta  t}  -  (L+{\cal  H}_t)F_t - {\cal
H}_t^{++} G_t = (i \frac{e^{-iL\delta t} - 1}{\delta t }T^{-1} -LT^{-1})
TF_t + i e^{-iLt} (\frac{f_{t+\delta t}-f_t}{\delta t}- \dot{f}_t), \\
- i \frac{G_{t+\delta  t}-G_t}{\delta  t}  -  (L^*+{\cal  H}^*_t)G_t - {\cal
H}_t^{--} F_t = (-i \frac{e^{-iL^*\delta t} - 1}{\delta t }T^{-1} -L^*T^{-1})
TF_t + i e^{-iL^*t} (\frac{g_{t+\delta t}-g_t}{\delta t}- \dot{g}_t).
\end{array}
$$
Since
$$
||(i\frac{e^{-iL\tau}-1}{\tau}T^{-1} - LT^{-1})\varphi||
\le \int_0^1 ds ||(e^{-iL\tau s}-1) LT^{-1}\varphi||\to_{\tau\to 0} 0,
$$
we obtain relations \r{bb22}.

Property \r{bb16c} is proved analogously  to  \c{MS-RJMP}:  one  should
consider the convergent in $||\cdot||$-norm series
$$
\left(
\begin{array}{cc}
G & F^* \\ F & G^*
\end{array}
\right)^{-1} =
\sum_{n=0}^{\infty}
\left(
\begin{array}{cc}
G_t^{(-n)} & F_t^{(-n)*} \\ F_t^{(-n)} & G_t^{(-n)*}
\end{array}
\right)
\left(
\begin{array}{cc}
e^{iL^*t} & 0 \\ 0 & e^{-iLt}
\end{array}
\right)
$$
with
$$
\left(
\begin{array}{cc}
G_t^{(-n)} & F_t^{(-n)*} \\ F_t^{(-n)} & G_t^{(-n)*}
\end{array}
\right)
= i \int_0^t d\tau
\left(
\begin{array}{cc}
G_{\tau}^{(-n+1)} & F_{\tau}^{(-n+1)*} \\ F_{\tau}^{(-n+1)} &
G_{\tau}^{(-n+1)*}
\end{array}
\right)
\left(
\begin{array}{cc}
Y_{\tau} & Z_{\tau} \\ -Z^*_{\tau} & -Y^*_{\tau}
\end{array}
\right)
$$
Lemma B.13 is proved.

{\bf Lemma  B.14.}  {\it Under conditions of lemma B.13 there exists a
solution to  the  Cauchy  problem  for  eq.\r{bb7}  with  the  initial
condition $M_0=0$.}

{\bf Proof.} It follows from \c{Ber} that the matrix $G$ is invertible
and $||G^{-1}||<1$. Consider the operator $M_t=F_tG_t^{-1}$. Note that
$||TM_t||_2 < \infty$, $||LM_t|| <\infty$. One has
$$
M_{t+\delta t}  - M_t = M_{t+\delta t} (G_t-G_{t+\delta t}) G_t^{-1} +
(F_{t+\delta t} - F_t) G_t^{-1},
$$
so that  $||T(M_{t+\delta  t}  -  M_t)||_2  \to_{\delta  t\to  0}  0$.
Therefore,
$$
\begin{array}{c}
\frac{M_{t+\delta t}-M_t}{\delta t} - \dot{F}_tG_t^{-1} + F_T G_t^{-1}
\dot{G}_t G_t^{-1} = \\
M_{t+\delta t}T T^{-1} (\frac{G_t-G_{t+\delta t}}{\delta t} - \dot{G}_t)
G_t^{-1} +   (M_{t+\delta   t}  -  M_t)T  T^{-1}\dot{G}_t  G_t^{-1}  +
(\frac{F_{t+\delta t}-F_t}{\delta t} - \dot{F}_t) G_t^{-1}.
\end{array}
$$
Ananlogously to lemmas B.12, B.13, one finds
$$
||(\frac{G^+_{t+\delta t}          -G^+_t}{\delta         t}         -
\dot{G}_t^+)T^{-1}\varphi|| \to_{\delta t \to 0} 0.
$$
Therefore,
$$
||\frac{M_{t+\delta t}-M_t}{\delta t}- \dot{M}_t||_2
\to_{\delta  t  \to 0} 0.
$$
Lemma B.14 is proved.

Therefore, we have proved the following theorem.

{\bf Theorem   B.15.}   {\it  Let  $T$,  $L$  be
self-adjoint operators in $L^2({\bf R}^d)$ such that
$$
\begin{array}{c}
||T^{-1/2}LT^{-1/2}|| <\infty, \qquad
||LT^{-1}|| <\infty, \qquad ||T^{1/2}e^{-iLt}T^{-1/2}|| \le C,
\\
||Te^{-iLt}T^{-1}|| \le C, \qquad t\in [0,t_1].
\end{array}
$$
Let $T-c$ be positively definite for some positive constant $c$,
${\cal   H}^{+-}_t   =   L+{\cal   H}_t$,   ${\cal   H}^{++}$   be
operator-valued functions such that
$||T({\cal H}^{++}_{t+\delta t} - {\cal H}^{++}_t)||_2 \to_{\delta t\to
0} 0$,  ${\cal H}_t$, $T{\cal H}_tT^{-1}$, $T^{1/2}{\cal H}_tT^{-1/2}$
are strongly  continous  operator  functions,  $\overline{H}_t$  be  a
continous function.  Then there exists a unique solution  $\Psi_t$  to
the Cauchy problem \r{bb3},  provided that $\Psi_0 \in {\cal D} \equiv
\{\Psi \in  {\cal  F}  |  ||\Psi||_1^T  <\infty\}$  It  satisfies  the
properties $\Psi_t    \in   {\cal   D}$   and   $||\Psi_t-\Psi_0||_1^T
\to_{\delta \to 0} 0$.
}

\section{Some properties of the Weyl symbol}

The purpose of this appendix is to investigate some properties of Weyl
symbols of operators which are useful in justification  of  properties
H1-H6 of Appendix A.

\subsection{Definition of Weyl symbol}

Firs of  all,  remind  the definition of Weyl symbol of operator (see,
for example,  \c{M1,KM}).  Let $A(x,k)$,  $x,k \in  {\bf  R}^d$  be  a
classical observable  depending on coordinates $x = (x_1,...,x_d)$ and
momenta $k = (k_1,...,k_d)$.  To  specify  the  corresponding  quantum
observable $\hat{A}$ (to "quantize" the observable $A$),
one  should substitute the coordinates $x_i$ by
operators $\hat{x}_i$  of  multiplication by $x_i$,  while the momenta
$k_j$ should  be  substituted  by  the  operators  $\hat{k}_j  =  -  i
\partial/\partial k_j$.  However,  it  is  not  easy  to determine the
operator $A(\hat{x},\hat{k})$ for arbitrary function  $A$,  since  the
coordinate and  momenta  operators  do not commute.  Different ways of
ordering operators $\hat{x}$ and $\hat{k}$  are  known.  In  the  Weyl
approach, one first considers the partial case
$$
A = e^{i\alpha k + i \beta x}
\l{k1}
$$
and sets
$$
\hat{A} = e^{i\alpha \hat{k} + i\beta \hat{x}}
\l{k2}
$$
The operator  \r{k2}  can  be  defined  as a transformation taking the
initial condition $f^0(x)$ for the Cauchy problem for the equation
$$
- i \frac{\partial f^t}{\partial t} = (\alpha \hat{k} + \beta \hat{x})
f^t(x)
\l{k3}
$$
to the solution $f^1(x)$ to the Cauchy problem at $t=1$.  Eq.\r{k3} is
exactly solvable:
$$
f^t(x) =  e^{i\beta x t} e^{\frac{i}{2} \alpha \beta t^2} f^0(x + \alpha
t),
$$
so that
$$
(e^{i\alpha \hat{k}  +  i\beta   \hat{x}}   f)(x)   =   e^{i\beta   x}
e^{\frac{i}{2} \alpha \beta} f(x+\alpha).
\l{k4}
$$
One finds
$$
e^{i\alpha\hat{k}+ i\beta\hat{x}}   =   e^{i\beta\hat{x}}   e^{i\alpha
\hat{k}} e^{\frac{i}{2} \alpha  \beta}.
$$
For an  arbitrary function $A$,  one presents it as a superposition of
exponents \r{k1},
$$
A(x,k) = \int d\alpha d\beta \tilde{A}(\alpha,\beta) e^{i\alpha k +  i
\beta x}
$$
one sets
$$
\hat{A} =   \int  d\alpha  d\beta  \tilde{A}(\alpha,\beta)  e^{i\alpha
\hat{k} +  i \beta \hat{x}}
$$
Applying the formula for inverse Fourier transformation and making use
of formula \r{k4}, we find
$$
(\hat{A} f)     (x)     =     \int     \frac{d\alpha     dp}{(2\pi)^d}
A(x+\frac{\alpha}{2}; p) e^{-i\alpha p} f(x+\alpha).
\l{k5}
$$
We denote the operatoe $\hat{A}$ of the form \r{k5} as $\hat{A}= {\cal
W}(A)$. We will also write $A= \overline{\cal W}(\hat{A})$ if $\hat{A}
= {\cal W}(A)$.

{\bf Definition  C.1.}  {\it The  operator  ${\cal  W}(A)$ is called a Weyl
quantization of  the  function  $A$.  The   function   $\overline{\cal
W}(\hat{A})$ is called as a Weyl symbol of the operator $\hat{A}$.}

\subsection{Some calsses of Weyl symbols}

\subsubsection{Classes ${\cal A}_N$ and ${\cal B}_N$}

For investigations  of QFT ultraviolet divergences,  we are interested
in behavior of Weyl symbols of operators at large values  of  momenta.
Let us    introduce   some   important   spaces.   Let   $\omega_k   =
\sqrt{k^2+m^2}$ for some $m$.

{\bf Definition C.2.} {\it 1.  We say that a smooth function  $A(x,k)$
is of the class ${\cal B}_N$ if and only if the functions
$$
\omega_k^{N+s} \frac{\partial^s   A}{\partial   k^{i_1}  ...  \partial
k^{i_s}}
\l{k6+}
$$
are bounded for all $s$, $i_1,...,i_s$.
\\
2. Let  $A_n  \in  {\cal  B}_N$,  $n=\overline{1,\infty}$,  $A\in {\cal
B}_N$. We say that ${\cal B}_N-\lim_{n\to\infty} A_n = A$ if and  only
if
$$
\lim_{n\to\infty} \max_{k,x}
\omega_k^{N+s} \frac{\partial^s   (A_n - A)}{\partial   k^{i_1}  ...
\partial k^{i_s}} = 0
$$
for all $s$, $i_1,...,i_s$.
\\
3. We  say  that  a  function $A\in {\cal B}_N$ is of the class ${\cal
A}_N$ if and only if
$$
x_{j_1} ...    x_{j_R}    \frac{\partial}{\partial    x_{s_1}}     ...
\frac{\partial}{\partial x_{s_P}} A \in {\cal B}_N
$$
for all $R$, $P$, $j_1,...,j_R$, $s_1,...,s_P$.
\\
4. Let  $A_n \in {\cal A}_N$,  $A \in {\cal A}_N$.  We say that ${\cal
A}_N-lim_{n\to\infty} A_n = A$ if and only if
$$
{\cal B}_N - \lim_{n\to\infty}
x_{j_1} ...    x_{j_R}    \frac{\partial}{\partial    x_{s_1}}     ...
\frac{\partial}{\partial x_{s_P}} (A_n-A) = 0
$$
for all $R$, $P$, $j_1,...,j_R$, $s_1,...,s_P$.
}

Let us  investigate some properties of introduced classes ${\cal A}_N$
and ${\cal B}_N$.

{\bf Lemma C.1.} {\it 1. ${\cal A}_{N+R} \subseteq {\cal A}$,
${\cal B}_{N+R} \subseteq {\cal B}$ for $R\ge 0$. \\
2. Let ${\cal A}_{N+R} - \lim_{n\to\infty} A_n =A$ and $R\ge 0$.
Then ${\cal A}_{N} - \lim_{n\to\infty} A_n =A$.\\
3. Let ${\cal B}_{N+R} - \lim_{n\to\infty} A_n =A$ and $R\ge 0$.
Then ${\cal B}_{N} - \lim_{n\to\infty} A_n =A$.
}

The proof is obvious:  it is sufficient to notice that $\omega_k^{-R}$
is a bounded function.

{\bf Lemma C.2.} {\it
1. Let $A \in {\cal B}_N$.  Then $\frac{\partial}{\partial k_i} A  \in
{\cal B}_{N+1}$.\\
2. Let $A \in {\cal A}_N$. Then $x_iA \in {\cal A}_N$,
$\frac{\partial A}{\partial       x_i}      \in      {\cal      A}_N$,
$\frac{\partial}{\partial k_i} A \in {\cal A}_{N+1}$, $f(x)A \in {\cal
A}_N$ for smooth bounded function $f(x)$. \\
3. Let ${\cal B}_N-\lim_{n\to\infty} A_n = A$. Then
${\cal B}_{N+1}-\lim_{n\to\infty} \frac{\partial}{\partial k_i} A_n =
\frac{\partial}{\partial k_i} A$.\\
4. Let ${\cal A}_N-\lim_{n\to\infty} A_n =A$. Then
${\cal A}_N- \lim_{n\to\infty} x_iA_N = x_i A$,
${\cal A}_N- \lim_{n\to\infty}
\frac{\partial A_n}{\partial       x_i}   =
\frac{\partial A}{\partial       x_i}$,
${\cal A}_{N+1}-\lim_{n\to\infty}
\frac{\partial}{\partial k_i} A_n =
\frac{\partial}{\partial k_i} A$,
${\cal A}_{N}-\lim_{n\to\infty}
f(x)A_n = f(x) A$
for smooth bounded function $f(x)$. \\
}

The proof is also obvious.

{\bf Lemma C.3.} {\it Let $A_1  \in  {\cal  B}_{N_1}$,  $A_2\in  {\cal
B}_{N_2}$. Then $A_1A_2 \in {\cal B}_{N_1+N_2}$. }

{\bf Proof.} It is sufficient to check that the expression
$$
\omega_k^{N_1} \omega_k^{N_2} \omega_k^s \frac{\partial^s}
{\partial k_{i_1} ... \partial k_{i_s} } (A_1A_2)
$$
is bounded. This statement is a corollary of properties $A_1 \in {\cal
B}_{N_1}$, $A_2 \in {\cal B}_{N_2}$ and formula
$$
\frac{\partial}{\partial k_i}  (fg)  =  \frac{\partial}{\partial  k_i}
f\cdot  g + f \cdot \frac{\partial}{\partial k_i} g.
$$
Lemma C.3 is proved.

{\bf Lemma C.4.} {\it The following properties are satisfied:
$k_i \in {\cal B}_{-1}$, $\omega_k^{\alpha} \in {\cal B}_{-\alpha}$.
}

{\bf Proof.} Since $|k_i/\omega_k| <1$,  we obtain
the property    $k_i    \in   {\cal   B}_{-1}$.   For   the   function
$\omega_k^{\alpha}$, one has
$$
\frac{\partial}{\partial k_{i_1}}
...
\frac{\partial}{\partial k_{i_s}}         \omega_k^{\alpha}          =
\omega_k^{\alpha} {\cal P}(k_i/\omega_k)
\l{k7}
$$
where $\cal P$ is a polynomial in $k_i/\omega_k$.  Property \r{k7}  is
checked by  induction.  Therefore,  functions  \r{k6+} are bounded for
$N=1$. Lemma C.4 is proved.

{\bf Lemma C.5.} {\it
1. Let $A \in {\cal B}_N$. Then
$$
k_iA \in   {\cal  B}_{N-1},  \qquad  \omega_k^{-\alpha}  A  \in  {\cal
B}_{N+\alpha}, \qquad  \frac{\partial  A}{\partial  k_i}   \in   {\cal
B}_{N+1}.
$$
2. Let $A \in {\cal A}_N$. Then
$$
k_iA \in   {\cal  A}_{N-1},  \qquad  \omega_k^{-\alpha}  A  \in  {\cal
A}_{N+\alpha}, \qquad  \frac{\partial  A}{\partial  k_i}   \in   {\cal
A}_{N+1}.
$$
}

{\bf Proof.} Property 1 is a corollary of lemmas C.2 and C.4. Property
1 implies property 2. Lemma is proved.

{\bf Lemma C.6.} {\it
1. Let ${\cal B}_N-\lim_{n\to\infty} A_n = A$. Then
$$
{\cal  B}_{N-1}-\lim_{n\to\infty} k_iA_n = k_i A, \qquad
{\cal B}_{N+\alpha}-\lim_{n\to\infty}
\omega_k^{-\alpha}  A_n = \omega_k^{-\alpha}  A,
\qquad
{\cal B}_{N+1}- \lim_{n\to\infty}
\frac{\partial  A_n}{\partial  k_i}
= \frac{\partial  A}{\partial  k_i}
$$
2. Let ${\cal A}_N-\lim_{n\to\infty} A_n = A$. Then
$$
{\cal  A}_{N-1}-\lim_{n\to\infty} k_iA_n = k_i A, \qquad
{\cal A}_{N+\alpha}-\lim_{n\to\infty}
\omega_k^{-\alpha}  A_n = \omega_k^{-\alpha}  A,
\qquad
{\cal A}_{N+1}- \lim_{n\to\infty}
\frac{\partial  A_n}{\partial  k_i}
= \frac{\partial  A}{\partial  k_i}.
$$
}

The proof is analogous to the proof of lemma C.3.

{\bf Lemma C.7.} {\it
1. Let $A_1 \in {\cal A}_{N_1}$, $A_2 \in {\cal A}_{N_2}$. Then
$A_1A_2 \in {\cal A}_{N_1+N_2}$.\\
2. Let
${\cal A}_{N_1}- \lim_{n\to\infty} A_{1,n} = A_1$,
${\cal A}_{N_2}- \lim_{n\to\infty} A_{2,n} = A_2$.
Then
${\cal A}_{N_1 + N_2}- \lim_{n\to\infty} A_{1,n} A_{2,n} = A_1 A_2$.
}

The proof is analogous to lemma C.3.

\subsubsection{Properties of operators and symbols}

{\bf Lemma C.8.} {\it
1. Let $A\in {\cal A}_0$.  Then the operator ${\cal W}(A)$  \r{k5}  is
bounded.\\
2. Let ${\cal A}_0-\lim_{n\to\infty} A_n =0$.  Then $\lim_{n\to\infty}
||{\cal W}(A_n)|| = 0$.
}

{\bf Proof} (cf.  \c{MS3}).  Let us obtain an estimation for the  norm
$||\hat{A}||$. One has
$$
\hat{A} =  \int  d\beta  e^{i\beta  \hat{x}}  \int  d\alpha e^{i\alpha
(\hat{k}+ \beta/2)} \tilde{A}(\alpha,\beta).
$$
The estimation  $||\int  d\beta  \hat{F}(\beta)||  \le   \int   d\beta
||\hat{F}(\beta)||$ implies
$$
||\hat{A}|| \le   \int  d\beta  ||\int  d\alpha  e^{i\alpha(\hat{k}  +
\beta/2)} \tilde{A}(\alpha,\beta).
$$
However, for operator $F(\hat{k})$ one has  $||F(\hat{k})||  =  \sup_k
||F(k)||$, since  in  the  momentum representation $F(\hat{k})$ is the
operator of multiplication be $F(k)$. Therefore,
$$
\begin{array}{c}
||\int d\alpha  e^{i\alpha  (\hat{k}+\beta/2)} \tilde{A}(\alpha,\beta)
|| =     \max_k     |\int     d\alpha     e^{i\alpha      (k+\beta/2)}
\tilde{A}(\alpha,\beta)| =   \max_k   |\int   d\alpha   e^{i\alpha  k}
\tilde{A}(\alpha,\beta)| = \\
\max_k |  \int  \frac{dx}{(2\pi)^d}  A(x,k)
e^{-i\beta x}|      =     \frac{1}{(\beta^2+1)^N}     \max_k     |\int
\frac{dx}{(2\pi)^d (x^2+1)^N} e^{-i\beta  x}  (x^2+1)^N  (-\Delta_x  +
1)^N A(x,k)|.
\end{array}
$$
Here $N$ is an arbitrary number such that $N> d/2$. Thus,
$$
||\hat{A}|| \le \frac{1}{(2\pi)^d} \int \frac{d\beta dx}{(\beta^2+1)^N
(x^2+1)^N} \max_{kx} |(x^2+1)^N (-\Delta_x+1)^N A(x,k)|.
$$
The first  statement  is  judtified.  Proof of the second statement is
analogous. Lemma C.8 is proved.

{\bf Lemma C.9.}{\it
1. Let   $A\in   {\cal   A}_N$,  $N>d/2$.  Then  ${\cal  W}(A)$  is  a
Hilbert-Schmidt operator.\\
2. Let   ${\cal   A}_N-\lim_{n\to\infty}   A_n  =  0$,  $N>d/2$.  Then
$\lim_{n\to\infty} ||{\cal W}(A_n)||_2 = 0$.
}

{\bf Proof.} Let us use the property \c{M1,KM}
$$
||\hat{A}||_2^2 = \int \frac{dx dk}{(2\pi)^d} |A(x,k)|^2
$$
whcih can be obtained from definition \r{k5}. One has
$$
||\hat{A}||_2^2 \le       \int      \frac{dx}{(2\pi)^d      (x^2+1)^N}
\frac{dk}{\omega_k^{2N}} \max_{xk} |(x^2+1)^{N/2} \omega_k^N A(x,k)|^2.
$$
The first statement is justified.  Proof of the  second  statement  is
analogous. Lemma C.9 is proved.

\subsection{Properties of *-product}

Remind that the Weyl sumbol of the product of operators
$$
A*B = \overline{\cal W} ({\cal W}(A) {\cal W}(B))
$$
can be presented as \c{M1,KM}
$$
(A*B)(x,k) = \int \frac{d\beta_1 d\beta_2 d\xi_1 d\xi_2}{(2\pi)^{2d}}
A(x+\xi_1, k + \frac{\beta_2}{2}) B(x+\xi_2,  k  -  \frac{\beta_1}{2})
e^{-i\beta_1\xi_1 - i\beta_2\xi_2}
\l{k8}
$$
Formula \r{k8} can be obtained from definition \r{k5}.

Let us investigate some properties of formula \r{k8}.  Let us find  an
expansion of formula \r{k8} as $|k| \to\infty$. Formally, one has
$$
\begin{array}{c}
A(x+\xi_1, k    +    \frac{\beta_2}{2})    =     \sum_{n_2=0}^{\infty}
\frac{1}{2^{n_2} n_2!}  \frac{\partial^{n_2} A(x+\xi_1,  k) }{\partial
k^{i_1} ... \partial k^{i_{n_2} }} \beta_2^{i_1} ... \beta_2^{i_{n_2}};
\\
B(x+\xi_2, k    -  \frac{\beta_1}{2})    =     \sum_{n_1 =0}^{\infty}
\frac{(-1)^{n_1}}
{2^{n_1} n_1!}  \frac{\partial^{n_1} B(x+\xi_2,  k) }{\partial
k^{j_1} ... \partial k^{j_{n_1} }} \beta_1^{j_1} ... \beta_1^{j_{n_1}}.
\end{array}
$$
Therefore,
$$
\begin{array}{c}
(A*B)(x,k) =   \sum_{n_1n_2=0}^{\infty}  \frac{(-1)^{n_1}}{2^{n_1+n_2}
n_1! n_2!} \int
\frac{d\beta_1 d\beta_2 d\xi_1 d\xi_2}{(2\pi)^{2d}}
e^{-i\beta_1\xi_1 - i\beta_2\xi_2}
\frac{\partial^{n_2} A(x+\xi_1,  k) }{\partial
k^{i_1} ... \partial k^{i_{n_2} }} \beta_2^{i_1} ... \beta_2^{i_{n_2}}
\\ \times
 \frac{\partial^{n_1} B(x+\xi_2,  k) }{\partial
k^{j_1} ... \partial k^{j_{n_1} }} \beta_1^{j_1} ... \beta_1^{j_{n_1}}
=
\sum_{n_1n_2 = 0}^{\infty} \frac{i^{n_1-n_2}}{2^{n_1+n_2} n_1! n_2!}
\frac{\partial^{n_1 + n_2} A(x,  k) }{\partial
k^{i_1} ... \partial k^{i_{n_2} }
\partial x^{j_1} ... \partial x^{j_{n_1}} }
\frac{\partial^{n_1 + n_2} B(x,  k) }{\partial
x^{i_1} ... \partial x^{i_{n_2} }
\partial k^{j_1} ... \partial k^{j_{n_1}} }
\end{array}
\l{k8a}
$$
Denote
$$
(A \stackrel{K}{*} B)(x,k) =
\sum_{n_1n_2 \ge 0, n_1+n_2 \le K}
\frac{i^{n_1-n_2}}{2^{n_1+n_2} n_1! n_2!}
\frac{\partial^{n_1 + n_2} A(x,  k) }{\partial
k^{i_1} ... \partial k^{i_{n_2} }
\partial x^{j_1} ... \partial x^{j_{n_1}} }
\frac{\partial^{n_1 + n_2} B(x,  k) }{\partial
x^{i_1} ... \partial x^{i_{n_2} }
\partial k^{j_1} ... \partial k^{j_{n_1}} }
$$
This is an asymptotic expansion in $1/|k|$ as $|k|\to \infty$.  Let us
estimate an accuracy of the asymptotic series.

Making use of the relation
$$
A(x+\xi_1, k   +   \frac{\beta_2}{2})   -   A(x+\xi_1,k)   =  \int_0^1
d(\alpha_2-1) \frac{\partial}{\partial \alpha_2} A(x+\xi_1, k+\alpha_2
\frac{\beta_2}{2})
$$
and integrating by parts $N_2$ times, we find
$$
\begin{array}{c}
A(x+\xi_1, k     +    \frac{\beta_2}{2})    =    \sum_{n_2=0}^{N_2}
\frac{1}{2^{n_2} n_2!} \frac{\partial^{n_2} A(x+\xi_1,k)  }{  \partial
k^{i_1} ... \partial k^{i_{n_2}} } \beta_2^{i_1} ... \beta_2^{i_{n_2}}
\\ + \int_0^1 d\alpha_2 \frac{(1-\alpha_2)^{N_2}}{2^{N_2+1} N_2!}
\frac{\partial^{N_2+1} A(x+\xi_1, k + \alpha_2 \frac{\beta_2}{2})}
{\partial k^{i_1}  ...  \partial  k^{i_{N_2+1}}  }  \beta_2^{i_1}  ...
\beta_2^{i_{N_2+1}}.
\end{array}
$$
Analogously,
$$
\begin{array}{c}
B(x+\xi_2, k     -    \frac{\beta_1}{2})    =    \sum_{n_1=0}^{N_1}
\frac{(-1)^{n_1}}{2^{n_1} n_1!}
\frac{\partial^{n_1} A(x+\xi_2,k)  }{  \partial
k^{i_1} ... \partial k^{i_{n_1}} } \beta_1^{i_1} ... \beta_1^{i_{n_2}}
\\ + \int_0^1 d\alpha_1 \frac{(1-\alpha_1)^{N_1} (-1)^{N_1}}
{2^{N_1+1} N_1!}
\frac{\partial^{N_1+1} B(x+\xi_2, k - \alpha_1 \frac{\beta_1}{2})}
{\partial k^{i_1}  ...  \partial  k^{i_{N_1+1}}  }  \beta_1^{i_1}  ...
\beta_1^{i_{N_1+1}}.
\end{array}
$$
Therefore,
$$
\begin{array}{c}
(A*B)(x,k) =
\sum_{n_1= 0}^{N_1}
\sum_{n_2= 0}^{N_2}
\frac{i^{n_1-n_2}}{2^{n_1+n_2} n_1! n_2!}
\frac{\partial^{n_1 + n_2} A(x,  k) }{\partial
k^{i_1} ... \partial k^{i_{n_2} }
\partial x^{j_1} ... \partial x^{j_{n_1}} }
\frac{\partial^{n_1 + n_2} B(x,  k) }{\partial
x^{i_1} ... \partial x^{i_{n_2} }
\partial k^{j_1} ... \partial k^{j_{n_1}} }
\\ + \sum_{n_1   =   0}^{N_1}  r^{(1)}_{n_1N_2}  +  \sum_{n_2=0}^{\infty}
r^{(2)}_{N_1n_2} + R_{N_1N_2}
\end{array}
$$
with the following remaining terms,
$$
\begin{array}{c}
r^{(1)}_{n_1N_2} =     \int     \frac{d\beta_1     d\beta_2     d\xi_1
d\xi_2}{(2\pi)^{2d}} e^{-i\beta_1 \xi_1 - i \beta_2 \xi_2}
\int_0^1 d\alpha_2 \frac{(1-\alpha_2)^{N_2}}{2^{N_2+1} N_2!}
\frac{\partial^{N_2+1} A(x+\xi_1, k + \alpha_2 \frac{\beta_2}{2})}
{\partial k^{i_1}  ...  \partial  k^{i_{N_2+1}}  }  \beta_2^{i_1}  ...
\beta_2^{i_{N_2+1}} \\ \times  \frac{(-1)^{n_1} }{2^{n_1} n_1! }
\frac{\partial^{n_1} B(x+\xi_1,   k)}{\partial  k^{j_1}  ...  \partial
k^{j_{n_1}} } \beta_1^{j_1} ... \beta_1^{j_{n_1}};
\end{array}
$$
$$
\begin{array}{c}
r^{(2)}_{N_1n_2} =     \int     \frac{d\beta_1     d\beta_2     d\xi_1
d\xi_2}{(2\pi)^{2d}} e^{-i\beta_1 \xi_1 - i \beta_2 \xi_2}
\int_0^1 d\alpha_1 \frac{(1-\alpha_1)^{N_1} (-1)^{N_1+1}}
{2^{N_1+1} N_1!}
\frac{\partial^{N_1+1} B(x+\xi_2, k - \alpha_1 \frac{\beta_1}{2})}
{\partial k^{j_1}  ...  \partial  k^{j_{N_1+1}}  }  \beta_1^{j_1}  ...
\beta_1^{j_{N_1+1}} \\ \times  \frac{1}{2^{n_2} n_2! }
\frac{\partial^{n_2} B(x+\xi_2,   k)}{\partial  k^{i_1}  ...  \partial
k^{i_{n_2}} } \beta_2^{i_1} ... \beta_2^{i_{n_2}};
\end{array}
$$
$$
\begin{array}{c}
R_{N_1N_2} =     \int     \frac{d\beta_1     d\beta_2     d\xi_1
d\xi_2}{(2\pi)^{2d}} e^{-i\beta_1 \xi_1 - i \beta_2 \xi_2}
\int_0^1 d\alpha_2 \frac{(1-\alpha_2)^{N_2}}{2^{N_2+1} N_2!}
\frac{\partial^{N_2+1} A(x+\xi_1, k + \alpha_2 \frac{\beta_2}{2})}
{\partial k^{i_1}  ...  \partial  k^{i_{N_2+1}}  }  \beta_2^{i_1}  ...
\beta_2^{i_{N_2+1}}
\\ \times \int_0^1 d\alpha_1 \frac{(1-\alpha_2)^{N_2} (-1)^{N_1+1}}
{2^{N_1+1} N_1!}
\frac{\partial^{N_1+1} B(x+\xi_2, k - \alpha_1 \frac{\beta_1}{2})}
{\partial k^{j_1}  ...  \partial  k^{j_{N_1+1}}  }  \beta_1^{j_1}  ...
\beta_1^{j_{N_1+1}}.
\end{array}
$$

Let us investigate the remaining terms.

\subsubsection{The $k$-independent case}

{\bf Definition C.3.} {\it We say that the function $f(x)$, $x\in {\bf
R}^d$ is  of  the class $\cal C$ if $f$ is a smooth function such that
for each set $(i_1,...,i_l)$ there exists $m>0$ such that the function
$$
(x^2+1)^{-m} \frac{\partial^l}{\partial x^{i_1} ... \partial x^{i_l}}f
$$
is bounded.
}

Let $A=f(x)$,$f  \in  {\cal  C}$.  Then  the  only  nontrivial term is
$r^{(2)}_{N_10}$ which is taken by integrating by parts to the form
$$
r^{(2)}_{N_10} =      \int      \frac{d\beta_1       d\xi_1}{(2\pi)^d}
e^{-i\beta_1\xi_1} \frac{\partial^{N_1+1}}{\partial     x^{j_1}    ...
\partial x^{j_{N_1+1}} } f(x+\xi_1)
\int_0^1 d\alpha_1 (\frac{i}{2})^{N_1+1} \frac{(1-\alpha_1)^{N_1}}{N_1!}
\frac{\partial^{N_1+1} B(x, k - \alpha_1 \frac{\beta_1}{2})}
{\partial k^{j_1} ... \partial k^{j_{N_1+1}} }.
\l{k9}
$$
Let us prove some auxiliary statements.

{\bf Lemma C.10.} {\it For some constent $A_1$ the estimation
$$
\omega_k \le A_1 \omega_p \omega_{k-p}
\l{k10}
$$
is satisfied.
}

{\bf Proof.} Let $p=(\frac{1}{2} + \alpha) k + p_{\perp} $, $\alpha\in
{\bf R}$, $p_{\perp} \perp k$. Then
$$
\frac{\omega_k}{\omega_p \omega_{k-p} } \le \frac{\omega_k}
{\omega_{(1/2+\alpha)k} \omega_{(1/2-\alpha)k}} \equiv f(\alpha,k),
$$
so that it is sufficient to check estimation \r{k10} for $p=\alpha  k$
only. For the function $1/f^2$, one has
$$
\frac{1}{f^2(\alpha,k)} =  \frac{1}{k^2+m^2}
[(\frac{1}{2} + \alpha)^2 k^2 + m^2]
[(\frac{1}{2} - \alpha)^2 k^2 + m^2].
$$
It has the following minimal value
$$
min_{\alpha}
\frac{1}{f^2(\alpha,k)} =
\left\{
\begin{array}{c}
\frac{(k^2/4+m^2)^2}{k^2+m^2}, \qquad k^2<4m^2, \qquad \alpha=0. \\
\frac{k^2m^2}{k^2+m^2}, \qquad      k^2      >      4m^2,       \qquad
\alpha=\sqrt{\frac{1}{4}- \frac{m^2}{k^2}}.
\end{array}
\right.
\l{k10*}
$$
The quantity \r{k10*} is bounded below. Thus, lemma is proved.

{\bf Corollary.} {\it For $0<\gamma <1$,
$$
\frac{\omega_k}{\omega_p \omega_{k-\gamma p}} \le A_1.
$$
}

{\bf Lemma C.11.} {\it Let $C\in {\cal A}_N$,  $\chi  \in  {\cal  C}$,
$\varphi \in C[0,1]$. Then for
$$
F(x,k) =    \int_0^1   d\alpha   \varphi(\alpha)   \int   \frac{d\beta
d\xi}{(2\pi)^d} e^{-i\beta  \xi}  \chi(x+\xi)  C(x,  k-   \frac{\alpha
\beta}{2})
\l{k11}
$$
the function $\omega_k^{N}F$ is bounded.
}

{\bf Proof.}  Inserting the identity
$$
e^{-i\beta \xi}  =  (\xi^2  +1)^{-L_1}  (-  \frac{\partial^2}{\partial
\beta^2} + 1)^{L_1} e^{-i\beta \xi}
\l{k11*}
$$
and integrating by parts, we obtain that
$$
F(x,k) =    \int_0^1   d\alpha   \varphi(\alpha)   \int   \frac{d\beta
d\xi}{(2\pi)^d} \frac{1}{(\xi^2 + 1)^{L_1}} e^{-i\beta\xi} \chi(x+\xi)
\left( 1    -   \frac{\alpha^2}{4}   \frac{\partial^2}{\partial   k^2}
\right)^{L_1} C(x; k - \frac{\alpha \beta}{2}).
$$
For the function $\omega_k^NF$, one has
$$
\begin{array}{c}
\omega_k^N F =  \int_0^1  d\alpha  \varphi(\alpha)  \int  \frac{d\beta
d\xi}{(2\pi)^d} \frac{1}{(\xi^2+1)^{L_1}        \omega_{beta/2}^{L_2}}
\chi(x+\xi) \left( - \frac{1}{4} \frac{\partial^2}{\partial  \xi^2}  +
m^2 \right)^{\frac{L_2+N}{2}}                           e^{-i\beta\xi}
\frac{\omega_k^N}{\omega_{\beta/2}^N} \\ \times
\left( 1    -   \frac{\alpha^2}{4}   \frac{\partial^2}{\partial   k^2}
\right)^{L_1} C(x; k - \frac{\alpha \beta}{2}).
\end{array}
\l{k12}
$$
Choose $L_2$  to  be  such a number that $\frac{L_2+N}{2}$ is integer,
$L_2>d$. The property $\chi \in {\cal C}$ implies  that  there  exists
such $K$ that
$$
\frac{\partial^m}{\partial \xi_{i_1}    ...   \partial   \xi_{i_n}   }
\chi(x+\xi) = ((x+ \xi)^2 + 1)^K f_{m. i_1...i_m} (x+\xi), \qquad
m = \overline{0, \frac{L_2+N}{2}},
$$
where $f_{m,  i_1...  i_m}$ are bounded functions.  Choose $L_1$ to be
integer and $L_1 > \frac{K+d}{2}$.  Integrating expression \r{k12}  by
parts, making  use of corollary of lemma C.10 and property $C\in {\cal
A}_N$, we obtain that $\omega_k^NF$ is a bounded function.  Lemma C.11
is proved.

{\bf Lemma  C.12.}  {\it  Under  conditions of lemma C.11 $F \in {\cal
A}_N$.}

{\bf Proof.} It is sufficient to consider the functions
$$
\omega_k^{N+I} \frac{\partial^I}{\partial   k_{i_1}    ...    \partial
k_{i_I}} x_{j_1}  ...  x_{j_R}  \frac{\partial}{\partial  x_{s_1}} ...
\frac{\partial}{\partial x_{s_P}}F
\l{k12*}
$$
which are  expressed  via linear combinations of integrals of the type
\r{k11}. Lemma C.12 is a corollary of lemma C.11.

{\bf Lemma C.13.} {\it
Let ${\cal  A}_N-\lim_{n\to\infty}  C_n  =C$,  $\chi  \in  {\cal  C}$,
$\varphi \in C[0.1]$. Then ${\cal  A}_N-\lim_{n\to\infty}  F_n  =F$.
}

The proof is analogous to lemmas C.11 and C.12.

We obtain therefore the following theorem.

{\bf Theorem C.14.}
{\it
1. Let $f \in {\cal C}$, $B \in {\cal A}_N$. Then
$$
f * B = f \stackrel{K}{*} B + {R}_K
$$
with $R_K \in {\cal A}_{N+K+1}$.\\
2. Let $f\in {\cal C}$, ${\cal A}_N - \lim_{n\to\infty} B_n = 0$. Then
${\cal A}_{N+K+1}
- \lim_{n\to\infty} (f * B_n - f \stackrel{K}{*}B_n) = 0$.
}

\subsubsection{The $x$-independent case}

Let $A=A(k)$,  $A\in {\cal B}_{M_1}$, $B \in {\cal A}_{M_2}$. The only
nontrivial term is taken to the form:
$$
r^{(1)}_{0N_2}(x,k) =                \int_0^1                d\alpha_2
\left(-\frac{i}{2}\right)^{N_2+1} \frac{(1-\alpha_2)^{N_2}     }{N_2!}
\int \frac{d\beta_2 d\xi_2}{(2\pi)^d} e^{-i\beta_2\xi_2}
\frac{\partial^{N_2+1} A(k  +  \frac{\alpha_2  \beta_2}{2}) }{\partial
k^{i_1} ... \partial k^{i_{N_2+1}} } \frac{\partial^{N_2+1} B(x+\xi_2;
k)}{\partial x^{i_1} ... \partial x^{i_{N_2+1}} }.
$$

{\bf Lemma C.15.} {\it $C=C(k)$,  $C \in {\cal  B}_{K_1}$,  $K_1  >0$,
$D\in {\cal A}_{K_2}$, $\varphi \in C[0,1]$. Then for
$$
F(x,k) =    \int_0^1   d\alpha   \varphi(\alpha)   \int   \frac{d\beta
d\xi}{(2\pi)^d} e^{-i\beta \xi} C(k+ \frac{\alpha \beta}{2})  D(x+\xi,
k) \xi_{j_1} ... \xi_{j_m}
$$
the function $\omega_k^{K_1+K_2}F$ is bounded.
}

{\bf Proof.} Inserting the identity \r{k11*} and integrating by parts,
we obtain that
$$
\begin{array}{c}
F(x,k) =    \int_0^1   d\alpha   \varphi(\alpha)   \int   \frac{d\beta
d\xi}{(2\pi)^d} \frac{1}{(\xi^2+1)^{L_1}}  e^{-i\beta\xi}   D(x+\xi,k)
\left( 1    -   \frac{\alpha^2}{4}   \frac{\partial^2}{\partial   k^2}
\right)^{L_1}
\\ \times
(-\frac{i\alpha}{2})^m \frac{\partial}{\partial k_{j_1}}
... \frac{\partial}{\partial k_{j_m}} C(k + \frac{\alpha \beta}{2}).
\end{array}
$$
For the function $\omega_k^{K_1+K_2}F$, one has
$$
\begin{array}{c}
\omega_k^{K_1+K_2} F(x,k)  =  \int_0^1  d\alpha  \varphi(\alpha)  \int
\frac{d\beta d\xi}{(2\pi)^d}                  \frac{1}{(\xi^2+1)^{L_1}
\omega_{\beta/2}^{L_2} } \omega_k^{K_2} D(x+\xi,k)  \left(
- \frac{1}{4} \frac{\partial^2}{\partial \xi^2} + m^2
\right)^{\frac{L_2+K_1}{2}}
\\
e^{-i\beta\xi}
\frac{\omega_k^{K_1}}{\omega_{\beta/2}^{K_1} }
\left( 1- \frac{\alpha^2}{4} \frac{\partial^2}{\partial k^2}
\right)^{L_1} (-i\frac{\alpha}{2})^m        \frac{\partial^m}{\partial
k_{j_1} ... \partial k_{j_m}} C(k + \frac{\alpha \beta}{2})
\end{array}
$$
Integrating by parts for sufficiently large $L_1$,  $L_2$,  making use
of lemmas C.10, we check proposition of lemma C.15.

{\bf Lemma C.16.} {\it Under conditions of lemma  C.15  $F  \in  {\cal
A}_{K_1+K_2}$. }

{\bf Lemma C.17.} {\it Let ${\cal A}_{K_2}-\lim_{n\to\infty} D_n = D$,
$C=C(k)$, $C\in {\cal B}_{K_1}$, $K_1>0$, $\varphi \in C[0,1]$.
Then
${\cal A}_{K_1+K_2}-\lim_{n\to\infty} F_n = F$.
}

The proof  is  analogous  to lemmas C.12 and C.13.  We obtain then the
following theorem.

{\bf Theorem C.18.} {\it
1. Let $A=A(k)$, $A \in {\cal B}_{M_1}$, $B \in {\cal A}_{M_2}$. Then
$$
A*B = A \stackrel{K}{*} B + R_K
$$
with $R_K \in {\cal A}_{M_1+M_2+K+1}$, provided that $K+M_1+1>0$.\\
2. Let $A=A(k)$, $A\in {\cal B}_{M_1}$,
${\cal A}_{M_2}-\lim_{n\to\infty} B_n = B$. Then
$$
{\cal A}_{M_1+M_2+K+1}-\lim_{n\to\infty}
(A*B_n - A \stackrel{K}{*} B_n) =0,
$$
provided that $K+M_1+1>0$.
}

{\bf Remark.}  If  the  proposition  of  theorem C.18 is satisfied for
$K=K_0$, it is satisfied for all $K\le K_0$.  Therefore, the condition
$K+M_1+1>0$ can be omitted.

The following lemma is a corollary of theorem C.18.

{\bf Lemma C.19.}{\it
1. Let $A\in {\cal A}_N$,  $N>d$.  Then ${\cal W}(A)$ is of the  trace
class.\\
2. Let ${\cal A}_N-\lim_{n\to\infty} A_n=0$,  $N>d$.  Then
$\lim_{n\to\infty} Tr  {\cal W}(A_n) = 0$.
}

{\bf Proof.} Consider the operator
$$
\hat{B} = {\cal W}(B) = \hat{\omega}^{N/2} (x^2+1)^{N/2} {\cal W}(A)
$$
with
$$
B = \omega_k^{N/2} * (x^2+1)^{N/2} * A
$$
Since $B \in {\cal  A}_{N/2}$,  ${\cal  W}(B)$  is  a  Hilbert-Schmidt
operator according to lemma C.9. Therefore, ${\cal W}(A)$ is a product
of two Hilbert-Schmidt operators $(x^2+1)^{-N/2}  \hat{\omega}^{-N/2}$
and ${\cal W}(B)$. Thus, ${\cal W}(A)$ is of the trace class.

One also has:
$$
|Tr {\cal  W}(A_n)|  =  |Tr  (x^2+1)^{-N/2}  \hat{\omega}^{-N/2} {\cal
W}(B_n)| \le    ||(x^2+1)^{-N/2}    \hat{\omega}^{-N/2}||_2    ||{\cal
W}(B_n)||_2.
$$
Making use of lemma C.9, we prove lemma C.19.

\subsubsection{The ${\cal A}_N$-case}

Let $A \in {\cal A}_{M_1}$,  $B \in {\cal A}_{M_2}$. The $r$-terms can
be investigated as follows.

1. We substitute $\beta_{1,2}^j e^{-i\beta_{1,2}\xi_{1.2}} \equiv
i \frac{\partial}{\partial \xi_{1,2}^j} e^{-i\beta_{1,2}\xi_{1.2}}$
and integrate the expressions for $r^{(1)}$,  $r^{(2)}$,  $R$ by parts
with respect to $\xi_1$, $\xi_2$.

2. We consider the quantities like
$$
\omega_k^{N_1+N_2+M_1+ M_2+1  + L} \frac{\partial^L}{ \partial k^{i_1}
... \partial k^{i_L}}  x_{j_1}  ...  x_{j_J}
\frac{\partial}{\partial x_{s_1}} ...
\frac{\partial}{\partial x_{s_P}} r
$$
for $r =r^{(1)},  r^{(2)},  R$ and show them to be bounded. We use the
following statement.

{\bf Lemma  C.20.}  {\it  Let  $F  \in  {\cal A}_{K_1}$,  $G \in {\cal
A}_{K_2}$, $K_1,K_2>0$. Then the function
$$
\int \frac{d\beta_1     d\beta_2      d\xi_1      d\xi_2}{(2\pi)^{2d}}
e^{-i\beta_1\xi_1 -  i\beta_2 \xi_2} \omega_k^{K_1+K_2} F(x+\xi_1,  k+
\alpha_2 \frac{\beta_2}{2}) G(x+\xi_2, k - \alpha_1 \frac{\beta_1}{2})
\xi_1^{j_1} ... \xi_1^{j_m}
$$
is uniformly bounded with respect to $\alpha_1, \alpha_2 \in [0,1]$.
}

This lemma is proved analogously to lemmas C.11 and C.15.

3. Analogously to previous subsubsections,
we prove the following theorem.

{\bf Theorem C.21.}
{\it
1. Let $A \in {\cal A}_{M_1}$, $B \in {\cal A}_{M_2}$. Then
$$
A*B = A \stackrel{K}{*} B + R_K
$$
with $R_K \in {\cal A}_{M_1+M_2+K+1}$.\\
2. Let $A_n \in {\cal A}_{M_1}$, $B_n \in {\cal A}_{M_2}$. Then
$$
{\cal A}_{M_1+M_2+K+1}-\lim_{n\to\infty}
(A_n*B_n      -      A_n \stackrel{K}{*} B_n)
=
A*B - A \stackrel{K}{*} B.
$$
}

\subsection{Properties of the exponent}

Let us investigate now the properties of the exponent of the  operator
$\exp {\cal  W}(A)  \equiv  {\cal  W}(*\exp  A)$.  It is convenient to
consider the Fourier transformations of Weyl symbols,
$$
\tilde{A}(\gamma,k) = \int \frac{dx}{(2\pi)^d} e^{-i\gamma x} A(x,k).
$$
Introduce the following norms for Weyl symbols,
$$
||A||_{I,K} =  \max_{J+M+N  \le  K}  \max_{\gamma,  K} |\omega_k^{I+J}
\frac{\partial^J}{\partial k_{j_1} ...  \partial k_{j_J}} \gamma_{m_1}
... \gamma_{m_M} \frac{\partial^N \tilde{A}}{\partial \gamma_{n_1} ...
\partial \gamma_{n_N} }|.
\l{k19}
$$

{\bf Lemma C.22.} {\it
$A\in {\cal  A}_I$  if  and only if $||A||_{I,K} < \infty$ for all $k=
\overline{0,\infty}$.
}

The proof is obvious.

Let $C=A*B$.  Then  the  Fourier  transformation  $\tilde{C}$  can  be
expressed via $\tilde{A}$ and $\tilde{B}$ as follows,
$$
\tilde{C}(\gamma,k) =   \int    d\alpha    \tilde{A}(\alpha,    k    +
\frac{\gamma-\alpha}{2}) \tilde{B}(\gamma-\alpha,          k         -
\frac{\alpha}{2}).
\l{k20}
$$
The following estimation is satisfied.

{\bf Lemma  C.23.} {\it
For  arbitrary integer numbers $K$,  $L>d/2$ there
exists such a constant $b_K$ that
$$
||A*B||_{0,K} \le b_K ||A||_{0, K+2L} ||B||_{0,K}.
\l{k21}
$$
}

To prove estimation \r{k21},  one should use  definition  \r{k19}  and
formula \r{k20}:\\
(i) the derivatives $\partial/\partial \gamma_n$ are applied as
$$
\begin{array}{c}
\frac{\partial}{\partial \gamma_n}     (\tilde{A}(\alpha,     k      +
\frac{\gamma-\alpha}{2}) \tilde{B}(\gamma-\alpha,          k         -
\frac{\alpha}{2})) =  \frac{1}{2}  \frac{\partial  \tilde{A}}{\partial
k_n}(\alpha, k + \\
\frac{\gamma-\alpha}{2}) \tilde{B}(\gamma-\alpha,  k-
\frac{\alpha}{2}) + \tilde{A}(\alpha, k + \frac{\gamma-\alpha}{2})
\frac{\partial}{\partial \gamma_n}    \tilde{B}(\gamma-\alpha,   k   -
\frac{\alpha}{2});
\end{array}
$$
(ii) the   derivatives   $\partial/\partial    k_j$    are    applied
analogously;\\
(iii) the  multiplicators  $\gamma_m$  are  written  as   $\alpha_m   +
(\gamma_m - \alpha_m)$;\\
(iv) the estimations
$$
\omega_k \le C \omega_{\alpha/2} \omega_{k-\alpha/2},
\qquad
\omega_k \le C \omega_{\frac{\gamma - \alpha}{2}}
\omega_{k+ \frac{\gamma - \alpha}{2}}
$$
(lemma C.10) are taken into account. \\
(v) the integrating measure is written as
$$
d\alpha = \frac{d\alpha}{(\alpha^2+1)^L} (\alpha^2+1)^L.
$$
We obtain the estimation \r{k21}.

Consider the Weyl symbol of the exponent
$$
*\exp At -1 = \sum_{n=1}^{\infty} \frac{A^{*n} t^n}{n!}
\l{k22}
$$
with $A^{*n} = A* ... *A$.

{\bf Lemma C.24.} {\it  Let  $A  \in  {\cal  A}_M$,$M>0$.  Then  the
estimation \r{k22}  is  convergent in the $||\cdot||_{0,K}$-norm.  The
estimation $||*\exp At -1||_{0,K} \le  C_K$  is  satisfied  for  $t\in
[0,T]$.}

{\bf Proof.}  One has
$$
||A^{*n}||_{0,K} \le  b_K^{n-1}  ||A||_{0,K+2L}^{n-1}  ||A||_{0,K} \le
b_K^{n-1} ||A||^n_{0,K+2L}.
$$
Therefore,
$$
||*\exp At -1||_{0,K} \le \sum_{n=1}^{\infty}  \frac{1}{b_K}  \frac{(t
||A||_{0,K+2L} b_K)^n}{n!} \le
\frac{e^{t||A||_{0, K+2L} b_K} - 1}{b_K} \le C_K
$$
on $t \in [0.T]$. Lemma C.24 is proved.

{\bf Lemma C.25.} {\it  Let $A \in {\cal A}_M$, $M>0$. Then
$$
\sum_{m=N}^{\infty} \frac{A^{*m}}{m!} \in {\cal A}_{MN}.
$$
}

{\bf Proof.} One has
$$
\sum_{m=N}^{\infty} \frac{A^{*m}}{m!} = A^{*N} \left(
\frac{1}{N!} + \int_0^1 d\tau
\frac{(1-\tau)^{N-1}}{(N-1)!} (*\exp A\tau -1)
\right)
\l{k23}
$$
Lemma C.24 implies that
$$
\int_0^1 d\tau  \frac{(1-\tau)^{N-1}}{(N-1)!}  (*\exp  A\tau  -1)  \in
{\cal A}_0.
$$
It follows  from theorem C.21 that the symbol \r{k23} is of the ${\cal
A}_{NM}$-class. Lemma C.25 is proved.

{\bf Lemma C.26.} {\it Let $A_n \in  {\cal  A}_M$,  $M>0$  and
${\cal A}_M-\lim_{n\to\infty} A_n = A$. Then
$$
{\cal A}_{MN}-\lim_{n\to\infty}
\sum_{m=N}^{\infty} \frac{A_n^{*m}}{m!}
= \sum_{m=N}^{\infty} \frac{A^{*m}}{m!}.
$$
}

{\bf Proof.}   Because   relation  \r{k23}  and  theorem  C.21  it  is
sufficient to prove that
$$
{\cal A}_0-\lim_{n\to\infty}  \int_0^1  dt  \frac{(1-t)^{N-1}}{(N-1)!}
(*\exp A_n t - *\exp At) = 0.
\l{k24}
$$
One has
$$
*\exp A_nt - *\exp At = \int_0^t d\tau *\exp A(t-\tau) * (A_n-A) *\exp
A_n\tau.
$$
Making use of lemma C.23, we obtain then estimation \r{k24}.

\subsection{Estimations for the commutator}

Let $\hat{A}=f(\hat{x})$,  $\hat{B} = g(\hat{k})$.  To investigate the
properties of the commutator $\hat{K} =  [\hat{A};  \hat{B}]$,  it  is
convenient to  introduce  the notion of $\hat{x}\hat{k}$-symbol of the
operator instead       of       Weyl       symbol.       For       the
$\hat{x}\hat{k}$-quantization, the     operator     $e^{i\beta\hat{x}}
e^{i\alpha\hat{k}}$ corresponds to the function \r{k1}. Therefore, the
function
$$
A(x,k) =  \int  d\alpha  d\beta \tilde{A}(\alpha,\beta) e^{i\alpha k +
i\beta x}
$$
corresponds to the operator
$$
\hat{A} =  \int  d\alpha  d\beta  \tilde{A}(\alpha,\beta)
e^{i\beta \hat{x}} e^{i\alpha \hat{k}}
$$
For $\hat{x}  \hat{k}$-quantization,  the  *-product  defined from the
relations $\hat{C} = \hat{A} \hat{B}$, $C=A*B$ has the form \c{M1,KM}
$$
(A*B)(x,k) = A(x, k - i \frac{\partial}{\partial y}) B(y,k) |_{y=x}.
$$

{\bf Lemma C.27.} {\it
1. Let $A(x,k) = \varphi_1(x)  \varphi_2(k)$  with  bounded  functions
$\varphi_1$, $\varphi_2$. Then $||\hat{A}|| < \infty$. \\
2. Let $A \in L^2({\bf R}^{2d})$.  Then $\hat{A}$ is a Hilbert-Schmidt
operator.
}

{\bf Proof.}    1.    One     has     $\hat{A}=     \varphi_1(\hat{x})
\varphi_2(\hat{k})$, $||\hat{A}||      \le      ||\varphi_1(\hat{x})||
||\varphi_2(\hat{k})|| = \max |\varphi_1| \max |\varphi_2| < \infty$.\\
2. One has
$$
Tr A^+A = \frac{1}{(2\pi)^{2d}} \int dx dk |A(x,k)|^2 < \infty.
$$
The commutator $\hat{K} = [f(\hat{x}),  g(\hat{k})]$ has the following
$\hat{x}\hat{k}$-symbol:
$$
\begin{array}{c}
K(x,k) = [g(k) - g(k - i \frac{\partial}{\partial x}) ] f(x) =
\sum_{n=0}^L \frac{\partial^n   g}{\partial   k^{i_1}   ...   \partial
k^{i_n}} (-i)^n  \frac{\partial^n  f}{\partial  x^{i_1}  ...  \partial
x^{i_n}}  \\ -   \int_0^1   d\alpha   \frac{(1-\alpha)^L}{L!}   (-i)^{L+1}
\frac{\partial^{L+1} g(k-i \alpha \frac{\partial}{\partial x})}
{\partial k^{i_1}  ...  \partial  k^{i_{L+1}}  }  \frac{\partial^{L+1}
f}{\partial x^{i_1} ... \partial x^{i_{L+1}} }.
\end{array}
$$

{\bf Lemma C.28.} {\it Let
$C(x,k) = A(k-i\alpha \partial/\partial x) B(x) $. Then $||C||_{L^2} =
||A||_{L^2} ||B||_{L^2}$.
}

{\bf Proof.} Consider the Fourier transformation of the function $A$:
$$
A(k) = \int d\gamma \tilde{A}(\gamma) e^{i\gamma k}.
$$
One has $||A||_{L^2} = (2\pi)^{d/2} ||\tilde{A}||_{L^2}$ and
$$
C(x,k) = \int d\gamma \tilde{A}(\gamma) e^{i\gamma k}
e^{\gamma \alpha
\frac{\partial}{\partial x}} B(x).
$$
Since $e^{\gamma \alpha
\frac{\partial}{\partial x}} B(x) = B(x+\gamma\alpha)$, one has
$$
\begin{array}{c}
||C||_{L^2}^2 = \int dk dx d\gamma_1 d\gamma_2
\tilde{A}^*(\gamma_1) e^{-i\gamma_1k} B^*(x+\gamma_1\alpha)
\tilde{A}(\gamma_2) e^{i\gamma_2k} B(x+\gamma_2\alpha) = \\
(2\pi)^d \int  d\gamma  |\tilde{A}(\gamma)|^2  \int   dx   |B(x+\gamma
\alpha)|^2 = ||A||_{L^2}^2 ||B||_{L_2}^2.
\end{array}
$$
Lemma C.28 is proved.

We have obtained the following important statement.

{\bf Lemma C.29.} {\it Let
$\frac{\partial^n f}{\partial x^{i_1} ... \partial x^{i_n}}$,
$\frac{\partial^n g}{\partial   k^{i_1}   ...  \partial  k^{i_n}}$  be
bounded functions, $m,n=\overline{1,L}$, while
$$
\frac{\partial^{L+1} f}{\partial x^{i_1} ...  \partial  x^{i_{L+1}  }}
\in L^2, \qquad
\frac{\partial^{L+1} g}{\partial k^{i_1} ...  \partial  k^{i_{L+1}  }}
\in L^2.
$$
Then $[f(\hat{x}), g(\hat{k})]$ is a bounded operator.
}

\subsection{Asymptotic expansions of Weyl symbol }

To check the property of  Poincare  invariance,  it  is  important  to
investigate the large-$k$ expansion of the Weyl symbols. Introduce the
correponding definitions.

{\bf Definition C.4.} {\it 1.  We say that a smooth function $A(x,n)$,
$x,n \in  {\bf  R}^d$,  $|n|<1$,  is  of  the  calss  $\cal  L$ if the
functions
$$
\frac{\partial^I}{\partial n_{i_1} ...  \partial n_{i_I}} x_{j_1}  ...
x_{j_J} \frac{\partial^M}{\partial x_{m_1} ... \partial x_{m_M} }A
\l{k26}
$$
are bounded.\\
2. Let $A_s \in {\cal L}$, $s=\overline{1,\infty}$. We say that ${\cal
L}-\lim_{s\to\infty} A_s = 0$ if
$$
\sup_{|n|\le 1}
\lim_{s\to\infty}
\left| \frac{\partial^I}{\partial n_{i_1} ...  \partial n_{i_I}}
x_{j_1}  ... x_{j_J} \frac{\partial^M}{\partial x_{m_1}  ...
\partial  x_{m_M}  }A
\right| = 0.
$$
}

Definitions C.2 and C.4 imply the following statement.

{\bf Lemma  C.30.  }  {\it 1.  Let $A\in {\cal L}$.  Then the function
$B(x,k) = A(x,k/\omega_k)$ is of the class ${\cal B}_0$.\\
2. Let ${\cal L}-\lim_{s\to\infty} A_s = 0$. Then
${\cal B}_0-\lim_{s\to\infty} A_s(x,k/omega_k) = 0$.
}

Making use  of definition C.2 and lemma C.25,  we obtain the following
corollary.

{\bf Corollary.} {\it 1.  Let  $A\in  {\cal  L}$.  Then  the  function
$\omega_k^{-\alpha} A(x,k/\omega_k)$    is   of   the   class   ${\cal
A}_{\alpha}$.\\
2. Let ${\cal L}-\lim_{s\to\infty} A_s = 0$. Then
${\cal A}_{\alpha}-\lim_{s\to\infty} \omega_k^{-\alpha}
A_s(x,k/omega_k) = 0$.
}

{\bf Definition C.5.} {\it 1.  A formal asymptotic expansion is a  set
$\check{A}$ of  $\alpha  \in  {\bf R}$ and functions $A_0,A_1,...  \in
{\cal L}$.
We say that the formal asymptotic expansions
$\check{A} = (\alpha, A_0,A_1,..)$ and
$\check{B} = (\beta, B_0,B_1,..)$ are equivalent if
$\alpha-\beta$ is an integer number and $A_{l-\alpha+\beta} = B_l$ for
all $l=\overline{-\infty,+\infty}$  (we assume $A_l = 0$ and $B_l = 0$
for $l<0$. We denote formal asymptotic expansions of Weyl symbols as
$$
\check{A} \equiv   \sum_{n=0}^{\infty}   \omega_k^{-n-\alpha}   A_n(x,
k/\omega_k).
$$
If $A_0=0$,  ..., $A_{l-1}=0$, $A_l\ne 0$, the quantity $deg \check{A}
\equiv \alpha + n$ is called  as  a  degree  of  a  formal  asymptotic
expansion $\check{A}$.
\\
2. Let  $\check{A}_s$,  $s=\overline{1,\infty}$  and  $\check{A}$   be
formal asymptotic expansions of Weyl symbols. We say that
$F.E-\lim_{s\to\infty} \check{A}_s   =  A$  if  $\alpha_s=\alpha$  and
${\cal L}-\lim_{s\to\infty} (A_{s,n} - A_s) = 0$.
}

The summation and multiplication by numbers are obviously defined:
$$
\check{A}+ \lambda        \check{B}        =       \sum_{n=0}^{\infty}
\omega_k^{-n-\alpha} (A_n(x, k\omega_k) + \lambda B_n(x,k/\omega_k)).
$$
The product of formal asymptotic expansions of Weyl symbols
$$
\check{A} \equiv   \sum_{n=0}^{\infty}   \omega_k^{-n-\alpha}   A_n(x,
k/\omega_k),
\qquad
\check{B} \equiv   \sum_{n=0}^{\infty}   \omega_k^{-n-\beta}
B_n(x, k/\omega_k)
$$
is defined as
$$
\check{A}\check{B} \equiv                          \sum_{n=0}^{\infty}
\omega_k^{-n-\alpha-\beta} \sum_{s,l \ge 0; s+l = n}
A_s(x, k/\omega_k) B_l(x, k/\omega_k).
$$
Let $f=f(x)$, $f \in {\cal C}$. Then
$$
f(x) \check{A} \equiv
\sum_{n=0}^{\infty}   \omega_k^{-n-\alpha}
f(x) A_n(x,k/\omega_k).
$$
One also defines
$$
\omega_k^{-\beta} \check{A} \equiv
\sum_{n=0}^{\infty}   \omega_k^{-n-\alpha-\beta}
A_n(x,k/\omega_k)
$$
and
$$
\frac{\partial \check{A}}{\partial k_s} =
\sum_{l=0}^{\infty}   \omega_k^{-l-\alpha-1}
\left[
- (l+\alpha)   A_l(x,n)  +  \frac{\partial  A_l}{\partial  n_p}  (x,n)
(\delta_{ps}- n_pn_s) \right]|_{n=k/\omega_k}
$$
The *-product of formal asymptotic expansions is introduced as
$$
\begin{array}{c}
\check{A} * \check{B} \equiv \sum_{K=0}^{\infty} \sum_{n_1n_2  \ge  0,
n_1 + n_2 = K} \frac{i^{n_1-n_2} }{n_1! n_2! 2^{n_1+n_2}}
\frac{\partial^{n_1+n_2}}{\partial x^{i_1} ... \partial x^{i_{n_2}}
\partial k^{j_1} ... \partial k^{j_{n_1}} }
\sum_{l_1=0}^{\infty} \omega_k^{-l_1-\alpha_1} A_{l_1} (x, k/\omega_k)
\\ \times
\frac{\partial^{n_1+n_2}}{\partial x^{j_1} ... \partial x^{j_{n_1}}
\partial k^{i_1} ... \partial k^{i_{n_2}} }
\sum_{l_2=0}^{\infty} \omega_k^{-l_2-\alpha_2} A_{l_2} (x, k/\omega_k)
\end{array}
$$
The formal  asymptotic  expansions  $\check{A}  *  \omega_k^{\alpha}$,
$\check{A} * f(x)$ are defined analogously. The *-exponent of a formal
asymptotic expansion $\check{A}$ is defined as
$$
*\exp \check{A} - 1 = \sum_{n=1}^{\infty} \frac{\check{A}^{*n}}{n!}
$$
provided that $deg A$ is a positive integer number.

{\bf Definition C.6.} {\it
1. An asymptotic expansion of the Weyl symbol is a set  $\underline{A}
\equiv (A,\check{A})$  of  the Weyl symbol $A$ and a formal asymptotic
expansion $\check{A}$ such that
$$
A(x,k) - \sum_{l=0}^{n-1}
\frac{A_l(x,k/\omega_k)}{\omega_k^{l+\alpha}} \in {\cal A}_{n+\alpha}
$$
for all $n=\overline{0,\infty}$.\\
2. We say that $E-\lim_{s\to\infty} \underline{A_s} = \underline{A}$ if
$F.E-\lim_{s\to\infty} \check{A}_s = \check{A}$ and
$$
{\cal A}_{n+\alpha} -\lim_{s\to\infty}
( A_s(x,k) - \sum_{l=0}^{n-1}
\frac{A_{s,l}(x,k/\omega_k)}{\omega_k^{l+\alpha}})
=
A(x,k) - \sum_{l=0}^{n-1}
\frac{A_l(x,k/\omega_k)}{\omega_k^{l+\alpha}}
$$
for all $n=\overline{0,\infty}$.
}

{\bf Remark.} For given Weyl symbol $A$,  the asymptotic expansion  is
not unique. For example, let
$$
A(x,k) = m^2f(x)/\omega_k.
$$
One can choose $\alpha=2$, $A_0(x,n) = m^2f(x)$ anf find
$A(x,k) = \omega_k^{-2} A_0(x,k/\omega_k)$. On the other hand, one can
set $\alpha=0$,  $A_0(x,n)  =  f(x)  (1-n_in_i)$  and obtain $A(x,k) =
A_0(x,k/\omega_k)$ since $\omega_k^2- k_ik_i = m^2$.  We  see  that  a
degree is  a  characteristic  feature of an expansion rather than of a
symbol.

Let $\underline{A}=  (A,\check{A})$,  $\underline{B} = (B,\check{B})$.
Denote
$\underline{A}  *   \underline{B}   \equiv   (A*B,   \check{A}*
\check{B})$,\\
$\omega^{\alpha}_k * \underline{A} \equiv   (\omega_k^{\alpha} * A,
\omega_k^{\alpha} * \check{A})$,\\
$f(x)* \underline{A}   \equiv   (f(x) * A,  f(x) *   \check{A} )$,\\
$*\exp \underline{A} - 1 \equiv (*\exp A - 1, * \exp \check{A} - 1)$.

Theorems C.14, C.18, C.21 and lemmas C.25 and C.26 imply the following
statements.

{\bf Theorem C.31.} {\it
1. Let $\underline{A}$be an asymptotic expansion  of  a  Weyl  symbol.
Then $\omega^{\alpha}_k  *  \underline{A}$  and $f(x) * \underline{A}$
are asymptotic expansions of Weyl symbols under conditions of  theorem
C.14, while  $*\exp \underline{A} - 1$ is an asymptotic expansion of a
Weyl symbol,  provided that $deg  \check{A}$  is  a  positive  integer
number.\\
2. Let $\underline{A}$ and $\underline{B}$ be asymptotic expansions of
Weyl symbols.  Then  $\underline{A}  * \underline{B}$ is an asymptotic
expansion af a Weyl symbol.
}

{\bf Theorem C.32.} {\it
1. Let $E-\lim_{n\to\infty} \underline{A}_n = \underline{A}$. Then: \\
(a) $E-\lim_{n\to\infty} \omega_k^{\alpha} * \underline{A}_n
\omega_k^{\alpha} * \underline{A}$; \\
(b) $E-\lim_{n\to\infty} f(x) * \underline{A}_n
f(x) * \underline{A}$ under conditions of theorem C.14; \\
(c) $E-\lim{n\to\infty}   (*\exp   \underline{A_n}   -  1)  =  *  \exp
\underline{A} -1$ if $deg \check{A}_n$,  $deg \check{A}$ are  positive
integer numbers.\\
2. Let $E-\lim_{n\to\infty} \underline{A}_n = \underline{A}$ and
$E-\lim_{n\to\infty} \underline{B}_n = \underline{B}$. Then
$E-\lim_{n\to\infty} \underline{A}_n * \underline{B}_n =  \underline{A}
* \underline{B}$.
}

The time derivative of  teh  asymptotic  expansion  $\underline{A}(t)$
with respect to $t$ is defined in a standard way
$$
E-\lim_{\delta t \to 0 } \frac{\underline{A}(t+\delta t) -
\underline{A}(t)}{\delta t} = \frac{d\underline{A}(t)}{dt}.
$$
The integral $\int_{t_1}^{t_2} \underline{A}(t) dt$ is also defined in
a standard way.

Theorem C.32 imply the following statement.

{\bf Theorem  C.33.}  {\it 1.  Let $\underline{A}(t)$ be a continously
differentiable asymptotic expansion of a Weyl symbol. Then \\
(a)
$\frac{d}{dt} (\omega_k^{\alpha} * \underline{A}) = \omega_k^{\alpha} *
\frac{d\underline{A}}{dt}$; \\
(b)
$\frac{d}{dt} (f(x) * \underline{A}) = f(x) *
\frac{d\underline{A}}{dt}$ under conditions of theorem C.14. \\
(c)
$\frac{d}{dt} (*\exp   \underline{A}   -   1)   =    \int_0^1    d\tau
e^{\underline{A} (t-\tau)}      *      \frac{d\underline{A}}{dt}     *
e^{\underline{A} \tau}$;\\
(d) $\frac{d}{dt}      (\underline{A}      *      \underline{B})     =
\frac{d}{dt}\underline{A} *   \underline{B}    +    \underline{A}    *
\frac{d}{dt}\underline{B}$.
}

The only nontrivial statement is  (c).  It  is  proved  by  using  the
identity \c{KM}
$$
*\exp \underline{A}_1  -  *\exp \underline{A}_2 = \int_0^1 d\tau *\exp
(\underline{A}_1(1-\tau)) *  (\underline{A}_1  -  \underline{A}_2)   *
\exp(\underline{A}_2\tau).
$$

\newpage
\pagestyle{empty}

\end{document}